\documentclass[aps,prx,twocolumn,superscriptaddress,raggedbottom,showpacs,floatfix,longbibliography]{revtex4-1}
\usepackage{amsmath,amsfonts,amssymb,amsthm,graphics}
\usepackage[next]{inputenc}
\usepackage[colorlinks=true,citecolor=blue,linkcolor=blue]{hyperref}
\usepackage{bbm}
\usepackage{booktabs}
\usepackage{multirow}
\usepackage{hhline}
\usepackage{comment}
\usepackage{float}
\usepackage{latexsym}
\usepackage[dvipsnames]{xcolor}
\usepackage{graphicx}
\usepackage{epstopdf}
\usepackage{bm}
\usepackage{verbatim}
\usepackage{physics}
\usepackage{mathrsfs}
\usepackage{tikz}
\usepackage{chemfig}
\usepackage{pifont}
\usepackage{manfnt}
\usepackage{ulem}
\usepackage{mathtools}

\usepgflibrary{patterns} 
\usepgflibrary[patterns] 
\usetikzlibrary{patterns} 
\usetikzlibrary[patterns]
\usetikzlibrary{shapes,shadows,arrows,positioning,graphs}

\newcommand{\cmark}{\ding{51}}%
\newcommand{\xmark}{\ding{55}}%
\newcommand{\hexagona}{
 \raisebox{-7pt}{\includegraphics[height=4.5ex]{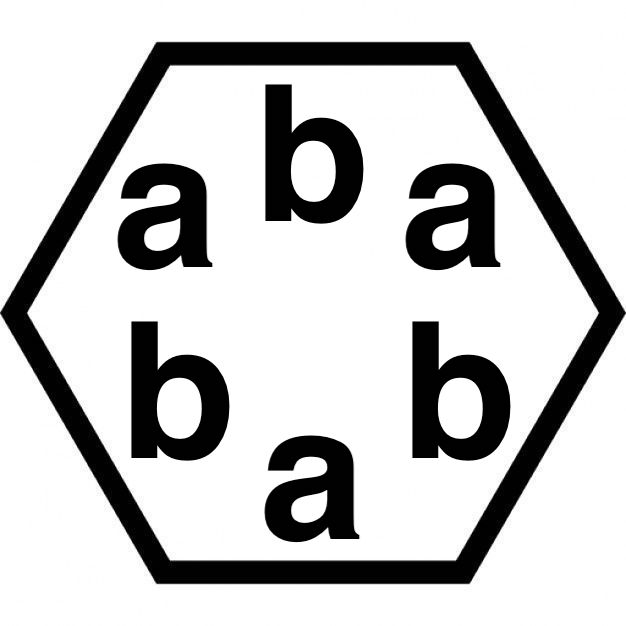}}}

\newcommand{\hexagonb}{
   \raisebox{-7pt}{\includegraphics[height=4.5ex]{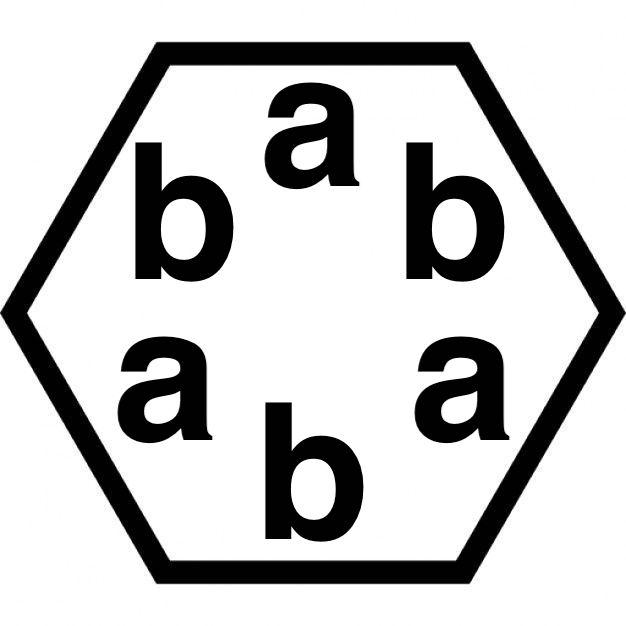}}}

\newcommand{\hexagon}{\mathord{\raisebox{-1pt}{\tikz{\node[draw,scale=.65,regular polygon, regular polygon sides=6,fill=none](){};}}}}
\newcommand{\octagon}{\mathord{\raisebox{-1pt}{\tikz{\node[draw,scale=.65,regular polygon, regular polygon sides=8,fill=none](){};}}}}

\begin{document}

\title{Classical vertex model dualities in a family of 2D frustrated quantum antiferromagnets}
\author{Shankar Balasubramanian}
\affiliation{Center for Theoretical Physics, Massachusetts Institute of Technology, Cambridge, MA 02139, USA}
\author{Victor Galitski}
\affiliation{Joint Quantum Institute and Condensed Matter Theory Center, Department of Physics, University of Maryland, College Park, Maryland 20742-4111, USA}
\author{Ashvin Vishwanath}
\affiliation{Department of Physics, Harvard University, Cambridge, MA 02138, USA}

\begin{abstract}

We study a general class of easy-axis spin models on a lattice of corner sharing even-sided polygons with all-to-all interactions within a plaquette. The low energy description corresponds to a quantum dimer model on a dual lattice of even coordination number with a multi dimer constraint.  At an appropriately constructed frustration-free Rokhsar-Kivelson (RK) point, the ground state wavefunction can be exactly mapped onto a classical vertex model on the dual lattice.  When the dual lattice is bipartite, the vertex models are bonded and are self dual under Wegner's duality, with the self dual point corresponding to the RK point of the original multi-dimer model.  We argue that the self dual point is a critical point based on known exact solutions to some of the vertex models.  When the dual lattice is non-bipartite, the vertex model is arrowed, and we use numerical methods to argue that there is no phase transition as a function of the vertex weights.  Motivated by these wavefunction dualities, we construct two other distinct families of frustration-free Hamiltonians whose ground states can be mapped onto these vertex models.  Many of these RK Hamiltonians provably host $\mathbb{Z}_2$ topologically ordered phases.

\end{abstract}

\maketitle

\section{Introduction \label{sec:introduction}}

Quantum spin models host a variety of strongly correlated phases of matter, from ordered states to featureless spin liquid phases.  The latter has turned into an active area of research over the past several  decades.  The simplest model of a gapped spin liquid is the resonating valence bond (RVB) state  proposed by Anderson \cite{Anderson1, Anderson2}. Further developments introduced the quantum dimer model, in which the singlet bond between pairs of spins was reduced to elementary dimer variables. The quantum dimer model on the square lattice was shown by Rokhsar and Kivelson to have an exactly solvable point (the RK point) \cite{RK, SRK}.  At such a point, the ground state is expressible as a uniform superpositions of dimer coverings.  Therefore, observables and correlations at the RK point are equivalent to those in a corresponding classical dimer model.  The combinatorics of classical dimers is a well-established problem in statistical mechanics: it was shown by Kasteleyn \cite{Kasteleyn} and Temperley and Fisher \cite{TemperleyFisher, Fisher, FS2} that the number of dimer coverings of a planar lattice can be written as a Pfaffian of an antisymmetric matrix.  Expressing the Pfaffian as a free fermion path integral, dimer-dimer correlation functions correspond to correlations in the free fermion theory.  For bipartite lattices, the fermions are massless, indicating power-law correlations, while for non-bipartite lattices, the free fermions are massive and correlations decay exponentially. 

An important distinction between quantum antiferromagnets on bipartite and non-bipartite lattices was highlighted in Refs. \cite{RS, FradkinKivelson, RSPRB90, RC89, JalabertSachdev91,  fradkin_2013, Sachdev2000,Wen91}. In the dimer limit, the former reduces to a U(1) gauge theory, which in its simplest form (except for special points)
is confining in 2+1D.  However, gapped spin liquids can occur in 2+1D quantum antiferromagnets on non-bipartite lattices \cite{Sachdev92}; explicit examples of this include the RVB phase of the quantum dimer model on the triangular   \cite{MS, FMS} and the Kagome   
\cite{MSP} lattices.  The concept of an RK ground state was further generalized by Henley to describe any classical statistical mechanics model satisfying detailed balance \cite{Henley}. 

\begin{table*}
 \begin{tabular}{||c | c| c| c||} 
 \hline
 Model description & Related integrable models & Spin Liquid? & Other notes \\ [0.5ex]
 \hline\hline
 Square (Section \ref{examplesq}) & 6- and 8-vertex & \xmark & special case of \cite{AFF}\\ 
 \hline
 Honeycomb (Section \ref{examplehoney}) & 20- and 32-vertex & \xmark & \\
 \hline
 BFG family (Section \ref{otherexamples}) & -- & \cmark & likely top. ordered\\
 \hline
 BFG spin-doubled family (Section \ref{otherexamples}) & -- & \xmark & \\
 \hline
 Ruby family (Section \ref{otherexamples}) & -- & \cmark & likely top. ordered\\
 \hline
 Ruby spin-doubled family (Section \ref{otherexamples}) & -- & \xmark & need extra ring exchanges\\
 \hline
 Quantum bond vertex model (Section \ref{pertvertex}) & 32-vertex & \cmark & both phases possible\\
 \hline
 Quantum arrow vertex model (Section \ref{pertvertex}) & -- & \cmark & only top. ordered phase\\
 \hline
\end{tabular}
\caption{A summary of some of the results; more such models can be constructed, which is further described in the text.}
\end{table*}

While dimer models were initially introduced to represent the singlet formation between spins on the lattice \cite{Anderson1, Anderson2}, which is made concrete using large-N generalizations of spin models \cite{RS}, a different viewpoint was put forward in \cite{FazekasAnderson}, where dimers represent the low energy manifold of a frustrated Ising model, and residual spin coupling generate quantum dynamics within this manifold. Here we will have such an origin in mind. In Ref. \cite{BFG}, Balents, Fisher, and Girvin (BFG) introduced a quantum spin model defined on a Kagome lattice (of corner-sharing hexagons) with further neighbor interactions (see \cite{MotrunichSenthil2D} for related `charging energy' models).  In the easy axis limit, the BFG model can be reduced to a dimer model on a triangular lattice with a three dimer per site constraint.  In this case, the dimer subspace arises from including Ising interactions between spins that takes the form of a ``charging energy'' on hexagonal plaquettes.  The fractionalized spinon and vison excitations were identified and numerics showed that the ground state features exponentially decaying correlations, strongly implying that the BFG model has a $\mathbb{Z}_2$ spin liquid phase \cite{SB}. 

\begin{figure}
    \centering
    \includegraphics[scale=0.225]{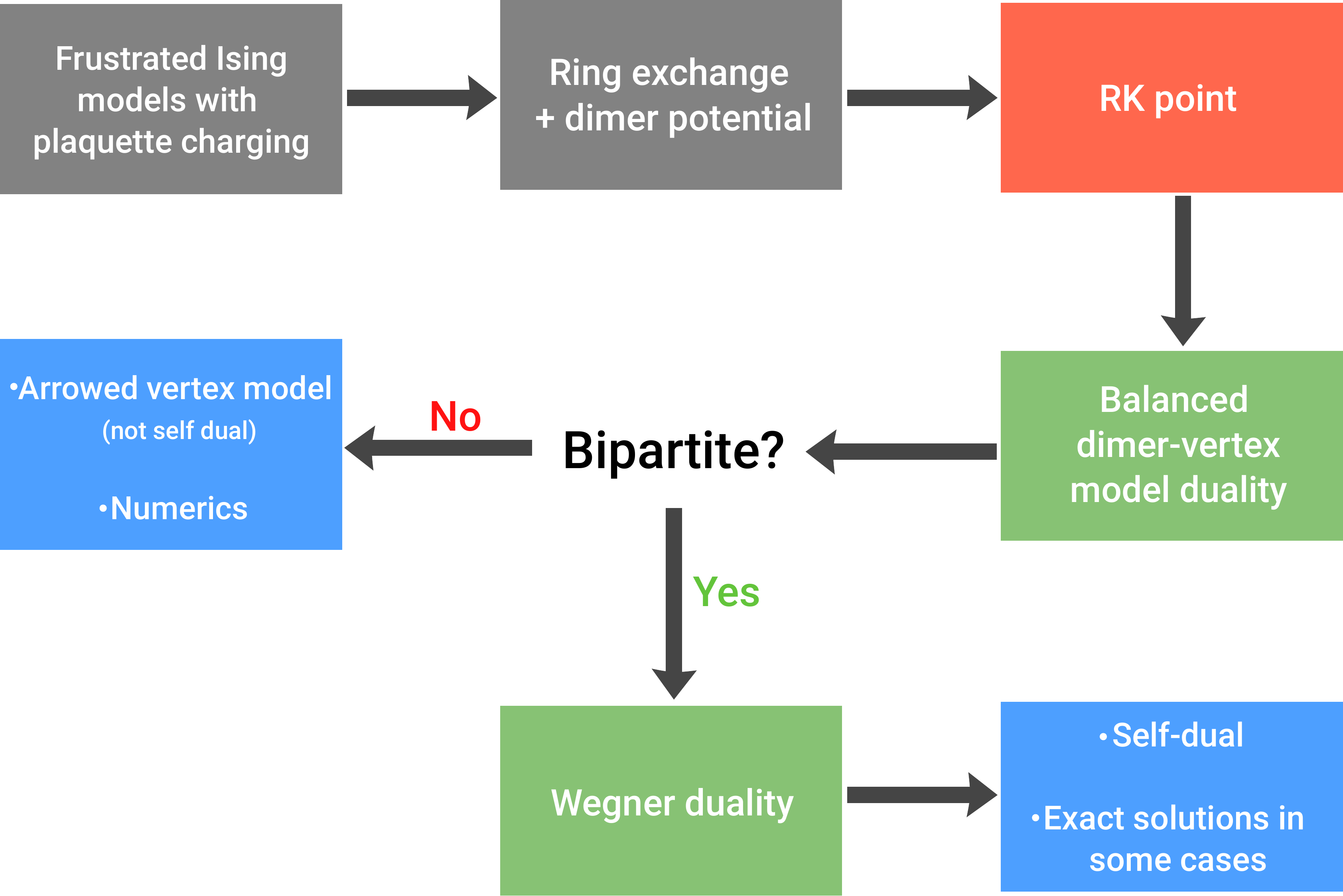}
    \caption{A schematic showing the flow of results in the paper.}
    \label{fig:layout}
\end{figure}

This paper considers a family of quantum spin models on lattices of corner sharing polygons in two dimensions, which are equivalent to dimer models with a multiple dimers per site constraint (see Figure \ref{fig:layout} for a layout) that generalize the BFG construction in various exactly solvable ways.  For such constraints, the usual Kasteleyn method for calculating various combinatorial properties of dimer coverings \emph{does not apply} \cite{Kasteleyn, Fisher, FS2, S1, S2, S3, FMS, FHMOS}. 
We first focus on an RK wavefunction which is a superposition of classical spin configurations with zero net magnetization around each polygon.  We use a Villain-type duality (which we call the balanced dimer-vertex model (BDVM) duality) to show that these classical statistical mechanics models are equivalent to arrowed vertex models on a dual lattice.  If the dual lattice is bipartite, we use Wegner's duality \cite{Wegner} to show that the vertex model is at a self dual point.  This loosely follows from the fact that the arrowed and bonded vertex models are equivalent for bipartite lattices.  We show that several of these self dual models are critical points of well-known integrable statistical mechanics models.  We discuss how to compute spin-spin correlators in the original RK wavefunctions by mapping $n$-point correlation functions to string correlation functions in the dual vertex models -- this allows for determination of the exact scaling behavior of certain correlation functions if the critical exponents in the dual vertex model are known.  The duality mappings we discuss bear no relation to the well-known height model representations of classical dimer models and the classical 6-vertex model \cite{henleyheight}. 

If the dual lattice is not bipartite, the arrowed vertex model can no longer be changed to a bonded vertex model.  As a result, Wegner's duality is no longer a self duality and the vertex model does not exhibit a self dual point.  Monte Carlo simulations indicate that the arrowed vertex models do not have a phase transition and that the vertex model is always in a topologically ordered phase.  We conjecture that this result holds for any non-bipartite lattice.

We use these duality results from classical statistical mechanics to construct several classes of parent Hamiltonians corresponding to these vertex models for general values of the vertex weights.  The first class of Hamiltonians is a frustrated antiferromagnetic model on corner-sharing polygons (see Table 1 for some examples) expanded in the easy axis limit, which are discussed in Section~\ref{examplesection}.  The dominant Ising couplings select classical spin configurations that satisfy the zero magnetization constraint on each polygon. Low energy processes within this manifold are ring exchanges, and an exact RK point can be constructed upon adding a ``dimer potential energy'' term.  In these models, the RK ground state corresponds to uniform superpositions of classical spin configurations satisfying the zero magnetization constraint on each polygon.  These models can be defined on both bipartite and non-bipartite dual lattices; the former is made possible with a ``spin-doubling'' trick discussed in the text. 

The second class of parent Hamiltonians we construct have RK wavefunction ground states that are exactly equivalent to bond vertex models on the dual lattice, which we discuss in Section~\ref{pertvertex}.  An example of such a model on the square lattice was constructed by Ardonne, Fendley, and Fradkin (AFF) \cite{AFF} and a different kind of 8-vertex model was constructed in \cite{Chakravarty}.  The Hamiltonians we construct can be considered to be generalizations of the AFF models on arbitrary planar lattices, whose phase diagrams can be inferred exactly through the Wegner dualities.  These models reduce to a commuting projector toric code Hamiltonian at a particular point, which indicates the existence of a topologically ordered phase.  We also construct parent Hamiltonians corresponding to the arrowed vertex models.  These models also possess an exactly solvable toric code-like point but do not exhibit a symmetry breaking transition unlike for the bonded vertex models.  However, some of the arrowed vertex models exhibit a hidden gapless point equivalent to a quantum coloring model (see Fig.~\ref{fig:kagomediagram} for details).

For ease of presentation, we devote Appendix~\ref{offcriticalHam} to the construction of the third class of parent Hamiltonians.  The RK wavefunction of these Hamiltonians are equivalent to a classical higher spin model which can be shown to exactly map onto an arbitrary point on the vertex model phase diagram.  Since this point in general is perturbed away from the self-dual point, the RK wavefunctions must be gapped.  These wavefunctions are then shown to be ground states of frustration-free parent Hamiltonians which correspond to multiple sheets of 2D easy axis antiferromagnets coupled together by particularly chosen XY exchanges.  We illustrate these explicit constructions when the dual vertex model is defined on a triangular lattice, where one can obtain exact results, but emphasize that our results can easily be generalized to any planar lattice.

\section{Duality mappings to vertex models for generalized RK 
wavefunctions}

We first start by considering a class of statistical mechanics models which will generate a Rokhsar-Kivelson wavefunction.  At the end of this section, we will discuss natural parent Hamiltonians which have such ground states.  Consider a lattice of {\it even}-sided corner sharing polygons $\{p\}$.  These polygons can be of mixed types in principle, and we will introduce examples of such models in Section \ref{examplesection}.  By corner sharing, we mean that no two polygons can share more than one corner and all corners of a polygon are shared.  Additionally, we place spins on each of the corners, and impose the constraint that $\sum_{i \in p} s_i = 0$.  The RK wavefunction is a uniform superposition over such configurations: $|\text{RK}\rangle \propto \sum_{c \in C} |c\rangle$.

We then define a \emph{dual lattice} $G_D$ to $G$ by allowing the sites of $G_D$ to be the centers of the polygons in $G$ and by connecting two sites of $G_D$ if the corresponding polygons in $G$ share a corner. Note, by the previous condition that the corner sharing polygons are even sided, $G_D$ must have {\em even} coordination number.  Further, the constraint described above can then be interpreted as a {\it balanced dimer constraint}: associating a dimer on $G_D$ with spin $s_i = +1$ on $G$ (and no dimer with $s_i = -1$), the constraint on spins $\sum_{i \in p} s_i = 0$ in a polygon implies that the number of dimers equals the number of non-dimers at a given site.  We will now consider the cases where $G_D$ is bipartite and non-bipartite separately.

\subsection{Bipartite $G_D$ and self-duality}

If we assume that $G_D$ is bipartite, then we may perform a duality which we call the {\it balanced dimer-vertex model duality} (BDVM duality) to map the partition function to that of a vertex model.  Then, we may use a duality similar to Wegner's weak graph transformation to show that this vertex model is at a self-dual point \cite{Wegner}.  The procedure is outlined below:

\subsubsection{Balanced dimer-vertex model (BDVM) duality}

\begin{figure}
    \centering
    \includegraphics[scale=0.7]{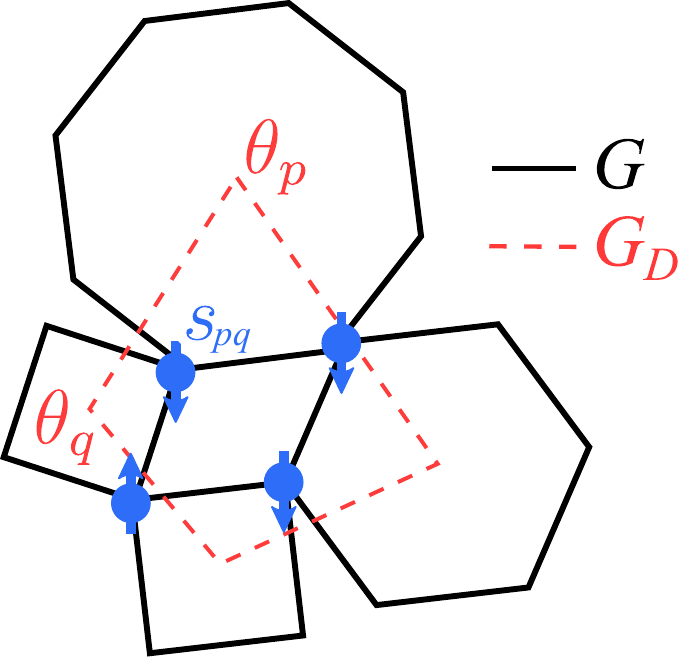}
    \caption{An example of four corner sharing polygons.  The spins are illustrated in blue, and the sites of the dual lattices are indicate in red and are on the centers of the polygon.  Part of the dual lattice is drawn with dashed lines.  Each spin corresponds to an edge in $G_D$.}
    \label{fig:dualitydiagram}
\end{figure}

We will derive a dual vertex model for the RK ground state wavefunction similar to the Villain duality for the two dimensional classical XY model.  Call $N$ the total number of sites and $n$ the total number of polygons.  We first note that the total number of valid configurations of spins in $G$ can be written as
\begin{equation}
    Z = \sum_{s \in \{-1,1\}^N} \prod_p \delta \left(\sum_{i \in p} s_i = 0\right),
\end{equation}
which is alternatively equal to the normalization of the RK wavefunction if each of the amplitudes are set to 1.  Here, $\{-1,1\}^N$ denotes a vector of $N$ $\pm 1$ entries.  Next, we write the $\delta$-function as a Fourier transform
\begin{equation}
    \delta \left(\sum_{i \in p} s_i = 0\right) = \frac{1}{2\pi}\int_0^{2\pi} d\theta_p\, e^{i \theta_p \sum_{i \in p} s_i},
\end{equation}
so that
\begin{equation}
    Z = \frac{1}{(2\pi)^n}\int_0^{2\pi} d^n\theta\sum_{s \in \{-1,1\}^N} \prod_p e^{i \theta_p \sum_{i \in p} s_i}.
\end{equation}
Next, we note that because two polygons share a corner (where spin $s_i$ is located), the following identity holds:
\begin{equation} \label{eqn:graphid}
    \sum_{s \in \{-1,1\}^N} \prod_p e^{i \theta_p \sum_{i \in p} s_i} = \prod_{(p,q)\in G_D} 2 \cos(\theta_p + \theta_q),
\end{equation}

where the notation $(p,q) \in G_D$ means that nodes $p$ and $q$ form an edge in $G_D$ (which has $n$ sites).  One may refer to Figure \ref{fig:dualitydiagram}.  Therefore, we may write the partition function as
\begin{equation}
    Z = \frac{2^N}{(2\pi)^n}\int_0^{2\pi} d^n\theta\prod_{(p,q)\in G_D} \cos(\theta_p + \theta_q).
\end{equation}
Next, we use the fact that $G_D$ is {\it bipartite} to perform a change of variables $\theta_p^A \to \theta_p^A$ and $\theta_p^B \to -\theta_p^B$ where the superscripts indicate the $A$ and $B$ sublattices.  This results in $\cos(\theta_p + \theta_q) \to \cos(\theta_p - \theta_q)$, and
\begin{equation}
    Z = \frac{2^N}{(2\pi)^n}\int_0^{2\pi} d^n\theta\prod_{(p,q)\in G_D} (\cos \theta_p  \cos \theta_q + \sin \theta_p  \sin \theta_q).
\end{equation}
Note, as a result of the bipartite nature of $G_D$, this model enjoys a U(1) global symmetry, corresponding to uniform rotation of the angles $\theta_p \rightarrow \theta_p+\epsilon$. 
We may re-express this partition function in terms of a vertex model.  On $G_D$, associate a bond with the two endpoints being cosines and no bond with the two endpoints being sines. Performing the product over all edges in $G_D$ in the integrand, one of the terms in the resulting sum can be written as
\begin{equation}
    Z(k_1, \cdots, k_n) = \frac{2^N}{(2\pi)^n} \int_0^{2\pi} d^n\theta \prod_{p \in G_D} \cos^{k_p} \theta_p \sin^{V_p- k_p} \theta_p , 
\end{equation}
where $Z = \sum_{k_1,\cdots, k_n} Z(k_1, k_2, \cdots, k_n)$ and $k_p$ runs from $0$ to $V_p$, where $V_p$ is the number of vertices of polygon $p$. Defining 
\begin{equation}\label{eq:vweights}
    W_p(k) = \langle \cos^k \theta \sin^{V_p-k} \theta \rangle,
\end{equation}
we find that
\begin{equation}
    Z = \sum_{k_1, k_2,\cdots, k_n} \prod_p W_p(k_p),
\end{equation}
which is a vertex model.  

For example, for a regular lattice of corner sharing square polygons, $V_p = 4$ and $W(0) = W(4) = 3/8$ while $W(2) = 1/8$ and $W(1) = W(3) = 0$.  This is a familiar 8-vertex model, which possesses an exact solution due to Baxter \cite{Baxter8V, BaxterBook}.  In general, we find that $W_p(n) = 0$ for $n$ odd.  We call this mapping the BDVM duality.

\subsubsection{Self-duality via Wegner duality}

It turns out that this choice of vertex weights leads to an additional property that the corresponding vertex model is at self-duality. To do this we need two steps - first extend the parameters of the vertex model, i.e. the vertex weights beyond the particular values obtained previously by the BDVM duality. Later, we will explore parent Hamiltonians for these generalized vertex models. Second,  we apply {\it another duality}, which is a specialized case of the weak graph transformation first introduced by Wegner \cite{Runnels, Wegner, WuWu} that apply to vertex models.  To start, consider a vertex model on the dual lattice $G_D$.  The spin variables $s_{pq}$ are still located on the edges of $G_D$ (i.e. the sites of the original lattice $G$, referring to Figure \ref{fig:dualitydiagram}), and we will assume that the vertex weights are only a function of the total number of bonds $b$ meeting at the sites of the dual lattice $G_D$ and are labelled as $W(b)$.  If $s_{pq} = 1$, then $(p,q)$ forms an occupied bond on the vertex model.  Then, the partition function can be written as
\begin{equation}
    Z = \sum_{\vec{s} \in \{-1,1\}^N} \prod_{p\in G_D} \sum_{b=0}^{V_p} \delta\left(S_{\text{tot}}(p) = 2b - V_p\right) W_p(b),
\end{equation}
where $S_{\text{tot}}(p) = \sum_{(p,q) \in G_D} s_{pq}$.  As before, we perform a Fourier transform of the $\delta$-functions
\begin{equation}
    \delta \left(\sum_{i \in p} s_i = c\right) = \frac{1}{2\pi}\int_0^{2\pi} d\theta_p\, e^{i \theta_p \left(\sum_{i \in p} s_i - c\right)},
\end{equation}
and use the identity from Equation \ref{eqn:graphid} to obtain
\begin{align}
    Z = \frac{1}{(2\pi)^n}\int_0^{2\pi} d^n\theta\sum_{\vec{s} \in \{-1,1\}^N} \prod_{(q,\ell)\in G_D} e^{i s_{q\ell} (\theta_q + \theta_\ell)} \nonumber\\ \times \prod_{p\in G_D} \sum_{b=0}^{V_p} e^{-i \theta_p (2b - V_p)} W_p(b).
\end{align}
Performing the sum over spins results in
\begin{align}
    Z = \frac{2^N}{(2\pi)^n}\int_0^{2\pi} d^n\theta \prod_{(q,\ell)\in G_D} \cos (\theta_q + \theta_\ell) \nonumber\\ \times \prod_{p\in G_D} \sum_{b=0}^{V_p} e^{-i \theta_p (2b - V_p)} W_p(b).
\end{align}
Next, we convert the product of cosines into a vertex model \emph{without} performing a sublattice change of variables.  This results in modified vertex rules whereby a bond corresponds to having cosines on either side of it while no bond corresponds to having $i$ times sine on either side of it.  Expanding the cosines and taking the product as before maps the partition function to
\begin{widetext}
\begin{equation}
    Z = \sum_{k_1, k_2,\cdots, k_n} \frac{2^N}{(2\pi)^n} \int_0^{2\pi} d^n\theta \prod_{p \in G_D} \sum_{b=0}^{V_p} i^{V_p-k_p} \cos^{k_p} \theta_p \sin^{V_p - k_p} \theta_p  e^{-i \theta_p (2b - V_p)} W_p(b).
\end{equation}
\end{widetext}
And thus we may again write
\begin{equation}
    Z = \sum_{k_1, k_2,\cdots, k_n} \prod_p W'_p(k_p),
\end{equation}
where
\begin{equation}\label{eqn:selfdualrel}
    W'_p(k_p) = \sum_{b=0}^{V_p} i^{V_p-k_p} \left \langle \cos^{k_p} \theta_p \sin^{V_p- k_p} \theta_p  e^{-i \theta_p (2b - V_p)} \right \rangle W_p(b).
\end{equation}
This defines a linear map $\mathcal{M}\colon W_p \to W'_p$ on the space of vertex weights to a dual space of vertex weights that leaves the partition function invariant.  The matrix $\mathcal{M}$ has elements:
\begin{equation}
    \mathcal{M}_{ab} = i^{V_p-a} \left \langle \cos^{a} \theta_p \sin^{V_p- a} \theta_p  e^{-i \theta_p (2b - V_p)} \right \rangle.
\end{equation}
$\mathcal{M}$ can be divided into disjoint eigenspaces corresponding to each of its distinct eigenvalues.  Any configuration of vertex weights which lives entirely in a given eigenspace will remain in the eigenspace under the application of $\mathcal{M}$.  Therefore, these eigenspaces define a self-dual manifold; if a parameterization of a vertex model pierces the self-dual manifold and exhibits a single phase transition, then the transition point occurs at the intersection with the self-dual manifold.  In the current context, we have analytically verified (up to $V_p=50$) that $W_p(k) = \langle \cos^k \theta \sin^{V_p-k} \theta \rangle$ is a vector residing in the self-dual manifold.  The duality transformation also maps the low temperature vertex models (where $W_p(0) = W_p(V_p) = 1$ and all other weights are zero) to the high temperature vertex models (where all weights equal $1$) and vice versa.  This is suggestive that this duality transformation maps between two phases, and indicates that self duality indicates the phase transition point.

An example of $\mathcal{M}_{ab}$ for $V_p=4$ (see figure \ref{fig:corrvert} and \ref{fig:hintmer} for the definition of this model) is
\begin{equation}
    \mathcal{M}_{ab} = \begin{pmatrix}
\frac{1}{4} & -1 & \frac{3}{2} & -1 & \frac{1}{4}\\
 -\frac{1}{4} & \frac{1}{2} & 0 & -\frac{1}{2} & \frac{1}{4}\\
 \frac{1}{4} & 0 & -\frac{1}{2} & 0 & \frac{1}{4}\\
 -\frac{1}{4} & -\frac{1}{2} & 0 & \frac{1}{2} & \frac{1}{4}\\
 \frac{1}{4} & 1 & \frac{3}{2} & 1 & \frac{1}{4}\\
\end{pmatrix},
\end{equation}
and explicitly, it can be seen that the eigenvalues of $\mathcal{M}_{ab}$ are $\pm 1$. The eigenspaces corresponding to these eigenvalues are
\begin{equation}
    \mathcal{V}_{-1} = \text{span}\left\{ \begin{pmatrix} 1\\0\\-1\\0\\1
    \end{pmatrix}, \begin{pmatrix} 4\\1\\-2\\1\\0
    \end{pmatrix}\right\}
\end{equation}
and 
\begin{equation}
    \mathcal{V}_{1} = \text{span}\left\{ \begin{pmatrix} -1\\1\\0\\0\\1
    \end{pmatrix}, \begin{pmatrix} 0\\-1\\0\\1\\0
    \end{pmatrix}, \begin{pmatrix} 6\\-3\\1\\0\\0
    \end{pmatrix}\right\}.
\end{equation}
The vector $W_p(k) = \langle \cos^k \theta \sin^{V_p-k} \theta \rangle$, or $\langle 3,0,1,0,3 \rangle$ explicitly, lies in the $\mathcal{V}_1$ eigenspace.

Later, we show that the phase transition is \emph{second order in some cases} by mapping the critical vertex models to spin models which have exact solutions.  A second order phase transition is a signature of a gapless system, which rules out the possibility that this RK point belongs to a gapped spin liquid phase.  In other cases, we cannot rule out that the phase transition is first order \cite{MJF} nor can we rule out the possibility of having multiple phase transitions.  Since the correlation length is generically finite at a first order phase transition, the quantum spin model should be gapped; if there are multiple phase transitions, self-duality can indicate that the model is in the middle phase, but more work would be required to characterize the intermediate phases.  

Finally, we note that although we restrict the dual lattice to be bipartite, we will show that the parent Hamiltonian generating such an RK wavefunction  will generically have frustrated interactions such as the charging term easy axis interactions.  

\subsection{Non-bipartite Dual Lattice $G_D$}

In the previous section, we assumed that the dual graph $G_D$ is bipartite.  If $G_D$ is not bipartite, there are two equivalent statistical mechanics models that one may write:
\begin{itemize}
    \item Bond vertex model: Since $G_D$ is non-bipartite, then we cannot perform the sublattice transformation $\theta_p^A \to \theta_p^A$ and $\theta_p^B \to -\theta_p^B$; therefore, the original vertex model suffers from a sign problem.  The signed vertex weights, which will be later useful to us in Section \ref{nonbipphase}, are $W_p(n) = i^{p-n}\langle \cos^n x \sin^{p-n} x \rangle$. 
    
    \item Arrow vertex model: there is a sign-free way we can write the vertex model: we perform a change of variables in the integral definition of the partition function, giving
\begin{align}
    \hspace{1cm}Z &= \frac{2^N}{(2\pi)^n}\int_0^{2\pi} d^n\theta  \prod_{(p,q)\in G_D} \sin (\theta_p + \theta_q) \\
    &= \frac{2^N}{(2\pi)^n}\int_0^{2\pi} d^n\theta  \prod_{(p,q)\in G_D} \left(\sin \theta_p \cos \theta_q + \cos \theta_p \sin \theta_q\right). \nonumber
\end{align}
This suggests a representation in terms of an \emph{arrowed} vertex model.  Denote an arrow from $p$ to $q$ for the term $\cos \theta_p \sin \theta_q$ and an arrow from $q$ to $p$ for the term $\sin \theta_p \cos \theta_q$.  The vertex weights $W_p(k)$ are the same as in Equation \ref{eq:vweights}, but now $k$ denotes the number of arrows pointing out of the vertex.
\end{itemize} 

The arrowed representation has the added benefit that it can be simulated in a sign-free manner via Monte Carlo by utilizing loop Metropolis updates, which we utilize for numerical simulations.  

In fact, both the arrowed and bond representations are related to one another via the Wegner duality.  To see this, map an arrowed edge to two separate edges with only one edge supporting a bond; this decorates each edge of $G_D$ with an additional node. The arrow points from the edge without a bond to the edge with a bond.  Performing the Wegner duality on this bond vertex model, the original nodes in $G_D$ satisfy Equation \ref{eqn:selfdualrel}, and the decorated nodes are mapped to:
\begin{equation}
    \begin{array}{c} W(1) = 1 \\ W(0) = W(2) = 0 \end{array} \xleftrightarrow[]{\text{Wegner}} \begin{array}{c} W(1) = 0 \\ W(0) = -W(2) = 1 \end{array}
\end{equation}
It is clear that this maps the arrowed vertex model at high temperatures (when all the arrowed weights are approximately equal) to the signed vertex model at low temperatures (where $W(0)$ and $W(V_p)$ are large in magnitude) and vice versa.  The fact that the arrowed vertex models are not self dual is likely a reflection of the absence of gapless RK points in these nonbipartite dual lattices.

\subsection{Parent Hamiltonians with RK wavefunction ground state}

\begin{figure}
    \centering
    \includegraphics[scale=0.65]{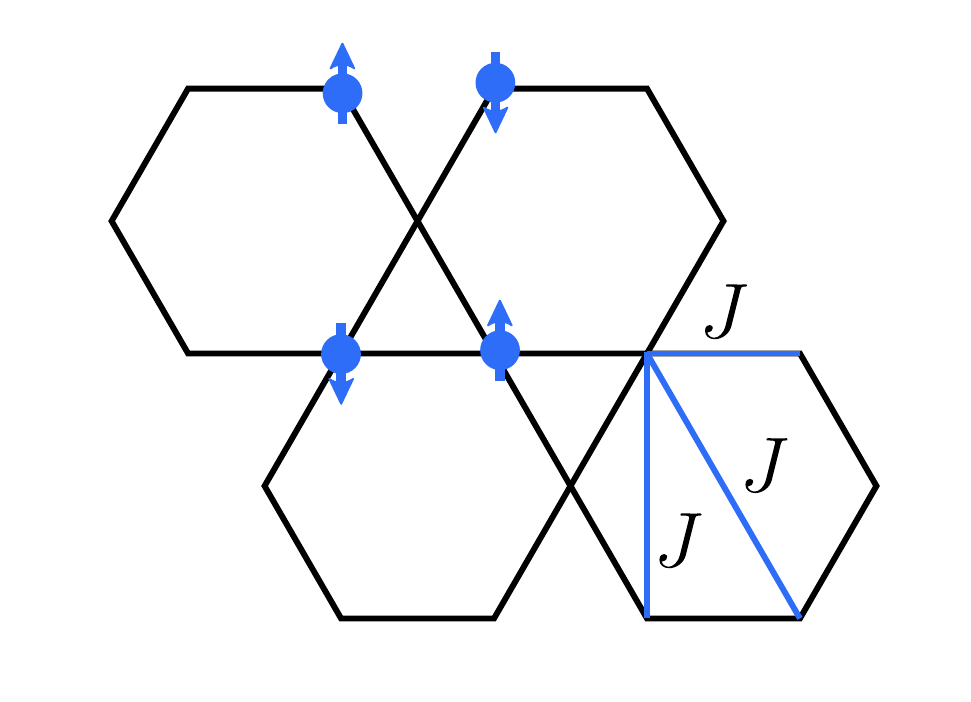}
    \caption{The BFG model is defined on the Kagome lattice, with second and third nearest-neighbor interactions denoted with the label $J$.  The ring exchange occurs on the bowties of the lattice, with an example of a flippable bowtie illustrated.}
    \label{fig:BFG}
\end{figure}

The balanced dimer constraint that we introduced can appear as the ground state of certain easy-axis Ising models with nearest neighbor and longer-range interaction in the easy axis limit.  In particular, consider the parent Hamiltonian
\begin{equation}
    \mathcal{H} = \sum_{p}  \sum_{\alpha} J_p^{\alpha} (S_p^{\alpha})^2,
\end{equation}
where $p$ labels a polygon, $S_p^{\alpha} = \sum_{i \in p} S_i^{\alpha}$ and $\alpha = x,y,z$.  In the easy axis limit, $J_z \gg J_x, J_y$ and we will assume $J_x = J_y = J_{\perp}$.  Treating the in-plane terms perturbatively, one generically finds that the leading order processes are ring exchanges that occur in the regions surrounding the polygons (particular examples will be illustrated later).  Denote by $\{r\}$ the regions in which the ring exchanges occur.  Then, the parent Hamiltonian becomes
\begin{equation}
    \mathcal{H} = \sum_{r} P_{\text{flip}}(r)\left(-J_r \prod_{i\in r} 2S_i^x\right),
\end{equation}
where $P(r)$ is a projection operator onto configurations involved in ring exchanges. The Hamiltonian can be deformed by adding a dimer potential energy term, thereby turning it frustration-free:
\begin{equation}
    \mathcal{H} = \sum_{r} P_{\text{flip}}(r)\left(-J_r \prod_{i\in r} 2S_i^x + u_r\right).
\end{equation}
This is a generalized quantum dimer model; here, we use a standard definition of a quantum dimer model $H = T + V$, where $T$ is a hopping operator between two dimer configurations connected by an elementary dimer move and $V$ is the energetic cost of having a given dimer configuration. The RK point corresponds to setting $J_r = u_r$, where the ground state is a uniform superposition of balanced dimer configurations.

A particular example of a balanced dimer model is the Balents-Fisher-Girvin (BFG) model, which is defined on the Kagome lattice.  The dual lattice $G_D$ is a triangular lattice, so the balanced dimer constraint amounts to 3 dimers per site.  The ring exchange processes occur at second order perturbation theory around bowties of the Kagome lattice (see Figure \ref{fig:BFG}):
\begin{equation}
    \mathcal{H}_{\text{ring}} = -J_{\text{ring}} \sum_{\bowtie} \left(S_1^+ S_2^- S_3^+ S_4^- + \text{h.c.}\right),
\end{equation}
Here, $J_{\text{ring}} = O(J_\perp^2/J_z)$.  The projection operator can be written as
\begin{equation}
    P_{\text{flip}}(\bowtie) = \sum_{\sigma = \pm 1} \prod_{(j \in \bowtie)=1}^4\left(\frac{1}{2} + \sigma (-1)^j S_j^z\right).
\end{equation}
The BFG model supports a $\mathbb{Z}_2$ spin liquid phase, which can be seen by performing a standard transformation to a $\mathbb{Z}_2$ gauge theory. It can be argued that the BFG model at the RK point is gapped by proving exponential decay of the spin-spin correlation function.  Exponential decay of the vison correlation function
\begin{equation}
    V_{ij} = \langle v_i v_j \rangle = \left |\langle \text{GS}| \prod_{k = i}^j 2S_k^z|\text{GS} \rangle \right |,
\end{equation}
where the product is along a path comprised of $\pm 60^\circ$ turns in the original Kagome lattice, justifies that the dual gauge theory is in a deconfined phase.  Both properties were verified via numerical QMC simulations \cite{BFG, SB}.

\section{Correlation functions}

\begin{figure}
    \centering
    \includegraphics[scale=0.5]{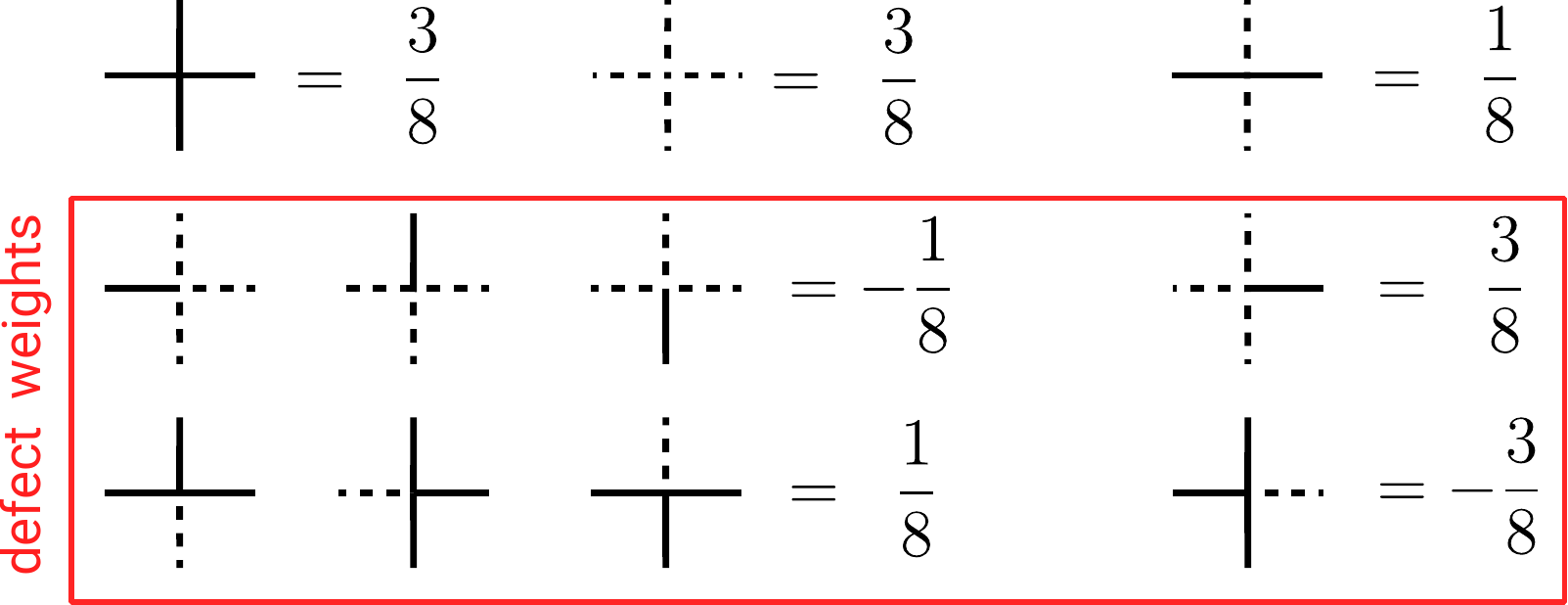}
    \caption{ The first row shows three of the vertex weights for the case of $V_p = 4$, the other weights can be generated by permuting the edges.  In the bottom two rows, the vertex weights for one of the defected nodes are shown, which are used in the computation of correlation functions.  To obtain the vertices for the other defected node, reflect the vertex configurations about the vertical axis.}
    \label{fig:corrvert}
\end{figure}

At the Rokhsar-Kivelson point, one may be interested in computing a classical correlation function which corresponds to some function of $S_i^z$.  In particular, we seek to compute
\begin{equation}
    \langle s_\alpha s_\beta \rangle = \frac{\sum_{s \in \{-1,1\}^N} s_\alpha s_\beta \prod_p \delta \left(\sum_{i \in p} s_i = 0\right)}{\sum_{s \in \{-1,1\}^N} \prod_p \delta \left(\sum_{i \in p} s_i = 0\right)}.
\end{equation}
Taking a Fourier transform of the numerator and denominator, we find that
\begin{equation}
    \langle s_\alpha s_\beta \rangle = \frac{\int_0^{2\pi} d^n\theta\sum_{s \in \{-1,1\}^N} s_\alpha s_\beta \prod_p e^{i \theta_p \sum_{i \in p} s_i}}{\int_0^{2\pi} d^n\theta\sum_{s \in \{-1,1\}^N} \prod_p e^{i \theta_p \sum_{i \in p} s_i}}.
\end{equation}
Performing the sum over $\vec{s}$ in the numerator gives a cosine term for all pairs of spins except at $\alpha$ and $\beta$, where we instead obtain a sine.  Therefore, we find that
\begin{widetext}
\begin{equation}
    \langle s_\alpha s_\beta \rangle = -\frac{\int_0^{2\pi} d^n\theta \sin (\theta_{p_\alpha} + \theta_{q_\alpha}) \sin (\theta_{p_\beta} + \theta_{q_\beta})\prod_{(p,q)\in G_D\setminus \{(p_\alpha, q_\alpha), (p_\beta, q_\beta)\}} \cos(\theta_p + \theta_q)}{\int_0^{2\pi} d^n\theta\prod_{(p,q)\in G_D} \cos(\theta_p + \theta_q)}.
\end{equation}
For simplicity, we will call $G_D\setminus \{(p_\alpha, q_\alpha), (p_\beta, q_\beta)\} = G_D'$.  Next, we perform the sublattice transformation $\theta_p^A \to \theta_p^A$ and $\theta_p^B \to -\theta_p^B$ (in the arrowed case, this step is not needed) to make all the weights positive:
\begin{equation}
    \langle s_\alpha s_\beta \rangle \propto \frac{1}{Z}\int_0^{2\pi} d^n\theta \sin (\theta_{p_\alpha} - \theta_{q_\alpha}) \sin (\theta_{p_\beta} - \theta_{q_\beta})\prod_{(p,q)\in G_D'} \cos(\theta_p - \theta_q).
\end{equation}
\end{widetext}
The denominator can be expressed in terms of the vertex model we have already considered.  The numerator can be expressed in terms of the same vertex model apart from modified (defected) vertex rules at the sites $\alpha$ and $\beta$.  In particular, for a given edge $\alpha$, denote a dimer for the configuration in which the left vertex gets a cosine and the right vertex gets a sine, and no dimer for the configuration in which the left vertex gets a sine and the right vertex gets a cosine.  In this case, the weights for the right vertex are the same as $W_p(k)$ identified in the previous section.  For the left vertex, the weight can be written as
\begin{equation}
    \widetilde{W_p}(k,m) = \left \langle \left(\cos^k \theta \sin^{V_p-1-k} \theta \right) \left((1-m)\cos \theta - m \sin\theta \right) \right \rangle,
\end{equation}
where $k$ is the number of bonds at the vertex apart from edge $\alpha$ and $m = 0$ if $\alpha$ is not a bond and $m=1$ when $\alpha$ is a bond.  A similar mapping can be made for the edge $\beta$, except that the defected vertex is the right site of $\beta$.  It is apparent that $\widetilde{W_p}(k,m)$ is non-zero only if the corresponding vertex admits an odd number of bonds (or if $k+m$ is odd).  Example vertex weights for the case of $V_p = 4$ are shown in Figure \ref{fig:corrvert}.  An alternative way of determining the vertex weights is to take an undefected vertex configuration in the original model, flip the defected edge and multiply the weight by $-1$ if the resulting edge does not have a dimer.

It is possible to write the partition function with two introduced defects in terms of a correlation function in the original vertex model without defects.  To do so, introduce a map $f: \mathcal{C}' \to \mathcal{C}$, where $\mathcal{C}'$ is the set of configurations satisfying the odd dimer constraint on the defected nodes and $\mathcal{C}$ is the set of configurations satisfying the original vertex constraint.  An element of these sets will be indicated by $c$ and $c'$ respectively.  The map $f$ constructs a consistent path $P$ connecting the two defected vertices and flips dimers along the path: i.e. for configuration $c$, if edge $e \in P$ is a dimer, then it will no longer be a dimer in $f(c)$ and vice versa.  By construction, this is a bijective map from $\mathcal{C}'$ to $\mathcal{C}$.  Therefore, denoting by $\Omega(c)$ the total weight of the vertex configuration $c$, we may write
\begin{equation}
    \langle s_\alpha s_\beta \rangle \propto \frac{\sum_{c' \in \mathcal{C}'}\Omega(c')}{\sum_{c \in \mathcal{C}}\Omega(c)} = \frac{\sum_{c: f(c) = c'}\left[\Omega(f(c))/\Omega(c)\right] \Omega(c)}{\sum_{c \in \mathcal{C}}\Omega(c)}
\end{equation}
Using the fact that $f$ is a bijective map, this is equal to $\left\langle \Omega(f(c))/\Omega(c) \right\rangle$.  As the weights $\Omega(f(c))$ and $\Omega(c)$ differ along a path connecting the two defected nodes, we may write that
\begin{equation}
    \langle s_\alpha s_\beta \rangle \propto \left\langle d_{\alpha} \left(\prod_{k = \alpha}^\beta S_k \right) d_{\beta} \right\rangle,
\end{equation}
where $d$ is a dimer variable equalling $-1$ if $\alpha$ has a dimer and $1$ otherwise.  These result from the endpoint conditions of the strings.  The operator $S$ is the ratio of the weights when flipping both bonds belonging to the path from $\alpha$ to $\beta$ to the weights without flipping both bonds.  To illustrate, for the case of $V_p = 4$, the weights are given by
\begin{equation}
    S_k = \begin{cases}
    \frac{1}{3} \hspace{0.5cm}\includegraphics[scale = 0.25, ]{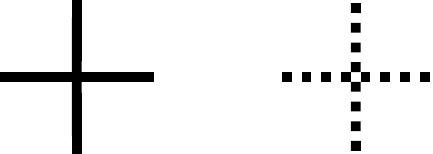}\\
    3 \hspace{0.55cm}\includegraphics[scale = 0.25, ]{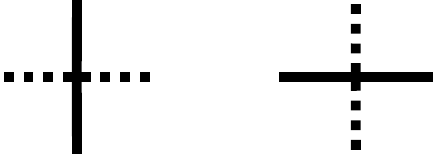}\\
    1 \hspace{0.5cm}\text{otherwise} \\
    \end{cases}
\end{equation}
We may then write the string operator as
\begin{equation}
    \prod_{k=\alpha}^\beta S_k = e^{ (n_1 - n_2) \log 3}
\end{equation}
where $n_1$ is the number of configurations with weight $3$ and $n_2$ is the number of configurations with weight $1/3$.  At the self dual point, which may correspond to a critical point, the string operator should decay algebraically, and thus the scaling behavior is dominated by fluctuations in $\langle n_1 - n_2 \rangle$.

The analysis above can be extended to other correlation functions.  For example, for correlation functions between vison/flux excitations (which is used to determine whether such a model supports deconfined spinons), one may need to compute the string operator
\begin{widetext}
\begin{equation}
\left\langle \prod_{k \in P(\alpha,\beta)} s_k \right\rangle \propto \frac{\int_0^{2\pi} d^n\theta \prod_{(p',q') \in P(\alpha, \beta)}\sin (\theta_{p'} + \theta_{q'}) \prod_{(p,q)\in G_D\setminus P(\alpha, \beta)} \cos(\theta_p + \theta_q)}{\int_0^{2\pi} d^n\theta\prod_{(p,q)\in G_D} \cos(\theta_p + \theta_q)}
\end{equation}
\end{widetext}
where $P(\alpha, \beta)$ denotes the set of edges on the dual graph $G_D$ along the path prescribed by the vison correlation function.  Along the edges in the path, assign a bond to sine-cosine assignments and no bond for a cosine-sine assignments.  This maps to a vertex model where all the vertices along the path $P(\alpha, \beta)$ admit an odd number of bonds.  To form a bijective map between such configurations (the set of which we call $\mathcal{C}''$) and the original vertex model configurations, define $g: \mathcal{C}'' \to \mathcal{C}$ to be the map which flips \emph{every other bond} along $P(\alpha, \beta)$.  The vison correlation function can then be expressed in terms of the correlation function of a string operator in the dual vertex model:
\begin{equation}
    \left \langle \prod_{k = \alpha}^\beta s_k \right \rangle \propto \left\langle d_{\alpha} \left(\prod_{k = \alpha}^\beta d_k S_k\right) d_{\beta} \right\rangle,
\end{equation}

\begin{figure}
    \centering
    \includegraphics[scale=0.325]{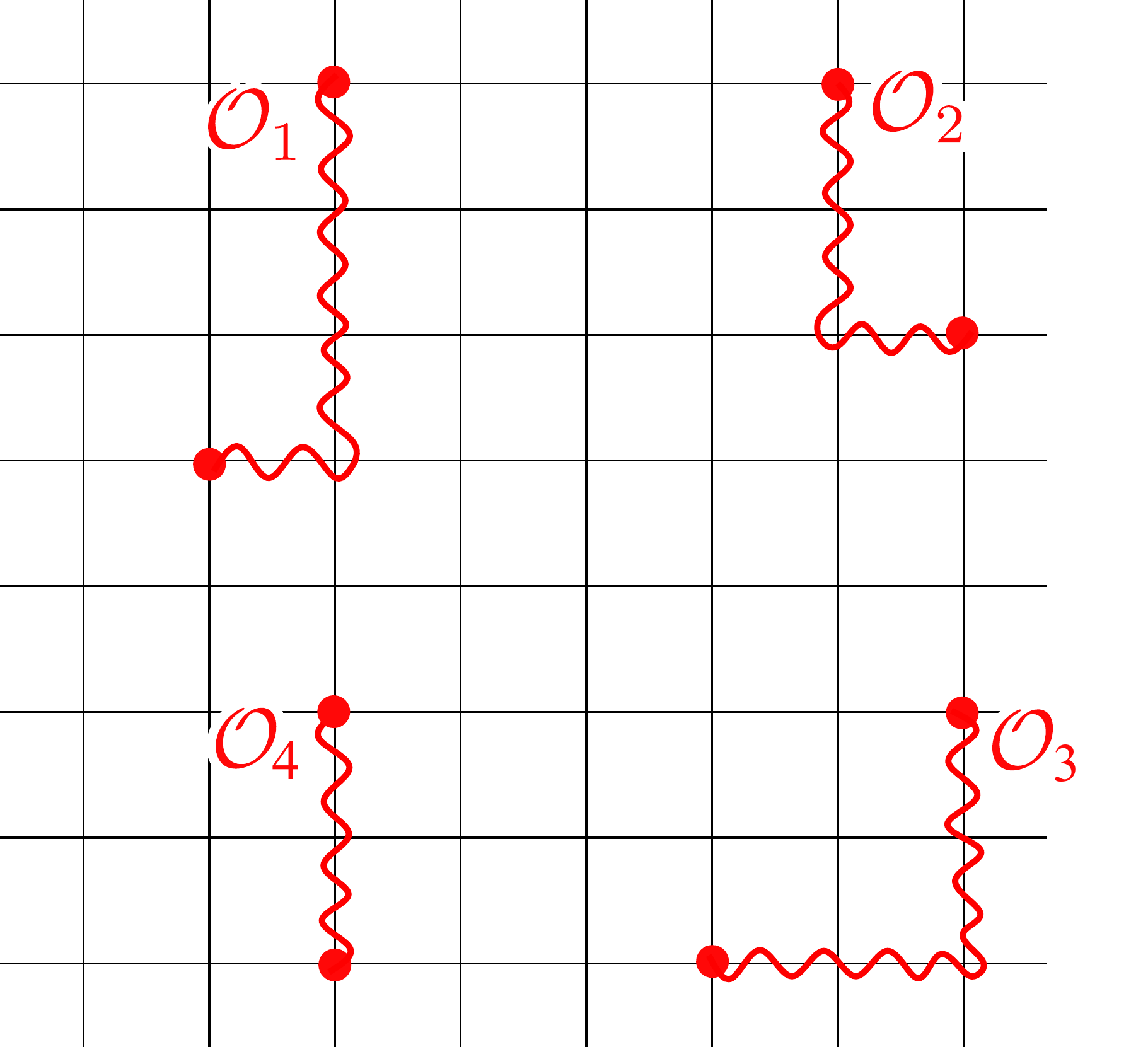}
    \caption{An illustration of the string operators for $V_p = 4$ (square lattice).  Note that correlation functions between two pairs of spins such that each pair is localized results in a two-point correlation function between local operators in the dual theory.  The illustrated 8-point correlator in the dual theory looks like $\langle\mathcal{O}_1\mathcal{O}_2\mathcal{O}_3\mathcal{O}_4 \rangle$.}
    \label{fig:correlators}
\end{figure}

Finally, we consider a generic $n$-point correlation function
\begin{equation}
    \left \langle s_1 s_2 \cdots s_n\right \rangle = \left \langle \prod_{s \in \text{strings}} \left(\prod_{i \in s} \mathcal{O}_i\right) \right \rangle,
\end{equation}
where ``strings'' denotes a set of paths connecting pairs of spins, with each spin being a member of one path.  The operator $\mathcal{O}_k = d_k S_k d_{k+1}$.  The large freedom by which we may choose paths follows from the gauge constraint $\langle \prod_{k \in \circlearrowleft} \mathcal{O}_k \rangle = 1$.

Because correlation functions of local operators in the dual theory decay as a power law, there are correlation functions in the original theory whose behavior we may immediately identify.  Consider the 4-point correlation function $\left \langle s_1 s_2 s_3 s_4\right \rangle$ where the pairs $(s_1, s_2)$ and $(s_3,s_4)$ are far separated from one another but elements within each pair are close.  If we choose strings connecting elements within each pair, then the string operators are local operators in the dual theory. Therefore, the correlator decays as a power law.  This procedure can be generalized and spins which are clustered together map to a single local operator in the dual theory.  An illustration is shown in Figure \ref{fig:correlators}.

\section{Numerical simulation of triangular and Kagome lattice vertex model}\label{numerics}

\begin{figure*}
    \centering
    \includegraphics[scale=0.51]{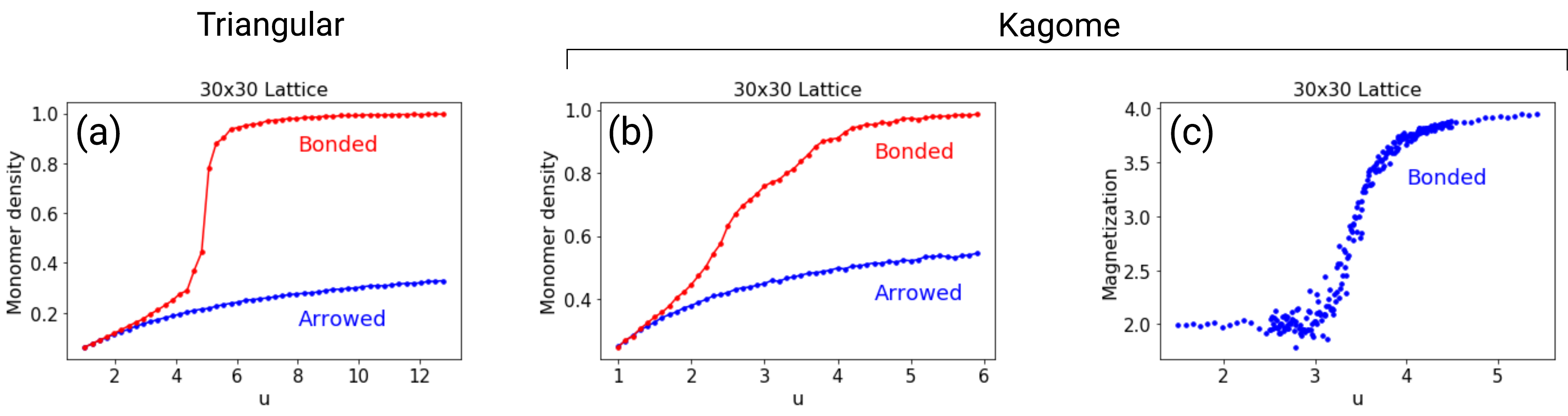}
    \caption{Panel (a) shows simulation results of the monomer density on a $30\times 30$ triangular lattice for the bonded and arrowed vertex model.  Panel (b) shows simulation results of the monomer density on a $30\times 30$ Kagome lattice for the bonded and arrowed vertex model, while panel (c) shows simulation results of the dimer magnetization on a $30\times 30$ Kagome lattice. }
   \label{fig:MCResults}
\end{figure*}

 As previously noted, the arrowed vertex model has the added advantage that it is sign free and can be simulated by Monte Carlo.  The results of a Monte Carlo simulation on a triangular lattice are in Figure \ref{fig:MCResults}[a].  We have simulated both an arrowed vertex model  with $W(0) = W(6) = u$ and $W(2) = W(4) = 1$ as well as a bonded vertex model with the same weights.  The low temperature phase corresponds to large $u$ while the high temperature phase corresponds to $u \approx 1$.  For the arrowed vertex model, we note that $u = 5$ is BDVM dual to the RK point of the BFG model; for the bonded vertex model, $u = 5$ is critical.  We plot a monomer density $f_0 + f_6$ where $f_n$ is the fraction of vertices with $n$ arrows pointing out (for arrowed vertex models) or $n$ bonds (for bond vertex models).  As the monomer density is equivalent to the internal energy, it must capture each phase transition.  The bonded vertex model has a critical point at $u = 5$, consistent with the Baxter-Wu model.  The arrowed vertex model thus has no phase transition; this means the vertex model is always in a disordered liquid phase.  At low temperatures (when $u$ is large), the monomer density for the arrowed vertex model approaches the stationary value of $\frac{1}{2}$.  Such configurations where half of the vertices are monomers can be explicitly constructed; there is an exponentially large degeneracy of them which indicates that no local symmetries are spontaneously broken.  The monomer density smoothly decreases as the temperature increases.
 
 We have also run Monte Carlo simulations of an arrowed and bonded vertex model on the Kagome lattice with $W(0) = W(4) = u$ and $W(2) = 1$, shown in Figure \ref{fig:MCResults}[b,c].  We compute the monomer density $f_0 + f_4$, and clearly find that the arrowed vertex model does not exhibit a phase transition.  For large values of $u$, the stationary value of the monomer density is $\frac{2}{3}$; again, one can construct an exponentially large number of such configurations with this property.  For the bonded vertex model, the monomer density seems to indicate features suggestive of multiple phase transitions, but the data is not conclusive and we leave further investigations to future work.  Instead, we plot the spontaneous dimer density of this system, which although cannot detect all phase transitions, detects a clear transition near $u = 3$, consistent with the self-duality property.  
 
 We also note that when $u = \infty$ one may obtain an exact result for the partition function of the arrowed Kagome vertex model.  When $u = \infty$, the allowed configurations of the statistical mechanics model corresponds to maximizing the number of 4-in or 4-out configurations.  Due to frustration this can be satisfied on two of three sites per each triangle.  One may draw bonds connecting the centers of the triangles in the Kagome lattice to form a honeycomb lattice.  Depending on whether vertex that a bond crosses satisfies the 4-in, 4-out, or 2-in-2-out rules, one assigns one of three possible colors.  This maps the partition function to that of a 3-coloring problem on the honeycomb lattice, which has an exact solution due to Baxter~\cite{baxtercoloring}.  In particular, in terms of effective height variables, the field theory description is that of a massless boson, and therefore $u = \infty$ is a gapless point.  Furthermore, when $u = 0$, duality maps the two-in-two-out ice rule to the point $u = 3$ in the bonded Kagome vertex model -- thus, we expect $u = 0$ in the arrowed Kagome vertex model to be a gapless point as well.
 
 The mapping to a three coloring problem in the $u \to \infty$ limit for the arrowed vertex models is a generic result that holds on \emph{any lattice} built from corner sharing triangles.  The proof of this is a simple generalization of the Kagome lattice case -- note that the three coloring problem is defined on the dual lattice where the sites are on the centers of the triangular plaquettes.

\section{Class I: Multi-dimer model parent Hamiltonians}\label{examplesection}

Thus far, we have studied classical statistical mechanics models which we claim to describe ground states of quantum spin models on corner sharing plaquettes which have zero magnetization per plaquette in the easy axis limit.  Here, we shall enumerate some explicit examples of quantum spin models, and describe a procedure for generating families of such models -- in particular, we introduce the BFG family and the ruby family.  We note that this is the first of three different parent Hamiltonian constructions we discuss in this paper.

\subsection{Warm-up (square lattice) example}\label{examplesq}
\begin{figure*}
    \centering
    \includegraphics[scale=0.45]{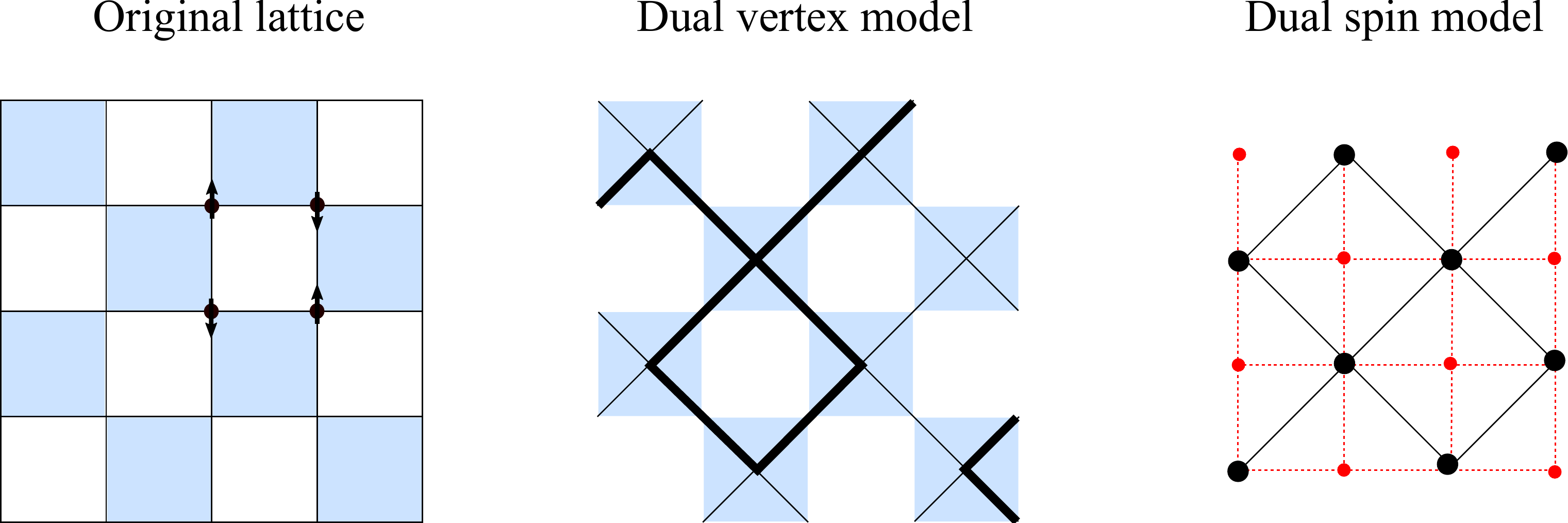}
    \caption{ The first panel shows the checkerboard lattice from whence the quantum spin model is defined.  The second panel shows the lattice where the dual vertex model is defined on (along with a representative vertex configuration).  The third panel shows the sites of the spin model (not discussed in the main text but in Appendix \ref{spinmodelphase}); the red sites are additional ``ghost spins'' added to simplify the spin model to a sum of 3-spin interactions.  The red spins are located on the sites of the dual vertex model.}
   \label{fig:hintmer}
\end{figure*}

The first model is one that is well-known, but we expound on it for illustrative purposes.  In fact, the phase diagram of the more general quantum 6- and 8-vertex model (of which this model is a special case) has been exhaustively studied \cite{AFF, Chakravarty}.  We consider an easy-axis Ising antiferromagnet on a checkerboard lattice shown in Figure \ref{fig:hintmer}.  Call the square plaquettes with the second nearest-neighbor bonds ``blue'' and the other plaquettes ``red'' -- the Hamiltonian can be written as
\begin{equation}
    \mathcal{H} = \sum_{\textcolor{blue}{\square_b}} J_{\alpha}\left(S_{\textcolor{blue}{\square_b}}^{\alpha}\right)^2
\end{equation}
where $S_{\textcolor{blue}{\square_b}}^{\alpha} = \sum_{i \in \textcolor{blue}{\square_b}} S_{i}^{\alpha}$.  In the easy-axis limit $J_x = J_y = 0$, the ground state configuration  corresponds to the sum of spins on the blue squares equalling zero.  Superimposing $G_D$ (which is a square lattice) on the checkerboard lattice, each ground state configuration on the checkerboard lattice corresponds to a configuration of a 6-vertex model on $G_D$.  At second order in perturbation theory in $J_x = J_y = J_\perp$, the dominant term is a four-spin ring exchange term on the red squares,
\begin{equation}
    \mathcal{H} = -J_{\text{ring}}\sum_{\textcolor{red}{\square_r}} P_{\text{flip}}(\textcolor{red}{\square_r}) \prod_{j=1}^4 2S_j^x,
\end{equation}
with the projection operator
\begin{align}
    P_{\text{flip}}(\textcolor{red}{\square_r}) = &\left(\frac{1}{2}-S_1^z\right)\left(\frac{1}{2}+S_2^z\right)\left(\frac{1}{2}-S_3^z\right)\left(\frac{1}{2}+S_4^z\right) 
    \nonumber \\ &+ \left(\frac{1}{2}+S_1^z\right)\left(\frac{1}{2}-S_2^z\right)\left(\frac{1}{2}+S_3^z\right)\left(\frac{1}{2}-S_4^z\right),
\end{align}
and the spin index is taken clockwise around $\textcolor{red}{\square_r}$.  This mimics an elementary dimer move on square plaquettes that mixes 6-vertex configurations within a topological sector.  Therefore, we can construct a Hamiltonian with an RK point by adding an artificial term corresponding to a self-energy of the dimers: 
\begin{equation}
    \mathcal{H} = \sum_{\textcolor{red}{\square_r}} P_{\text{flip}}(\textcolor{red}{\square_r})\left(-J_{\text{ring}} \prod_{j=1}^4 2S_j^x + u\right),
\end{equation}
At this RK point where $J_{\text{ring}} = u$, the ground state is a uniform superposition over all 6-vertex model configurations in $G_D$.  Owing to the exact solution to the 6-vertex model \cite{Lieb, BaxterBook} we know that dimer-dimer correlation functions decay algebraically.  Under the BDVM duality mapping, the corresponding 6-vertex model is a sign free 8-vertex model (due to $G_D$ being bipartite) with weights $W_4(n) = \langle \cos^n x \sin^{4-n} x \rangle$, which gives $W_4(4) = W_4(0) = 3 W_2(2)$.

We may also derive an equivalent classical spin model noting that $G_P$ is also a square lattice (the reader is deferred to Appendix \ref{spinmodelphase} for details on general spin model constructions).  A convenient representation is a 3-spin model constructed in Appendix \ref{spinmodelphase} for the triangular lattice vertex model \cite{BaxterBook, HintermannMerlini}.  For each square plaquette, impose the weight
\begin{equation}
    W_{\square} = \cosh J(s_1s_2 + s_2s_3 + s_3s_4 + s_4s_1),
\end{equation}
which takes values $\cosh 4J$ for all and no dimer configurations, and $1$ for two dimer configurations.  The value of $J$ solves $\cosh 4J = 3$.  The partition function for the vertex model can be written as
\begin{equation}
    Z = \sum_{\vec{s} = \{-1,1\}^N} \prod_{s_1^q,\ldots,s_4^q \in \square} \cosh J(s_1^{q}s_2^{q} + s_2^{q}s_3^{q} + s_3^{q}s_4^{q} + s_4^{q}s_1^{q})
\end{equation}
Next, we place additional spins in the centers of the square plaquettes (forming the union jack lattice) and use the identity
\begin{equation}
    \cosh J\left(s_1s_2 + s_2s_3 + s_3s_4 + s_4s_1\right) = \sum_{\sigma} \exp\left(J \sum_{\triangle} \sigma s_i s_j\right)
\end{equation}
where $\sigma$ is the newly added ghost spin, and $\triangle$ indicates that the sum is taken over triangles in each plaquette.  Then
\begin{equation} \label{eq:baxterwupartition}
    Z = \sum_{\vec{s} = \{-1,1\}^N} \exp\left(J \sum_{\triangle} s_i s_j s_k\right),
\end{equation}
This Hamiltonian has been exactly solved by Hintermann and Merlini \cite{HintermannMerlini}.  At zero temperature, there is a four-fold degenerate ground state as well as a second order phase transition at $\cosh 4J = 3$ associated with $\mathbb{Z}_2 \times \mathbb{Z}_2$ symmetry breaking.  This is consistent with algebraically decaying correlation functions \cite{DWZG}.

\subsection{Honeycomb model and unified Hamiltonian}\label{examplehoney}
\begin{figure*}
    \centering
    \includegraphics[scale=0.25]{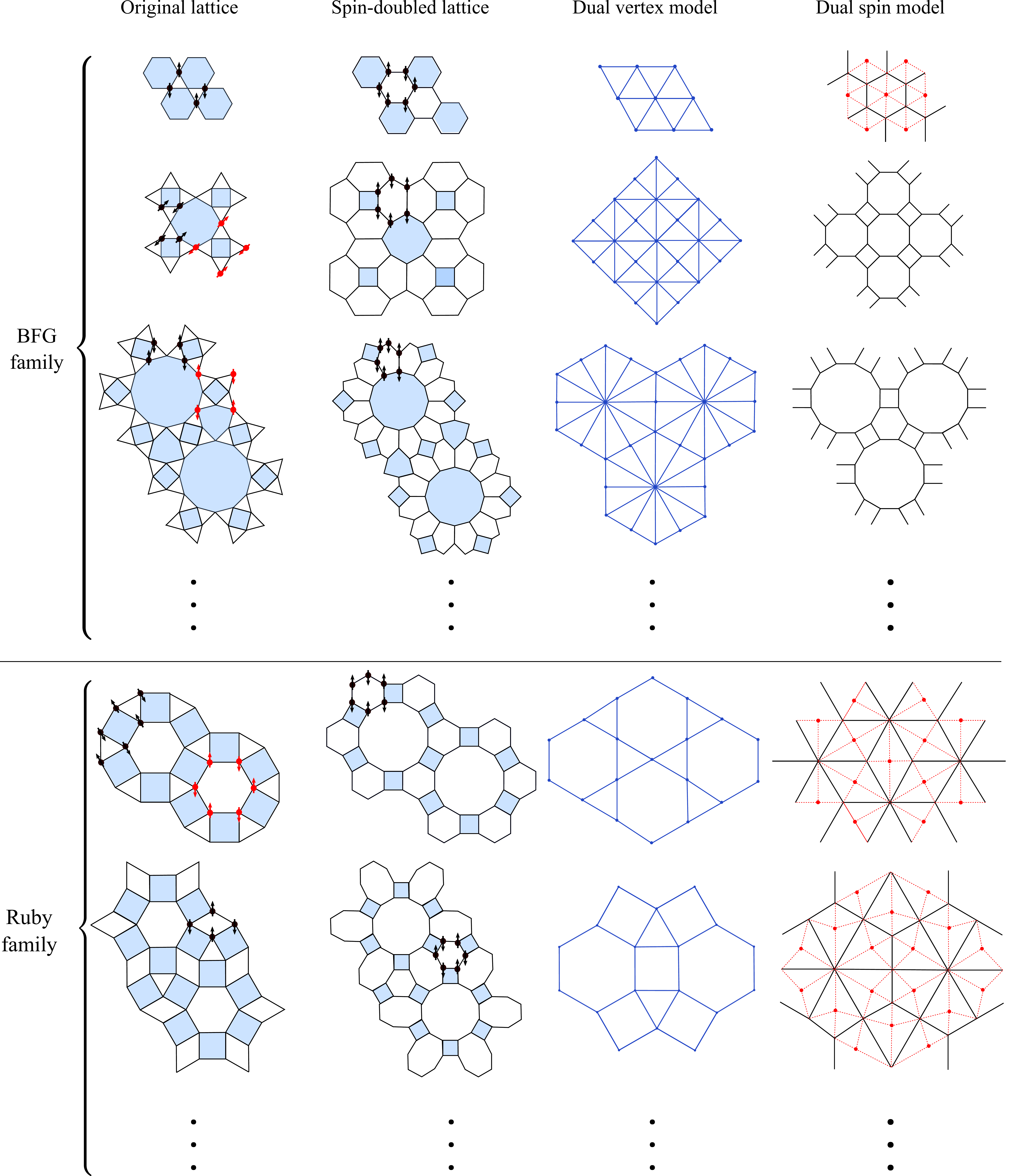}
    \caption{A summary of the models we consider in Class I. The first column indicates the original spin model and the ring exchanges.  The second column indicates the spin-doubled models.  In these columns, a shaded polygon indicates all-to-all interactions within the polygon.  The third column indicates the graphs $G_D$, and the fourth column indicates the graphs $G_P$ (which are discussed in Appendix \ref{spinmodelphase}).  Additionally, red dashed lines are present in the fourth column if the spin model can be written as sums of effective 3-spin interactions.  Some members of the BFG and ruby family are shown.}
    \label{fig:baxterwu}
\end{figure*}

One can enumerate regular and planar lattices for $G_D$ and derive a quantum spin model, but invariably one runs into two issues: first, that $G_D$ is not bipartite, and second that $G_D$ has vertices with odd degree.  The square lattice is the only exception \emph{unless} we allow for decoration nodes with degree 2.  The simplest such quantum spin model with degree 2 decoration nodes is the frustrated honeycomb lattice shown in Figure \ref{fig:baxterwu}.  The hexagonal plaquettes of the honeycomb lattice can be separated into three sublattices; we include frustrated interactions between all pairs of spins for hexagons in one of the sublattices, which we denote as red hexagons (the other hexagons will be blue hexagons).  The Hamiltonian is
\begin{equation}\label{eq:honeycombmodel}
    \mathcal{H} = \sum_{\textcolor{red}{\hexagon_r}} J_{\alpha} \left(S^{\alpha}_{\textcolor{red}{\hexagon_r}}\right)^2 + \sum_{\textcolor{blue}{\langle i, j\rangle_b}} J_{\alpha} S^{\alpha}_{\textcolor{blue}{i}}S^{\alpha}_{\textcolor{blue}{j}},
\end{equation}
where the blue edges indicate the bonds connecting between two red hexagons.  In the easy-axis limit where $S_z \gg S_x=S_y=S_\perp$, the classical ground state corresponds to zero total spin on each red hexagon and opposite spins on each blue bond.  Drawing an arrow from an up to a down spin on each blue bond, this is equivalent to an arrowed ice-rule vertex model on a triangular lattice that has been previously solved using Bethe ansatz methods \cite{Kelland}.  To third order in perturbation theory, the Hamiltonian is
\begin{equation}
    \mathcal{H} = J_{\text{ring}} \sum_{\textcolor{blue}{\hexagon_b}} P_{\text{flip}}(\textcolor{blue}{\hexagon_b}) \prod_{j=1}^6 2S_j^x,
\end{equation}
where $J_{\text{ring}} = O(J_\perp^3/J_z^2)$ and the projection operator is defined as
\begin{widetext}
\begin{align}
    P_{\text{flip}}(\textcolor{blue}{\hexagon_b}) = \left(\frac{1}{2}-S_1^z\right)\left(\frac{1}{2}+S_2^z\right)&\left(\frac{1}{2}-S_3^z\right)\left(\frac{1}{2}+S_4^z\right)\left(\frac{1}{2}-S_5^z\right)\left(\frac{1}{2}+S_6^z\right)
  \nonumber \\&+\left(\frac{1}{2}+S_1^z\right)\left(\frac{1}{2}-S_2^z\right)\left(\frac{1}{2}+S_3^z\right)\left(\frac{1}{2}-S_4^z\right)\left(\frac{1}{2}+S_5^z\right)\left(\frac{1}{2}-S_6^z\right),
\end{align}
\end{widetext}
where spin indices taken clockwise about the blue hexagons.  $J_{\text{ring}}$ has the wrong sign, but this can be amended by a $\mathbb{Z}_2$ transformation on the spins of one sublattice which preserves the spin algebra.  One can extend this Hamiltonian to one which has an RK point:
\begin{equation}\label{eq:honeyham}
    \mathcal{H} = \sum_{\textcolor{blue}{\hexagon_b}} P_{\text{flip}}(\textcolor{blue}{\hexagon_b})\left(-J_{\text{ring}} \prod_{j=1}^6 2S_j^x + u\right),
\end{equation}
The graph $G_D$ connects the centers of the red hexagons with the centers of the blue bonds, and is equivalent to a triangular lattice with additional decorated nodes on the centers of edges.  Since $G_D$ is bipartite, the bond vertex model will have weights $W_6(6) = W_6(0) = 5 W_6(2) = 5 W_6(4)$ for the degree 6 nodes and $W_2(2) = W_2(0)$ for the degree 2 nodes. This is equivalent to the bond vertex model on the triangular lattice and therefore this model is dual to the critical point of the Baxter-Wu model (see Appendix \ref{spinmodelphase} for a derivation). 

One important point to address is whether the honeycomb model might break ergodicity.  One piece of evidence to support this is the large number of nontrivial symmetries: these can be constructed from drawing a line starting and ending at infinity $\ell$ that avoids the red polygons and crosses some number of blue bonds.  For each blue bond crossed, associate one of the two spins to the line.  Constructing the operator
\begin{equation}
    \mathcal{O}_{\ell} = \sum_{\alpha \in \ell} S_{\alpha}^z,
\end{equation}
one can show that $[\mathcal{O}_{\ell}, H] = 0$ for all lines $\ell$.  If $\ell$ is chosen to be a loop, the corresponding conserved $\mathcal{O}_{\ell}$ is equivalent to the zero magnetization constraint on the blue bonds and red hexagons; thus, two operators $\mathcal{O}_{\ell}$ and $\mathcal{O}_{\ell'}$ differing by a loop are equivalent.  Due to this structure of conserved quantities, there are a large number of disconnected sectors, and the RK wavefunction is a uniform superposition within each sector.  To decrease the number of ergodic sectors, one can include ring exchange processes at higher order in perturbation theory, which decreases the number of conserved charges.  As is usually done, we assume that the equal-time quantities within each ergodic sector can be approximated by thermodynamic averages in the Baxter-Wu model.  

The honeycomb lattice Hamiltonian above and the BFG model can be unified into the following Hamiltonian
\begin{widetext}
\begin{equation}
    \boxed{\mathcal{H}(J'_z) = \sum_{\textcolor{red}{\hexagon_r}} J_{\alpha} \left(S^{\alpha}_{\textcolor{red}{\hexagon_r}}\right)^2 + \sum_{\textcolor{blue}{\langle i, j\rangle_b}} J'_{\alpha} S^{\alpha}_{\textcolor{blue}{i}}S^{\alpha}_{\textcolor{blue}{j}} \nonumber + u_{\hexagon}\sum_{\textcolor{blue}{\hexagon_b}} P_{\text{flip}}(\textcolor{blue}{\hexagon_b}) + u_{\bowtie}\sum_{\textcolor{blue}{\bowtie_b}} P_{\text{flip}}(\textcolor{blue}{\bowtie_b})}
\end{equation}
\end{widetext}
where $J'_{z}$ can be positive or negative, and $u_{\hexagon} = O(J_\perp^3/J_z^2)$ and $u_{\bowtie} = O(J_\perp^4/J_z^3)$.  For $J'_z$ positive, the Hamiltonian is equivalent to Equation \ref{eq:honeyham} at third order in perturbation theory (the last bowtie term may be neglected at this order).  If $J'$ is negative, then the blue bonds are ferromagnetic and the effective Hamiltonian has a different ring exchange process at fourth order perturbation theory:
\begin{equation}
    \mathcal{H} = \sum_{\textcolor{blue}{\bowtie_b}} P_{\text{flip}}(\textcolor{blue}{\bowtie_b})\left(-J'_{\text{ring}} \prod_{j=1}^8 2S_j^x + u_{\bowtie}\right),
\end{equation}
where the ring exchange is illustrated in Figure \ref{fig:baxterwu} and $J'_{\text{ring}}=O(J_\perp^4/J_z^3)$.  If spins on the blue bonds are identified with each other, then this is equivalent to the BFG model and is gapped at the RK point.  Note that due to the easy axis configurations of the above Hamiltonian not satisfying the constraint imposed by $P_{\text{flip}}(\textcolor{blue}{\hexagon_b})$, one will need to flip additional pairs of spins before $P_{\text{flip}}(\textcolor{blue}{\hexagon_b})$ acts non-trivially -- thus any process involving a hexagonal ring exchange occurs at higher order.  When $J_z' = 0$, the system is gapped: the leading order contributions come at first order in perturbation theory and do not couple spins in different red hexagons.  Thus, the above Hamiltonian interpolates between the honeycomb model and the BFG model by tuning the sign of the easy-axis interaction on the blue bonds.

This method is an application of what we call \emph{``spin-doubling''} -- by splitting a shared corner in a lattice with non-bipartite $G_D$ into two antiferromagnetically coupled spins, the spin model yields a gapless critical point rather than a gapped phase.

\subsection{Other examples}\label{otherexamples}

We have enumerated a few more examples of this procedure in Figure \ref{fig:baxterwu}.  The first few lattices below the honeycomb lattice are a few members of the BFG family, which are comprised of corner-sharing polygons separated by triangles.  The square kagome lattice is the simplest such example, with the Hamiltonian
\begin{equation}
    \mathcal{H} = \sum_{\octagon} J_o^{\alpha} \left(S^{\alpha}_{\octagon}\right)^2 + \sum_{\square} J_s^{\alpha} \left(S^{\alpha}_{\square}\right)^2.
\end{equation}
A slightly different square Kagome $SU(N)$ Heisenberg model has been previously considered; that model was mapped onto an 8-vertex model in the large $N$ limit \cite{SG}.  Treating the in-plane couplings in the above Hamiltonian perturbatively, we find conventional bowtie exchanges, ``distorted'' bowtie exchanges (indicated with a subscript $d$), as well as a competing diamond exchange:
\begin{align}
    \mathcal{H} = \sum_{\bowtie} &P_{\text{flip}}(\bowtie)\left(-J_1 \prod_{j=1}^4 2S_j^x + u_1\right) \nonumber \\ &+ \sum_{\bowtie_d} P_{\text{flip}}(\bowtie_d)\left(-J_2 \prod_{j=1}^4 2S_j^x + u_2\right) \nonumber \\ &+ \sum_{\Diamond} P_{\text{flip}}(\Diamond)\left(-J_3 \prod_{j=1}^4 2S_j^x + u_3\right),
\end{align}
where the RK point corresponds to $J_1 = u_1$, $J_2 = u_2$, and $J_3 = u_3$.  The effective exchange coefficients are
\begin{align}
    J_1 &= -\frac{1}{4}\left(\frac{{J_o^{\perp}}^2}{J_s^z} + \frac{{J_s^{\perp}}^2}{J_o^z}\right) \nonumber \\
    J_2 &= -\frac{1}{4}\left(\frac{J_o^{\perp}J_s^{\perp}}{J_o^z} + \frac{2 {J_o^{\perp}}^2}{J_o^z + J_s^z}\right) \nonumber \\
    J_3 &= -\frac{{J_o^{\perp}}^2}{2 J_o^z}
\end{align}
In the limit where $J_s^z \gg J_o^z$ and $J_s^\perp \ll J_o^\perp$, the dominant process is the ring exchange on the diamonds, reducing this model to the square lattice model.  Presumably, perturbing about this point by turning on bowtie exchanges will gap the system.  When spin-doubling is performed, the Hamiltonian is 
\begin{equation}
    \mathcal{H} = \sum_{\textcolor{red}{\octagon_r}} J^{\alpha} \left(S^{\alpha}_{\textcolor{red}{\octagon_r}}\right)^2 + \sum_{\textcolor{red}{\square_r}} J^{\alpha} \left(S^{\alpha}_{\textcolor{red}{\square_r}}\right)^2 + \sum_{\textcolor{blue}{\langle i, j\rangle_b}} J_{\alpha} S^{\alpha}_{\textcolor{blue}{i}}S^{\alpha}_{\textcolor{blue}{j}},
\end{equation}
where the blue bonds and red plaquettes are illustrated in the second column.  Like in the previous subsection, one can construct a Hamiltonian which interpolates between the antiferromagnetic and ferromagnetic spin-doubled models.  Another model in the BFG family is explicitly given in the third row and one may proceed similarly and produce more elements of the family.  Understanding the nature of the fractionalized excitations in these models is left to a future work; for the square kagome lattice, we constructed its dual $\mathbb{Z}_2$ gauge theory and found multiple types of flux excitations.  This indicates the presence of a finer symmetry that will not be the focus of this paper.

A different class of models that one may construct is the ruby family.  Consider the ruby lattice in the first column of Figure \ref{fig:baxterwu}.  This Hamiltonian involves ring exchanges on two different types of hexagonal plaquettes, and is therefore ergodically mixes over configurations satisfying a two dimer constraint on the dual Kagome lattice.  Because this is dual to an arrowed 8-vertex model on the Kagome lattice we believe the Hamiltonian is gapped and should support a $\mathbb{Z}_2$ spin liquid phase.

When the spin-doubling process is performed, the Hamiltonian has ring-exchange terms on the denoted hexagonal plaquettes in the second column of Figure \ref{fig:baxterwu}.  However, these ring exchange terms commute with each other and do not ergodically connect classical spin configurations.  The resulting ground state is the product state
\begin{equation}
    |\text{GS}\rangle = \bigotimes_{\hexagon} \frac{1}{\sqrt{2}}\left(\left|\scalebox{0.2}{\chemfig{*6(=-=-=-)}}\right \rangle + \left|\scalebox{0.2}{\chemfig{*6(-=-=-=)}}\right\rangle\right),
\end{equation}
and thus the Hamiltonian is trivially gapped.  If we add a ring exchange term on the 12-sided polygon by hand, then the dimer moves are ergodic and the RK point will mimic an ice rule model on the dual lattice and will therefore be gapless.  The dual spin model (see Appendix \ref{spinmodelphase} for details) is defined on the dice lattice, a lattice of two interpenetrating honeycomb sublattices.  The partition function of the statistical mechanics model can be written in the form of Equation \ref{eq:baxterwupartition} -- the triangles are denoted in the last column of Figure \ref{fig:baxterwu}.  Thus the high temperature expansion introduced in Appendix \ref{spinmodelphase} applies.

In a similar way, one may construct other lattices in the ruby family by taking a lattice in the BFG family and constructing the edge dual lattice.  This is illustrated in Figure \ref{fig:baxterwu}.  One will need to add an additional ring exchange term by hand to obtain a nontrivial RK ground state.  However, these models have the convenient property that the effective statistical mechanics described by the RK wavefunction can be described in terms of a classical Hamiltonian with 3-spin interactions, which is not the case for the BFG family.

\section{Class II: Parent Hamiltonians for dual vertex models}\label{pertvertex}

In this section, we construct a parent Hamiltonian whose ground state is a weighted superposition of configurations of the dual vertex model -- i.e. a ``quantum vertex model''.  The parent Hamiltonians we construct utilizes the spin-doubling construction introduced in Section \ref{otherexamples} and allows for fluctuating plaquette charging energies.  We start by briefly reviewing the relevant dualities we observed in the paper:
\begin{itemize}
    \item Consider a classical statistical mechanics model with Ising degrees of freedom on a lattice of corner-sharing polygons, where the total spin on each polygon is zero.  If the dual lattice $G_D$ is bipartite (note that the zero-total spin constraint implies that $G_D$ must have even coordination number), then this model can be mapped onto a bond vertex model on $G_D$ with vertex weights $W_p(n) = 
    \langle \cos^{n} x \sin^{V_p - n} x \rangle$.  This is the BDVM duality.
    \item If the dual lattice $G_D$ is not bipartite, then this can be mapped onto an arrowed vertex model on $G_D$ with vertex weights $W_p(n) = 
    \langle \cos^{n} x \sin^{V_p - n} x \rangle$ and $n$ denotes the number of arrows pointing out of the vertex.
\end{itemize}
Therefore, we will be constructing Hamiltonians for both the bond and arrow type vertex models with general weights $W_p(n)$.  The ground states of these Hamiltonians take the form
\begin{equation}
    \ket{\text{GS}} = \sum_{C} \sqrt{W(C)} \ket{C}
\end{equation}
where $W(C)$ is the Boltzmann weight of spin configuration $C$ in the classical vertex model (either arrowed or bonded).  The ground state is designed such that its normalization is the partition function of the classical vertex model.

We will first work with non-bipartite $G_D$ (in particular, a triangular lattice).  In particular, the RK point of the BFG model is BDVM dual to an arrowed vertex model where $W_6(0) = W_6(6) = 5$ and $W_6(2) = W_6(4) = 1$.  To extend this to a one parameter family of models, we will call $W_6(0) = W_6(6) = u$; note that this model was simulated in Figure \ref{fig:MCResults}.  For each arrow, assign two spins (this is analogous to the spin-doubling method) which point in opposite directions.  The direction of the arrow corresponds to the direction from the up spin to the down spin.  Next, consider the unperturbed Hamiltonian (the reader is referred to the top panel of the second column in Figure~\ref{fig:baxterwu} for the locations of the red hexagons and blue bonds)
\begin{equation}
    \mathcal{H}_0 = -J_z \sum_{\textcolor{red}{\hexagon_r}} \prod_{i \in \textcolor{red}{\hexagon_r}} S_i^z + J_z \sum_{\textcolor{blue}{\langle i, j \rangle_b}} S_{\textcolor{blue}{i}}^z S_{\textcolor{blue}{j}}^z.
\end{equation}
The first term of the Hamiltonian enforces that the ground state configurations have an even number of arrows pointing out at each vertex, and the second term enforces that the two spins associated to an arrow are antiparallel.  The perturbing Hamiltonian is an XY term associated to each pair of spins on an arrow, and the full Hamiltonian is
\begin{equation}
    \mathcal{H} = \mathcal{H}_0 + J_\perp \sum_{\textcolor{blue}{\langle i, j \rangle_b}} \left(S_{\textcolor{blue}{i}}^+ S_{\textcolor{blue}{j}}^- + \text{h.c.}\right)
\end{equation}
Identifying each red hexagon as a site on a triangular sublattice, this describes an easy-axis Ising model with an antiferromagnetic interaction on blue bonds and a constraint on the parity of the total spin around each red hexagon.  To third order in perturbation theory, the Hamiltonian is
\begin{equation}
    \mathcal{H}_{\text{eff}} =  -c \frac{J_{\perp}^3}{J_z^2}\sum_{\textcolor{blue}{\hexagon_b}} \prod_{\textcolor{blue}{\langle i, j \rangle_b} \in \textcolor{blue}{\hexagon_b}} \left(S_{\textcolor{blue}{i}}^+ S_{\textcolor{blue}{j}}^- + S_{\textcolor{blue}{i}}^- S_{\textcolor{blue}{j}}^+ \right),
\end{equation}
which has been projected on the space of spin configurations satisfying $\prod_{i \in \textcolor{red}{\hexagon_r}} S_i^z = 1$ for all red hexagons and the arrow constraint $S_{\textcolor{blue}{i}}^z S_{\textcolor{blue}{j}}^z = -1$.  In the original arrowed vertex model on the triangular lattice, these terms corresponding to reversing arrows around principal triangular loops: in particular,
\begin{equation}
    \mathcal{O}_{ij} = S_{\textcolor{blue}{i}}^+ S_{\textcolor{blue}{j}}^- + S_{\textcolor{blue}{i}}^- S_{\textcolor{blue}{j}}^+
\end{equation}
is an operator that reverses an arrow at link $\langle i, j \rangle$, assuming that the spins at $\langle i, j\rangle$ satisfy $Z_i Z_j = -1$. 

To engineer an RK ground state, we need to add dimer potential energy terms (this procedure is further discussed and used in Appendix~\ref{offcriticalHam}) such that the Hamiltonian becomes frustration-free.  Denote $C_1$ to be a given arrow configuration of the vertex model and $C_2$ to be another arrow configuration which differs from $C_1$ by an arrow reversal (i.e. flipping all spins) around a single plaquette.  Also call $\mathcal{W}(C)$ to be the Boltzmann weight of the vertex configuration $C$.  We define the four projectors (we will use the notation $S_{\hexagon}^z$ to indicate the total spin around a red hexagon):
\begin{align}
    \mathcal{P}_{\hexagon, i, j}(0) &= \sqrt{u}^{-1} \delta(S_{\hexagon}^z = 0), \nonumber\\
    \mathcal{P}_{\hexagon, i, j}(2) &= \left(\frac{\sqrt{u}-1}{4} \left(1-S_i^z\right)\left(1-S_j^z\right) + 1\right) \delta(S_{\hexagon}^z = 2), \nonumber \\
    \mathcal{P}_{\hexagon, i, j}(4) &= \left(\frac{\sqrt{u}-1}{4} \left(1+S_i^z\right)\left(1+S_j^z\right) + 1\right) \delta(S_{\hexagon}^z = 4), \nonumber \\
    \mathcal{P}_{\hexagon, i, j}(6) &= \sqrt{u}^{-1} \delta(S_{\hexagon}^z = 6),
\end{align}
where the $\delta$-function is defined as
\begin{equation}
 \delta(S_{\hexagon}^z = n) = \frac{\prod_{m \neq n} (S_{\hexagon}^z - m)}{\prod_{m \neq n} (n - m)}.
\end{equation}
The spins $S_i^z$ and $S_j^z$ correspond to the two spins that are part of the red hexagon $\hexagon$ and are involved in an elementary ring exchange process around a blue hexagon (the other four spins in the ring exchange are paired with two other red hexagons).  

These projectors are defined in the following way: each spin configuration on a red hexagon corresponds to a configuration of arrows for this hexagon; $i$ and $j$ refer two spins belonging to a particular red hexagon that participate in a triangular arrow reversal process (which is equivalent to a ring exchange around the blue hexagon).  The projectors defined above compute the contribution to the Boltzmann weight ratio $\sqrt{\mathcal{W}(C_2)/\mathcal{W}(C_1)}$ before and after the arrow reversal that corresponds to the change in Boltzmann weight of the vertex/red hexagon that arrows with spins $i$ and $j$ are associated to.  

Next, define $\mathcal{P}_{\hexagon, i, j} = \sum_{n=0}^6 \mathcal{P}_{\hexagon, i, j}(n)$. Each triangular arrow reversal involves three red hexagons $\hexagon_1$, $\hexagon_2$, and $\hexagon_3$ and the Boltzmann weight ratio $\sqrt{\mathcal{W}(C_2)/\mathcal{W}(C_1)}$ will therefore a product of the contributions from each of these hexagons.  Thus, we define the quantities
\begin{equation}
   \mathcal{P}_{\triangle} = \mathcal{P}_{\hexagon_1, i, j} \mathcal{P}_{\hexagon_2, j, k} \mathcal{P}_{\hexagon_3, k, i}.
\end{equation}
We caution the reader that $\mathcal{P}_{\hexagon_1, i, j}$ and $\mathcal{P}_{\hexagon_2, j, k}$ do not involve the same spin $S_j^z$, because each arrow is associated with two spins.  Rather, $j$ indicates an arrow (i.e., two spins) and the spin operators always correspond to the spin that lies on the designated hexagon.  Deforming the Hamiltonian to 
\begin{equation}\label{eq:arrowedvertham}
    \mathcal{H} = \mathcal{H}_{\text{eff}} + c \frac{J_{\perp}^3}{J_z^2} \sum_{\triangle} \mathcal{P}_{\triangle},
\end{equation}
results in a frustration-free point, as the ground state is equivalent to the classical arrowed 32-vertex model, $\ket{\text{GS}} \propto \sum_C \sqrt{\mathcal{W}(C)} \ket{C}$.  An equivalent and more convenient way of writing this Hamiltonian is
\begin{equation}\label{eq:newarrowedvertham}
    \mathcal{H} = \sum_{\triangle, C} \mathcal{P}\left(\frac{\ket{C_\triangle}}{\omega^{1/2}(C, \overline{C})} - \omega^{1/2}(C, \overline{C})\ket{\overline{C}_\triangle}\right)
\end{equation}
where $C$ is a configuration of arrows on a triangle of the lattice, and $\overline{C}$ is a configuration where the arrows are reversed.  The notation $\mathcal{P}(\ket{\psi}) = \ketbra{\psi}{\psi}$.  Further, $\omega(C, \overline{C})$ is the square root of the ratio of Boltzmann weights of $C$ and $\overline{C}$.  To show the equivalence of Eqn.~\ref{eq:arrowedvertham} (at third order in perturbation theory) and Eqn.~\ref{eq:newarrowedvertham}, one can directly verify that the matrix elements of $\mathcal{H}$ with the states $\ket{C_\triangle}$ and $\ket{\overline{C}_\triangle}$ agree.  From Eqn.~\ref{eq:newarrowedvertham}, it is clear that the ground state is indeed as advertised.

\begin{figure}
    \centering
    \includegraphics[scale=0.035]{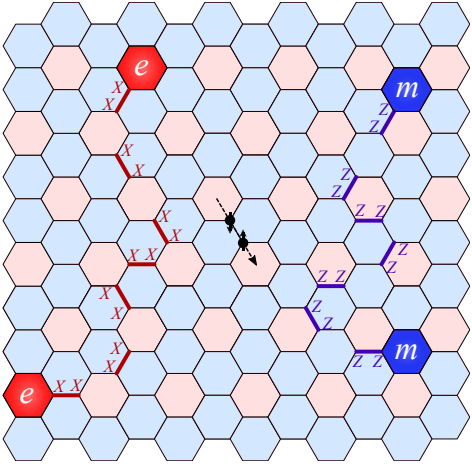}
    \caption{The quantum arrowed vertex model is defined on a honeycomb lattice with the red hexagons corresponding to sites of a triangular lattice.  The ring exchanges around blue hexagons correspond to an elementary arrow reversal around triangles if the correspondence between spins and arrows is as indicated in the figure.  The anyon excitations at $u = 1$ are shown, as well as the operators constructing them.}
    \label{fig:anyonexcitations}
\end{figure}

Based on numerical simulations from Section \ref{numerics}, this model does not have a phase transition and the above Hamiltonian is in a single phase \emph{regardless} of the value of $u$.  In particular, this is a disordered phase corresponding to a gapped quantum spin liquid.  To see this, we note that the point $u = 1$ corresponds to the quantum Hamiltonian (with $J_z \gg J_{\perp}$)
\begin{equation}
    \mathcal{H} \sim -J_z\sum_{\textcolor{red}{\hexagon_r}} ZZZZZZ - c\frac{J_{\perp}^3}{J_z^2}\sum_{\textcolor{blue}{\hexagon_b}} XXXXXX,
\end{equation}
which is a sum of commuting terms, where we have projected onto the arrow subspace $Z_{\textcolor{blue}{i}} Z_{\textcolor{blue}{j}} = -1$.  One can show that, like the toric code, the spectrum can be deduced exactly: it is gapped and supports $e$ and $m$ excitations \cite{kitaev}.  The excitations are shown in Figure \ref{fig:anyonexcitations}; pairs of $e$ excitations correspond to string operators connecting two defected red hexagons, while pairs of $m$ excitations correspond to string operators connecting two defected blue hexagons:
\begin{equation}
    E_{\textcolor{red}{i},\textcolor{red}{j}} = \prod_{k \in \ell(\textcolor{red}{i},\textcolor{red}{j})} X_k \hspace{0.75cm} M_{\textcolor{blue}{i},\textcolor{blue}{j}} = \prod_{k \in \ell(\textcolor{blue}{i},\textcolor{blue}{j})} Z_k.
\end{equation}
As there are twice as many blue hexagons as red hexagons, there are twice as many $m$ excitations as there are $e$ excitations.  Apart from this difference, the fusion rules and mutual statistics are the same as that of the toric code.  For any $u < \infty$, the model is still topologically ordered because of a lack of a transition in the arrowed vertex model; however, the model takes a more unusual form in that the ground state approaches a uniform superposition of $2$ and $4$ arrow configurations with a fixed fraction of monomers ($0$ and $6$ arrow configurations) equal to $\frac{1}{2}$ when $u = \infty$, thus constituting an emergent $U(1)$ symmetry which fixes the number of monomers.  Due to the arrow constraint, this is the maximum possible monomer fraction.  The nature of the excitations at this point is a subject of further investigation.

We note that this construction can be straightforwardly generalized to other lattices in the BFG and Ruby families, and support gapped phases which are topologically ordered due to the existence of a commuting projector model at a point on the phase diagram.  For example, for spins on sites of the ruby lattice, we must work with an arrowed vertex model on the dual Kagome lattice.  To write down a frustration-free Hamiltonian, we start with a spin doubled ruby lattice (so that arrows are identified with two spins) and define the projectors
\begin{align}
    \mathcal{P}_{\square, i, j}(0) &= \sqrt{u}^{-1} \delta(S_{\square}^z = 0), \nonumber\\
    \mathcal{P}_{\square, i, j}(2) &= \left(\frac{\sqrt{u}-1}{2} \left(1-S_i^z S_j^z\right) + 1\right) \delta(S_{\square}^z = 2), \nonumber \\
    \mathcal{P}_{\square, i, j}(4) &= \sqrt{u}^{-1} \delta(S_{\square}^z = 4),
\end{align}
where $\square$ correspond to squares in the spin-doubled ruby lattice.  As the elementary arrow reversal processes must be over triangles and hexagons on the Kagome lattice to ensure ergodicity, we define the operators 
\begin{align}
   \mathcal{P}_{\triangle} &= \mathcal{P}_{\square_1, i, j} \mathcal{P}_{\square_2, j, k} \mathcal{P}_{\square_3, k, i},\\
   \mathcal{P}_{\hexagon} &= \mathcal{P}_{\square_1, i, j} \mathcal{P}_{\square_2, j, k} \mathcal{P}_{\square_3, k, i} \mathcal{P}_{\square_4, i, j} \mathcal{P}_{\square_5, j, k} \mathcal{P}_{\square_6, k, i},
\end{align}
where the labels $1,\cdots,4$ and $1,\cdots,6$ index the vertices of the Kagome lattice that partake in the particular arrow reversal process.  Treating this model perturbatively is less straightforward than for the triangular lattice, because the hexagonal arrow reversal terms come at sixth order perturbation theory along with undesirable non-local terms corresponding to two decoupled triangular exchanges.   An alternative approach would be to directly write down the ring exchange terms by hand.  Such a Hamiltonian is
\begin{equation}
    \mathcal{H}_{\text{eff}} = -\sum_{p = \textcolor{blue}{\triangle_b},\textcolor{blue}{\hexagon_b} }\prod_{\textcolor{blue}{\langle i, j \rangle_b} \in p} \mathcal{O}_{ij},
\end{equation}
where the reader is referred to the fourth panel of the second column in Figure~\ref{fig:baxterwu} to understand the various terms in the Hamiltonian above.  We have also assumed that this model is projected onto both the even parity subspace ($\prod_{i \in \textcolor{red}{\hexagon_r}} S_i^z = 1$) and the arrow constrained subspace $S_i^z S_j^z = -1$.  We may deform this Hamiltonian by adding the dimer potential terms, yielding:
\begin{equation}
    \mathcal{H} = \mathcal{H}_{\text{eff}} + \sum_{\triangle} \mathcal{P}_{\triangle} + \sum_{\hexagon} \mathcal{P}_{\hexagon},
\end{equation}
which becomes frustration-free -- the ground state corresponds to superpositions of configurations of an arrowed vertex model on the Kagome lattice.  The point $u = 1$ corresponds to a commuting projector model with $\mathbb{Z}_2$ topological order.  As before, one may construct $e$ particles by string operators emanating from a red square and $m$ particles by string operators emanating from the blue hexagons and blue dodecagons in the spin-doubled lattice.  

When $u = \infty$, we argued that the classical arrowed vertex model is equivalent to a 3-coloring model on the honeycomb lattice.  The quantum vertex model also approaches the quantum 3-coloring model, so long as one adds additional terms to the quantum vertex model above.  In particular, one adds a term by hand: 
\begin{equation}
   \mathcal{H}_{\text{eff}} \to  \mathcal{H}_{\text{eff}} - \sum_{p = \textcolor{blue}{\left(\triangle_1, \cdots, \triangle_6\right)}}\prod_{\textcolor{blue}{\langle i, j \rangle_b} \in p} \mathcal{O}_{ij},
\end{equation}
where the six triangles surround a hexagon in the Kagome lattice.  In addition, one adds an analogous dimer potential energy such that the Hamiltonian remains frustration free.  The ground state of this model still coincides with classical configurations of the arrowed vertex model.  With this additional term added to the Hamiltonian, it can be seen that when $u = \infty$ the allowed ring exchange processes coincide with those of the quantum 3-coloring model~\cite{chamoncoloring}:
\begin{equation}
    \mathcal{H}_{\text{3-color}} = \sum_{\hexagon, a \neq b} \left(\ket{\hexagona} - \ket{\hexagonb}\right)\left(\bra{\hexagona}-\bra{\hexagonb}\right)
\end{equation}
where the Hamiltonian is defined on the honeycomb lattice, and $a,b = 1,2,3$ denote colors of bonds.  In this limit, the model becomes gapless.  This quantum three coloring model emerges for \emph{any quantum vertex model defined on a lattice of corner sharing triangles}.

The quantum vertex model constructions support the claim that a balanced dimer model such as the BFG model or the two dimer model on a Kagome lattice supports a $\mathbb{Z}_2$ spin liquid phase.  This is due to the fact that the BDVM duality maps the balanced dimer models to a point in the disordered phase of the arrowed vertex model.  Through numerical evidence of a lack of a phase transition (see Section~\ref{numerics}) as well as the existence of an exactly solvable point exhibiting toric code topological order, we can claim that balanced dimer models on the triangular and Kagome lattices support spin liquid phases and conjecture that this holds for a balanced dimer model on any non-bipartite lattice.  

Suppose now that we want to define a quantum vertex model whose ground state configurations are comprised of bonded vertex model configurations.  Then, one can replace pairs of spins corresponding to an arrow with a single spin which corresponds to a dimer/bond variable, at the expense of writing a Hamiltonian that breaks the $U(1)$ symmetry corresponding to the total magnetization.  Alternatively, one can keep both spins associated with an arrow but force a \emph{ferromagnetic} coupling between them, similar to what was done in Section \ref{examplehoney}).  The ring exchange terms are identical and the interaction term takes the same form as before.  However, these models have a markedly different phase diagram.  The square lattice model is the Ardonne-Fendley-Fradkin quantum 8-vertex model analyzed in Ref.~\cite{AFF}.  On a triangular lattice, Wegner's duality indicates an ordering transition at $u = 5$ to a ferromagnetic phase, while below $u = 5$, the model is in a disordered phase.  To see that the disordered phase is topologically ordered, at $u = 1$, this model reduces to the toric code model on a triangular lattice:
\begin{equation}
    \mathcal{H} \sim -\sum_{\includegraphics[scale = 0.08]{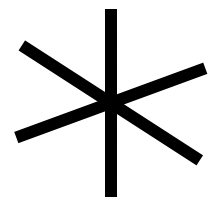}} ZZZZZZ - \sum_{\triangle} XXX,
\end{equation}
which is $\mathbb{Z}_2$ topologically ordered.  On the Kagome lattice, there is a transition at $u = 3$, potentially unknown transitions at other values of $u \neq 3$ (which are related to each other by Wegner's duality), and at $u = 1$ the model reduces to a toric code model:
\begin{equation}
    \mathcal{H} \sim -\sum_{\includegraphics[scale = 0.10]{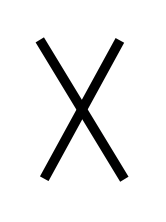}} ZZZZ - \sum_{\triangle} XXX - \sum_{\hexagon} XXXXXX
\end{equation}
that also exhibits $\mathbb{Z}_2$ topological order.  Furthermore, when $u = 0$, this model reduces to the two-dimer model on the ruby lattice.  Numerics indicate that this point is smoothly connected to the $u = 1$ point, and thus $u = 0$ should also be $\mathbb{Z}_2$ topologically ordered, which provides an alternate proof that the ruby lattice model is topologically ordered.  Note that this argument cannot be used for the BFG model on the triangular lattice, as the three dimer constraint cannot be obtained from any suitable limit of the bonded vertex model.  This procedure can be generalized to all of the lattices in the BFG and ruby families, which enjoy the self duality property but may also contain intermediate ordered phases.  These are left to future works. 


\subsection{Phase diagram and beyond the RK point}

\begin{figure}
    \centering
    \includegraphics[scale=0.48]{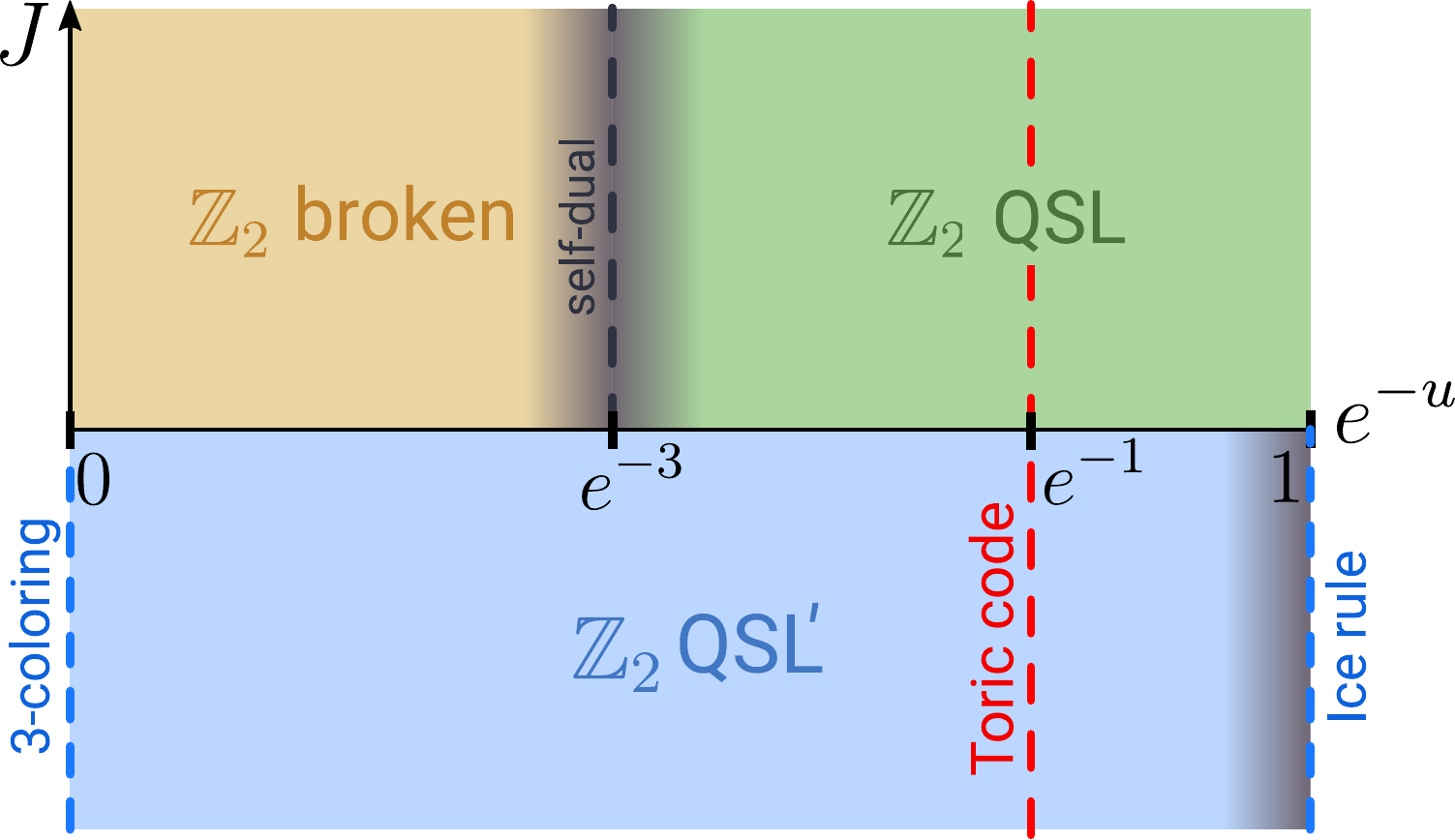}
    \caption{The phase diagram for the Kagome lattice vertex model.  The dark gray region indicates potentially unknown ordered phases that may emerge in the vicinity of the self dual and ice rule lines.  For the triangular lattice, the ice rule and coloring lines are no longer critical lines, and the self dual line occurs at $u = 5$.}
    \label{fig:kagomediagram}
\end{figure}

In this subsection, we will explicitly illustrate both the RK phase diagrams including the self-dual and gapless points, as well as provide a discussion of some additional limits of the model beyond the RK point.  In the previous subsection, we discussed how to obtain the quantum vertex model in a perturbative limit of an easy-axis Ising model, but in the present, we will work with the more presentable Hamiltonian which is similar to the Hamiltonian in Eqn.~\ref{eq:newarrowedvertham}:
\begin{equation}
    \mathcal{H} = \sum_{\ell, C_{\ell}} \mathcal{P}\left(\omega(\overline{C}_{\ell}, C_{\ell})^{1/2} \ket{C_{\ell}} - \omega(\overline{C}_{\ell}, C_{\ell})^{-1/2} \ket{\overline{C}_{\ell}}\right),
\end{equation}
where $\mathcal{P}(\ket{\psi}) = \ketbra{\psi}{\psi}$, $\ell$ is a minimal length loop on the lattice (e.g. triangles in the triangular lattice), and $C$ is an assignment of spin variables around each vertex involved in loop $\ell$, with $\overline{C}$ denoting the configuration where spin variables \emph{only on the edges of $\ell$} are negated.  Again, we assume the Hamiltonian is projected onto the even dimer subspace, which can be implemented energetically by adding another term to the Hamiltonian.  The quantity $\omega(\overline{C}_{\ell}, C_{\ell})$ is the square root of the ratio of the Boltzmann weights of $\overline{C}_{\ell}$ to $C_{\ell}$ in the classical vertex model.  In addition, to tune between the arrowed and bonded vertex models, we have two spins per link and add the fusion term
\begin{equation}
    \mathcal{H} \to \mathcal{H} - J \sum_{\langle i,j\rangle} Z_i Z_j.
\end{equation}
In the triangular and kagome lattices, $\omega(\overline{C}_{\ell}, C_{\ell})$ only depends on $u$ -- for the kagome lattice, we display the RK phase diagram as a function of $(\exp(-u), J)$ in Figure~\ref{fig:kagomediagram} (we use $\exp(-u)$ to compactify the interval $[0,\infty)$ to $[0,1]$).  The triangular lattice phase diagram is mentioned in the caption of Figure~\ref{fig:kagomediagram}.

Though most of our discussion has been focused on the RK manifold, there are some natural deformations away from the RK manifold which can be studied.  We choose to split $\mathcal{H} = \mathcal{H}_A + \mathcal{H}_B$, with
\begin{equation}
    \mathcal{H}_A = \sum_{\ell, C_{\ell}} \omega(\overline{C}_{\ell}, C_{\ell})\ketbra{C_{\ell}}{C_{\ell}} + \frac{1}{\omega(\overline{C}_{\ell}, C_{\ell})}\ketbra{\overline{C}_{\ell}}{\overline{C}_{\ell}}
\end{equation}
and
\begin{equation}
    \mathcal{H}_B = -\sum_{\ell, C_{\ell}} \left(\ketbra{C_{\ell}}{\overline{C}_{\ell}} + \text{h.c.}\right)
\end{equation}
This resembles the canonical dimer model, and we consider the deformation $\mathcal{H}(r) = \mathcal{H}_A + r \mathcal{H}_B$.  When $r = 1$, this is the RK quantum vertex model.  Let us choose $J > 0$.  Other limits of interest include:
\begin{itemize}
    \item $r = 0$:  The Hamiltonian becomes diagonal in the dimer basis, and the ground states are classical configurations of the vertex model.  The ground state configuration(s) is found by minimizing:
    \begin{equation}
        C_{\text{gs}} = \text{argmin}_C \left( \frac{\sum_{\ell} \sqrt{\mathcal{W}(\overline{C}_{\ell})}}{\sqrt{\mathcal{W}(C)}}\right).
    \end{equation}
    where $\mathcal{W}(C)$ denotes the Boltzmann weight of the classical configuration $C$ in the classical vertex model.  When $u > 1$, the minimizing configuration is the ferromagnetic configurations (the all dimer or no dimer configurations), rendering this region a $\mathbb{Z}_2$ broken phase.  When $u < 1$, the ground state degeneracy is likely extensive, spanned by certain configurations of 2 and 4 dimer configurations upon a preliminary analysis. 
    
    \item $u = 1$:  When $u = 1$, the contribution $\mathcal{H}_A$ is proportional to the identity and what remains is $\mathcal{H}_B$ projected onto the even dimer subspace.  This therefore gives toric code topological order, regardless of the value of $r$ (notably, changing the sign of $r$ changes the charge sector of the toric code that one works in).

    \item $r = -1$:  At this point, a ``dual'' RK point emerges only for the triangular lattice vertex model.  The Hamiltonian can be written in the form
    \begin{equation}
    \mathcal{H} = \sum_{\ell, C_{\ell}} \mathcal{P}\left(\omega(\overline{C}_{\ell}, C_{\ell})^{1/2} \ket{C_{\ell}} + \omega(\overline{C}_{\ell}, C_{\ell})^{-1/2} \ket{\overline{C}_{\ell}}\right),
    \end{equation}
    which is positive semidefinite.  Further, one can show the existence of a zero energy ground state
    \begin{equation}
        \ket{\text{GS}} = \sum_{C} (-1)^{d(C)}\sqrt{\mathcal{W}(C)} \ket{C} 
    \end{equation}
    where $d(C)$ counts the number of dimers in $C$, and $\mathcal{W}(C)$ is the Boltzmann weight of classical configuration $C$.  This follows from the fact that the loop updates are around triangles -- on the Kagome or square lattice where there are hexagonal and square loops respectively, this point does not admit an exact ground state.  Since $\braket{\text{GS}}{\text{GS}}$ is the partition function of the classical vertex model on a triangular lattice, the phase diagram is identical to the one we have already discussed.
    \item $r = \pm \infty$:  At this point, the dominant term (assuming it is much larger than the energetic term enforcing the even dimer Hilbert space constraint) is $\mathcal{H}_B$.  The ground state of this Hamiltonian at $r = \infty$ (for example on the triangular lattice) are product states in the $X$-eigenstate bases $\ket{\pm}$ satisfying the constraint that the number of $\ket{-}$ states for each triangle is even.  When $r = -\infty$, the number of $\ket{+}$ states for each triangle is even.  When both the Hilbert space constraint and $\mathcal{H}_A$ are treated as perturbations, one may expect different results based on their relative magnitudes.  The ground state becomes topologically ordered when the Hilbert space constraint is dominant, and likely ordered otherwise.  This is a subject of future work.
\end{itemize}
Further numerical investigations are needed to understand various aspects of this phase diagram, particularly in the regime where $|r| < 1$ and $u < 1$.  Furthermore, the location of the transitions (both for fixed $r$ varying $u$ and fixed $u$ varying $r$) are also subjects of future work.

\section{Discussion}

We first would like to recapitulate the main points of this paper:
\begin{itemize}
    \item We consider easy-axis antiferromagnets on lattices of corner-sharing polygons with all-to-all interactions within each polygon (examples being the BFG and Ruby families).  We tune these Hamiltonians to a frustration-free RK point where the ground state is a uniform superposition of spin configurations satisfying a zero magnetization constraint on each polygon.
    \item Using wavefunction dualities, we show that if the dual lattice is bipartite, then the wavefunction above can be mapped onto the critical point of a bonded classical vertex model. For example, the corner sharing square lattice antiferromagnet has an RK wavefunction which can be mapped onto an 8 vertex model where the vertex parameters are known to be critical.
    \item If the dual lattice is not bipartite, the wavefunction is mapped onto a point on the phase diagram of an arrowed classical vertex model.
    \item Analytical and numerical arguments indicate that the arrowed classical vertex model has no phase transition as a function of the vertex weights, and we argue that this means the corresponding quantum spin model is in a topologically ordered phase.
    \item We present three different constructions (one of which is discussed in Appendix~\ref{offcriticalHam}) of Hamiltonians corresponding to perturbations about these points on the classical vertex model phase diagram. 
\end{itemize}

Our analysis has been restricted mostly to characterizing properties of the ground states of the aforementioned spin models.  By a variational method one can study the nature of excitations in this theory (see Ref \cite{RK}).  Furthermore, since the dualities we presented are at the wavefunction level, we inquire whether it can be possible to construct duality on the operator level, which would be also be useful for identifying excitations.

One would benefit from having exact solutions of ice-rule models on a general lattice via Bethe ansatz.  Such a solution was found by Kelland \cite{Kelland} for the triangular lattice by observing a $U(1)$ symmetry corresponding to arrow continuity; this symmetry exists on other lattices, so it seems plausible for these models to be integrable.  One point we did not discuss is that these ice rule models possess a representation in terms of height fields -- in the continuum limit, we expect the effective field theory for these height fields to be a massless gaussian theory.  Going beyond the ice rule constraints to consider a general arrowed vertex possibly breaks integrability, though we think it may still be tractable to prove that these models do not exhibit a phase transition.

Another unexplored question relates to the exact nature of the phase diagram of vertex models for larger values of the coordination number $V_p$.  For $V_p = 8$, we had mentioned that the vertex model is parameterized by two parameters: $W_8(0)/W_8(4) = u$ and $W_8(2)/W_8(4) = v$.  Wegner's duality gives us a self-dual line occurring at $u - 4v = 5$.  The low ($u \gg v$) and high ($u \approx v \approx 1$) temperature phases are ferromagnetic and disordered, and the weights $W_8(n) = \langle \cos^n x \sin^{8-n} x \rangle$ lie in the self-dual boundary. However, it is possible that either there are additional critical lines in this model, or that there is a gapless intermediate phase and the self-dual line is not a critical point.  We conjecture that critical exponents vary continuously along the self-dual boundary, in correspondence with the behavior in the 8-vertex model \cite{AFF}.  The quantum Hamiltonians corresponding to such vertex models may possess similarly interesting behavior.  A follow-up paper discusses these quantum vertex models in 3+1D \cite{Bala}.

One point to emphasize is that although most of these models do not have experimental applicability, one promising model is the easy-axis Ising model on the ruby lattice: $\mathcal{H}_{\text{ruby}} = \sum_{\square} \vec{S}_{\square} \cdot \vec{S}_{\square}$, discussed in Section \ref{otherexamples}.  In the easy-axis limit, and tuned to the RK point, we argue that this model can support a $\mathbb{Z}_2$ spin liquid phase.  This model only requires second nearest neighbor interactions on square plaquettes, unlike in the BFG model, where third nearest neighbor interactions are needed \cite{BFG}.  However, further work is necessary to understand whether the spin liquid phase persists in the pure ring exchange limit. Note that these models feature two dimers on the dual kagome lattice; in contrast, the case of a single dimer on the same dual lattice was discussed in  the easy axis limit \cite{Pollmann} and Ising limits \cite{Ruben}, where spin liquid phases were identified even away from fine tuned limits.

Finally, moving beyond the RK point is an important direction that has only been minimally addressed in this paper.  Getting a better analytical understanding of the transitions and the interplay of phases is an important direction to pursue.  Furthermore, we can consider Hamiltonians of the form $\mathcal{H} = \mathcal{H}_A + r \mathcal{H}_B$, where
\begin{equation}
    \mathcal{H}_A = \sum_{\ell, C_{\ell}} \xi(C_{\ell}) \left(\omega(\overline{C}_{\ell}, C_{\ell})\ketbra{C_{\ell}}{C_{\ell}} + \frac{1}{\omega(\overline{C}_{\ell}, C_{\ell})}\ketbra{\overline{C}_{\ell}}{\overline{C}_{\ell}}\right)
\end{equation}
and
\begin{equation}
    \mathcal{H}_B = -\sum_{\ell, C_{\ell}} \left(\ketbra{C_{\ell}}{\overline{C}_{\ell}} + \text{h.c.}\right)
\end{equation}
with $\xi(C_{\ell})$ positive parameters depending on the local spin configuration around a loop $C_{\ell}$.  When $r = 1$, an RK point for the quantum vertex model is still achieved.  However, for different choices of $\xi(C_{\ell})$, one can realize unusual ordered phases at $r = \pm \infty$ and $r = 0$.  Characterizing these phases and their transitions into topologically ordered phases would likely be a interesting pursuit.

\section{Acknowledgments}

We would like to thank Daniel Bulmash, Olexei Motrunich (S.B. and V.G.), Steven Kivelson (V.G.), and Ruben Verresen (S.B.) for useful discussions.  A.V. and V. G. were supported by the Simons Collaboration on Ultra Quantum Matter, a grant from the Simons Foundation (651440, A.V.).  S.B. was supported by the National Science Foundation Graduate Research Fellowship under Grant No. 1745302. V.G. was also supported by NSF DMR-2037158 and US-ARO Contract No.W911NF1310172.

\bibliography{cit}

\begin{thebibliography}{55}%
\makeatletter
\providecommand \@ifxundefined [1]{%
 \@ifx{#1\undefined}
}%
\providecommand \@ifnum [1]{%
 \ifnum #1\expandafter \@firstoftwo
 \else \expandafter \@secondoftwo
 \fi
}%
\providecommand \@ifx [1]{%
 \ifx #1\expandafter \@firstoftwo
 \else \expandafter \@secondoftwo
 \fi
}%
\providecommand \natexlab [1]{#1}%
\providecommand \enquote  [1]{``#1''}%
\providecommand \bibnamefont  [1]{#1}%
\providecommand \bibfnamefont [1]{#1}%
\providecommand \citenamefont [1]{#1}%
\providecommand \href@noop [0]{\@secondoftwo}%
\providecommand \href [0]{\begingroup \@sanitize@url \@href}%
\providecommand \@href[1]{\@@startlink{#1}\@@href}%
\providecommand \@@href[1]{\endgroup#1\@@endlink}%
\providecommand \@sanitize@url [0]{\catcode `\\12\catcode `\$12\catcode
  `\&12\catcode `\#12\catcode `\^12\catcode `\_12\catcode `\%12\relax}%
\providecommand \@@startlink[1]{}%
\providecommand \@@endlink[0]{}%
\providecommand \url  [0]{\begingroup\@sanitize@url \@url }%
\providecommand \@url [1]{\endgroup\@href {#1}{\urlprefix }}%
\providecommand \urlprefix  [0]{URL }%
\providecommand \Eprint [0]{\href }%
\providecommand \doibase [0]{http://dx.doi.org/}%
\providecommand \selectlanguage [0]{\@gobble}%
\providecommand \bibinfo  [0]{\@secondoftwo}%
\providecommand \bibfield  [0]{\@secondoftwo}%
\providecommand \translation [1]{[#1]}%
\providecommand \BibitemOpen [0]{}%
\providecommand \bibitemStop [0]{}%
\providecommand \bibitemNoStop [0]{.\EOS\space}%
\providecommand \EOS [0]{\spacefactor3000\relax}%
\providecommand \BibitemShut  [1]{\csname bibitem#1\endcsname}%
\let\auto@bib@innerbib\@empty
\bibitem [{\citenamefont {Anderson}(1973)}]{Anderson1}%
  \BibitemOpen
  \bibfield  {author} {\bibinfo {author} {\bibfnamefont {P.}~\bibnamefont
  {Anderson}},\ }\href {\doibase https://doi.org/10.1016/0025-5408(73)90167-0}
  {\bibfield  {journal} {\bibinfo  {journal} {Materials Research Bulletin}\
  }\textbf {\bibinfo {volume} {8}},\ \bibinfo {pages} {153} (\bibinfo {year}
  {1973})}\BibitemShut {NoStop}%
\bibitem [{\citenamefont {Anderson}(1987)}]{Anderson2}%
  \BibitemOpen
  \bibfield  {author} {\bibinfo {author} {\bibfnamefont {P.~W.}\ \bibnamefont
  {Anderson}},\ }\href {\doibase 10.1126/science.235.4793.1196} {\bibfield
  {journal} {\bibinfo  {journal} {Science}\ }\textbf {\bibinfo {volume}
  {235}},\ \bibinfo {pages} {1196} (\bibinfo {year} {1987})}\BibitemShut
  {NoStop}%
\bibitem [{\citenamefont {Rokhsar}\ and\ \citenamefont {Kivelson}(1988)}]{RK}%
  \BibitemOpen
  \bibfield  {author} {\bibinfo {author} {\bibfnamefont {D.~S.}\ \bibnamefont
  {Rokhsar}}\ and\ \bibinfo {author} {\bibfnamefont {S.~A.}\ \bibnamefont
  {Kivelson}},\ }\href {\doibase 10.1103/PhysRevLett.61.2376} {\bibfield
  {journal} {\bibinfo  {journal} {Phys. Rev. Lett.}\ }\textbf {\bibinfo
  {volume} {61}},\ \bibinfo {pages} {2376} (\bibinfo {year}
  {1988})}\BibitemShut {NoStop}%
\bibitem [{\citenamefont {Kivelson}\ \emph {et~al.}(1987)\citenamefont
  {Kivelson}, \citenamefont {Rokhsar},\ and\ \citenamefont {Sethna}}]{SRK}%
  \BibitemOpen
  \bibfield  {author} {\bibinfo {author} {\bibfnamefont {S.~A.}\ \bibnamefont
  {Kivelson}}, \bibinfo {author} {\bibfnamefont {D.~S.}\ \bibnamefont
  {Rokhsar}}, \ and\ \bibinfo {author} {\bibfnamefont {J.~P.}\ \bibnamefont
  {Sethna}},\ }\href {\doibase 10.1103/PhysRevB.35.8865} {\bibfield  {journal}
  {\bibinfo  {journal} {Phys. Rev. B}\ }\textbf {\bibinfo {volume} {35}},\
  \bibinfo {pages} {8865} (\bibinfo {year} {1987})}\BibitemShut {NoStop}%
\bibitem [{\citenamefont {Kasteleyn}(1963)}]{Kasteleyn}%
  \BibitemOpen
  \bibfield  {author} {\bibinfo {author} {\bibfnamefont {P.~W.}\ \bibnamefont
  {Kasteleyn}},\ }\href {\doibase 10.1063/1.1703953} {\bibfield  {journal}
  {\bibinfo  {journal} {Journal of Mathematical Physics}\ }\textbf {\bibinfo
  {volume} {4}},\ \bibinfo {pages} {287} (\bibinfo {year} {1963})},\ \Eprint
  {http://arxiv.org/abs/https://doi.org/10.1063/1.1703953}
  {https://doi.org/10.1063/1.1703953} \BibitemShut {NoStop}%
\bibitem [{\citenamefont {Temperley}\ and\ \citenamefont
  {Fisher}(1961)}]{TemperleyFisher}%
  \BibitemOpen
  \bibfield  {author} {\bibinfo {author} {\bibfnamefont {H.~N.~V.}\
  \bibnamefont {Temperley}}\ and\ \bibinfo {author} {\bibfnamefont {M.~E.}\
  \bibnamefont {Fisher}},\ }\href {\doibase 10.1080/14786436108243366}
  {\bibfield  {journal} {\bibinfo  {journal} {The Philosophical Magazine: A
  Journal of Theoretical Experimental and Applied Physics}\ }\textbf {\bibinfo
  {volume} {6}},\ \bibinfo {pages} {1061} (\bibinfo {year} {1961})},\ \Eprint
  {http://arxiv.org/abs/https://doi.org/10.1080/14786436108243366}
  {https://doi.org/10.1080/14786436108243366} \BibitemShut {NoStop}%
\bibitem [{\citenamefont {Fisher}(1961)}]{Fisher}%
  \BibitemOpen
  \bibfield  {author} {\bibinfo {author} {\bibfnamefont {M.~E.}\ \bibnamefont
  {Fisher}},\ }\href {\doibase 10.1103/PhysRev.124.1664} {\bibfield  {journal}
  {\bibinfo  {journal} {Phys. Rev.}\ }\textbf {\bibinfo {volume} {124}},\
  \bibinfo {pages} {1664} (\bibinfo {year} {1961})}\BibitemShut {NoStop}%
\bibitem [{\citenamefont {Fisher}\ and\ \citenamefont
  {Stephenson}(1963)}]{FS2}%
  \BibitemOpen
  \bibfield  {author} {\bibinfo {author} {\bibfnamefont {M.~E.}\ \bibnamefont
  {Fisher}}\ and\ \bibinfo {author} {\bibfnamefont {J.}~\bibnamefont
  {Stephenson}},\ }\href {\doibase 10.1103/PhysRev.132.1411} {\bibfield
  {journal} {\bibinfo  {journal} {Phys. Rev.}\ }\textbf {\bibinfo {volume}
  {132}},\ \bibinfo {pages} {1411} (\bibinfo {year} {1963})}\BibitemShut
  {NoStop}%
\bibitem [{\citenamefont {Read}\ and\ \citenamefont {Sachdev}(1991)}]{RS}%
  \BibitemOpen
  \bibfield  {author} {\bibinfo {author} {\bibfnamefont {N.}~\bibnamefont
  {Read}}\ and\ \bibinfo {author} {\bibfnamefont {S.}~\bibnamefont {Sachdev}},\
  }\href {\doibase 10.1103/PhysRevLett.66.1773} {\bibfield  {journal} {\bibinfo
   {journal} {Phys. Rev. Lett.}\ }\textbf {\bibinfo {volume} {66}},\ \bibinfo
  {pages} {1773} (\bibinfo {year} {1991})}\BibitemShut {NoStop}%
\bibitem [{\citenamefont {Fradkin}\ and\ \citenamefont
  {Kivelson}(1990)}]{FradkinKivelson}%
  \BibitemOpen
  \bibfield  {author} {\bibinfo {author} {\bibfnamefont {E.}~\bibnamefont
  {Fradkin}}\ and\ \bibinfo {author} {\bibfnamefont {S.}~\bibnamefont
  {Kivelson}},\ }\href@noop {} {\bibfield  {journal} {\bibinfo  {journal}
  {Modern Physics Letters B}\ }\textbf {\bibinfo {volume} {4}},\ \bibinfo
  {pages} {225} (\bibinfo {year} {1990})}\BibitemShut {NoStop}%
\bibitem [{\citenamefont {Read}\ and\ \citenamefont {Sachdev}(1990)}]{RSPRB90}%
  \BibitemOpen
  \bibfield  {author} {\bibinfo {author} {\bibfnamefont {N.}~\bibnamefont
  {Read}}\ and\ \bibinfo {author} {\bibfnamefont {S.}~\bibnamefont {Sachdev}},\
  }\href {\doibase 10.1103/PhysRevB.42.4568} {\bibfield  {journal} {\bibinfo
  {journal} {Phys. Rev. B}\ }\textbf {\bibinfo {volume} {42}},\ \bibinfo
  {pages} {4568} (\bibinfo {year} {1990})}\BibitemShut {NoStop}%
\bibitem [{\citenamefont {Read}\ and\ \citenamefont
  {Chakraborty}(1989)}]{RC89}%
  \BibitemOpen
  \bibfield  {author} {\bibinfo {author} {\bibfnamefont {N.}~\bibnamefont
  {Read}}\ and\ \bibinfo {author} {\bibfnamefont {B.}~\bibnamefont
  {Chakraborty}},\ }\href {\doibase 10.1103/PhysRevB.40.7133} {\bibfield
  {journal} {\bibinfo  {journal} {Phys. Rev. B}\ }\textbf {\bibinfo {volume}
  {40}},\ \bibinfo {pages} {7133} (\bibinfo {year} {1989})}\BibitemShut
  {NoStop}%
\bibitem [{\citenamefont {Jalabert}\ and\ \citenamefont
  {Sachdev}(1991)}]{JalabertSachdev91}%
  \BibitemOpen
  \bibfield  {author} {\bibinfo {author} {\bibfnamefont {R.~A.}\ \bibnamefont
  {Jalabert}}\ and\ \bibinfo {author} {\bibfnamefont {S.}~\bibnamefont
  {Sachdev}},\ }\href {\doibase 10.1103/PhysRevB.44.686} {\bibfield  {journal}
  {\bibinfo  {journal} {Phys. Rev. B}\ }\textbf {\bibinfo {volume} {44}},\
  \bibinfo {pages} {686} (\bibinfo {year} {1991})}\BibitemShut {NoStop}%
\bibitem [{\citenamefont {Fradkin}(2013)}]{fradkin_2013}%
  \BibitemOpen
  \bibfield  {author} {\bibinfo {author} {\bibfnamefont {E.}~\bibnamefont
  {Fradkin}},\ }\href {\doibase 10.1017/CBO9781139015509} {\emph {\bibinfo
  {title} {Field Theories of Condensed Matter Physics}}},\ \bibinfo {edition}
  {2nd}\ ed.\ (\bibinfo  {publisher} {Cambridge University Press},\ \bibinfo
  {year} {2013})\BibitemShut {NoStop}%
\bibitem [{\citenamefont {Sachdev}\ and\ \citenamefont
  {Vojta}(2000)}]{Sachdev2000}%
  \BibitemOpen
  \bibfield  {author} {\bibinfo {author} {\bibfnamefont {S.}~\bibnamefont
  {Sachdev}}\ and\ \bibinfo {author} {\bibfnamefont {M.}~\bibnamefont
  {Vojta}},\ }\href {arXiv:cond-mat/9910231} {\bibfield  {journal} {\bibinfo
  {journal} {J. Phys. Soc. Jpn.}\ }\textbf {\bibinfo {volume} {69}},\ \bibinfo
  {pages} {1} (\bibinfo {year} {2000})}\BibitemShut {NoStop}%
\bibitem [{\citenamefont {Wen}(1991)}]{Wen91}%
  \BibitemOpen
  \bibfield  {author} {\bibinfo {author} {\bibfnamefont {X.~G.}\ \bibnamefont
  {Wen}},\ }\href {\doibase 10.1103/PhysRevB.44.2664} {\bibfield  {journal}
  {\bibinfo  {journal} {Phys. Rev. B}\ }\textbf {\bibinfo {volume} {44}},\
  \bibinfo {pages} {2664} (\bibinfo {year} {1991})}\BibitemShut {NoStop}%
\bibitem [{\citenamefont {Sachdev}(1992)}]{Sachdev92}%
  \BibitemOpen
  \bibfield  {author} {\bibinfo {author} {\bibfnamefont {S.}~\bibnamefont
  {Sachdev}},\ }\href {\doibase 10.1103/PhysRevB.45.12377} {\bibfield
  {journal} {\bibinfo  {journal} {Phys. Rev. B}\ }\textbf {\bibinfo {volume}
  {45}},\ \bibinfo {pages} {12377} (\bibinfo {year} {1992})}\BibitemShut
  {NoStop}%
\bibitem [{\citenamefont {Moessner}\ and\ \citenamefont {Sondhi}(2001)}]{MS}%
  \BibitemOpen
  \bibfield  {author} {\bibinfo {author} {\bibfnamefont {R.}~\bibnamefont
  {Moessner}}\ and\ \bibinfo {author} {\bibfnamefont {S.~L.}\ \bibnamefont
  {Sondhi}},\ }\href {\doibase 10.1103/PhysRevLett.86.1881} {\bibfield
  {journal} {\bibinfo  {journal} {Phys. Rev. Lett.}\ }\textbf {\bibinfo
  {volume} {86}},\ \bibinfo {pages} {1881} (\bibinfo {year}
  {2001})}\BibitemShut {NoStop}%
\bibitem [{\citenamefont {Fendley}\ \emph {et~al.}(2002)\citenamefont
  {Fendley}, \citenamefont {Moessner},\ and\ \citenamefont {Sondhi}}]{FMS}%
  \BibitemOpen
  \bibfield  {author} {\bibinfo {author} {\bibfnamefont {P.}~\bibnamefont
  {Fendley}}, \bibinfo {author} {\bibfnamefont {R.}~\bibnamefont {Moessner}}, \
  and\ \bibinfo {author} {\bibfnamefont {S.~L.}\ \bibnamefont {Sondhi}},\
  }\href {\doibase 10.1103/PhysRevB.66.214513} {\bibfield  {journal} {\bibinfo
  {journal} {Phys. Rev. B}\ }\textbf {\bibinfo {volume} {66}},\ \bibinfo
  {pages} {214513} (\bibinfo {year} {2002})}\BibitemShut {NoStop}%
\bibitem [{\citenamefont {Misguich}\ \emph {et~al.}(2002)\citenamefont
  {Misguich}, \citenamefont {Serban},\ and\ \citenamefont {Pasquier}}]{MSP}%
  \BibitemOpen
  \bibfield  {author} {\bibinfo {author} {\bibfnamefont {G.}~\bibnamefont
  {Misguich}}, \bibinfo {author} {\bibfnamefont {D.}~\bibnamefont {Serban}}, \
  and\ \bibinfo {author} {\bibfnamefont {V.}~\bibnamefont {Pasquier}},\ }\href
  {\doibase 10.1103/PhysRevLett.89.137202} {\bibfield  {journal} {\bibinfo
  {journal} {Phys. Rev. Lett.}\ }\textbf {\bibinfo {volume} {89}},\ \bibinfo
  {pages} {137202} (\bibinfo {year} {2002})}\BibitemShut {NoStop}%
\bibitem [{\citenamefont {Henley}(2004)}]{Henley}%
  \BibitemOpen
  \bibfield  {author} {\bibinfo {author} {\bibfnamefont {C.~L.}\ \bibnamefont
  {Henley}},\ }\href {\doibase 10.1088/0953-8984/16/11/045} {\bibfield
  {journal} {\bibinfo  {journal} {Journal of Physics: Condensed Matter}\
  }\textbf {\bibinfo {volume} {16}},\ \bibinfo {pages} {S891} (\bibinfo {year}
  {2004})}\BibitemShut {NoStop}%
\bibitem [{\citenamefont {Ardonne}\ \emph {et~al.}(2004)\citenamefont
  {Ardonne}, \citenamefont {Fendley},\ and\ \citenamefont {Fradkin}}]{AFF}%
  \BibitemOpen
  \bibfield  {author} {\bibinfo {author} {\bibfnamefont {E.}~\bibnamefont
  {Ardonne}}, \bibinfo {author} {\bibfnamefont {P.}~\bibnamefont {Fendley}}, \
  and\ \bibinfo {author} {\bibfnamefont {E.}~\bibnamefont {Fradkin}},\ }\href
  {\doibase https://doi.org/10.1016/j.aop.2004.01.004} {\bibfield  {journal}
  {\bibinfo  {journal} {Annals of Physics}\ }\textbf {\bibinfo {volume}
  {310}},\ \bibinfo {pages} {493} (\bibinfo {year} {2004})}\BibitemShut
  {NoStop}%
\bibitem [{\citenamefont {Fazekas}\ and\ \citenamefont
  {Anderson}(1974)}]{FazekasAnderson}%
  \BibitemOpen
  \bibfield  {author} {\bibinfo {author} {\bibfnamefont {P.}~\bibnamefont
  {Fazekas}}\ and\ \bibinfo {author} {\bibfnamefont {P.~W.}\ \bibnamefont
  {Anderson}},\ }\href {\doibase 10.1080/14786439808206568} {\bibfield
  {journal} {\bibinfo  {journal} {The Philosophical Magazine: A Journal of
  Theoretical Experimental and Applied Physics}\ }\textbf {\bibinfo {volume}
  {30}},\ \bibinfo {pages} {423} (\bibinfo {year} {1974})},\ \Eprint
  {http://arxiv.org/abs/https://doi.org/10.1080/14786439808206568}
  {https://doi.org/10.1080/14786439808206568} \BibitemShut {NoStop}%
\bibitem [{\citenamefont {Balents}\ \emph {et~al.}(2002)\citenamefont
  {Balents}, \citenamefont {Fisher},\ and\ \citenamefont {Girvin}}]{BFG}%
  \BibitemOpen
  \bibfield  {author} {\bibinfo {author} {\bibfnamefont {L.}~\bibnamefont
  {Balents}}, \bibinfo {author} {\bibfnamefont {M.~P.~A.}\ \bibnamefont
  {Fisher}}, \ and\ \bibinfo {author} {\bibfnamefont {S.~M.}\ \bibnamefont
  {Girvin}},\ }\href {\doibase 10.1103/PhysRevB.65.224412} {\bibfield
  {journal} {\bibinfo  {journal} {Phys. Rev. B}\ }\textbf {\bibinfo {volume}
  {65}},\ \bibinfo {pages} {224412} (\bibinfo {year} {2002})}\BibitemShut
  {NoStop}%
\bibitem [{\citenamefont {Senthil}\ and\ \citenamefont
  {Motrunich}(2002)}]{MotrunichSenthil2D}%
  \BibitemOpen
  \bibfield  {author} {\bibinfo {author} {\bibfnamefont {T.}~\bibnamefont
  {Senthil}}\ and\ \bibinfo {author} {\bibfnamefont {O.}~\bibnamefont
  {Motrunich}},\ }\href {\doibase 10.1103/physrevb.66.205104} {\bibfield
  {journal} {\bibinfo  {journal} {Physical Review B}\ }\textbf {\bibinfo
  {volume} {66}} (\bibinfo {year} {2002}),\
  10.1103/physrevb.66.205104}\BibitemShut {NoStop}%
\bibitem [{\citenamefont {Sheng}\ and\ \citenamefont {Balents}(2005)}]{SB}%
  \BibitemOpen
  \bibfield  {author} {\bibinfo {author} {\bibfnamefont {D.~N.}\ \bibnamefont
  {Sheng}}\ and\ \bibinfo {author} {\bibfnamefont {L.}~\bibnamefont
  {Balents}},\ }\href {\doibase 10.1103/PhysRevLett.94.146805} {\bibfield
  {journal} {\bibinfo  {journal} {Phys. Rev. Lett.}\ }\textbf {\bibinfo
  {volume} {94}},\ \bibinfo {pages} {146805} (\bibinfo {year}
  {2005})}\BibitemShut {NoStop}%
\bibitem [{\citenamefont {Samuel}(1980{\natexlab{a}})}]{S1}%
  \BibitemOpen
  \bibfield  {author} {\bibinfo {author} {\bibfnamefont {S.}~\bibnamefont
  {Samuel}},\ }\href {\doibase 10.1063/1.524404} {\bibfield  {journal}
  {\bibinfo  {journal} {Journal of Mathematical Physics}\ }\textbf {\bibinfo
  {volume} {21}},\ \bibinfo {pages} {2806} (\bibinfo {year}
  {1980}{\natexlab{a}})},\ \Eprint
  {http://arxiv.org/abs/https://doi.org/10.1063/1.524404}
  {https://doi.org/10.1063/1.524404} \BibitemShut {NoStop}%
\bibitem [{\citenamefont {Samuel}(1980{\natexlab{b}})}]{S2}%
  \BibitemOpen
  \bibfield  {author} {\bibinfo {author} {\bibfnamefont {S.}~\bibnamefont
  {Samuel}},\ }\href {\doibase 10.1063/1.524405} {\bibfield  {journal}
  {\bibinfo  {journal} {Journal of Mathematical Physics}\ }\textbf {\bibinfo
  {volume} {21}},\ \bibinfo {pages} {2815} (\bibinfo {year}
  {1980}{\natexlab{b}})},\ \Eprint
  {http://arxiv.org/abs/https://doi.org/10.1063/1.524405}
  {https://doi.org/10.1063/1.524405} \BibitemShut {NoStop}%
\bibitem [{\citenamefont {Samuel}(1980{\natexlab{c}})}]{S3}%
  \BibitemOpen
  \bibfield  {author} {\bibinfo {author} {\bibfnamefont {S.}~\bibnamefont
  {Samuel}},\ }\href {\doibase 10.1063/1.524406} {\bibfield  {journal}
  {\bibinfo  {journal} {Journal of Mathematical Physics}\ }\textbf {\bibinfo
  {volume} {21}},\ \bibinfo {pages} {2820} (\bibinfo {year}
  {1980}{\natexlab{c}})},\ \Eprint
  {http://arxiv.org/abs/https://doi.org/10.1063/1.524406}
  {https://doi.org/10.1063/1.524406} \BibitemShut {NoStop}%
\bibitem [{\citenamefont {Fradkin}\ \emph {et~al.}(2004)\citenamefont
  {Fradkin}, \citenamefont {Huse}, \citenamefont {Moessner}, \citenamefont
  {Oganesyan},\ and\ \citenamefont {Sondhi}}]{FHMOS}%
  \BibitemOpen
  \bibfield  {author} {\bibinfo {author} {\bibfnamefont {E.}~\bibnamefont
  {Fradkin}}, \bibinfo {author} {\bibfnamefont {D.~A.}\ \bibnamefont {Huse}},
  \bibinfo {author} {\bibfnamefont {R.}~\bibnamefont {Moessner}}, \bibinfo
  {author} {\bibfnamefont {V.}~\bibnamefont {Oganesyan}}, \ and\ \bibinfo
  {author} {\bibfnamefont {S.~L.}\ \bibnamefont {Sondhi}},\ }\href {\doibase
  10.1103/PhysRevB.69.224415} {\bibfield  {journal} {\bibinfo  {journal} {Phys.
  Rev. B}\ }\textbf {\bibinfo {volume} {69}},\ \bibinfo {pages} {224415}
  (\bibinfo {year} {2004})}\BibitemShut {NoStop}%
\bibitem [{\citenamefont {Wegner}(1973)}]{Wegner}%
  \BibitemOpen
  \bibfield  {author} {\bibinfo {author} {\bibfnamefont {F.}~\bibnamefont
  {Wegner}},\ }\href {\doibase https://doi.org/10.1016/0031-8914(73)90381-9}
  {\bibfield  {journal} {\bibinfo  {journal} {Physica}\ }\textbf {\bibinfo
  {volume} {68}},\ \bibinfo {pages} {570} (\bibinfo {year} {1973})}\BibitemShut
  {NoStop}%
\bibitem [{\citenamefont {Henley}(1997)}]{henleyheight}%
  \BibitemOpen
  \bibfield  {author} {\bibinfo {author} {\bibfnamefont {C.~L.}\ \bibnamefont
  {Henley}},\ }\href@noop {} {\bibfield  {journal} {\bibinfo  {journal}
  {Journal of statistical physics}\ }\textbf {\bibinfo {volume} {89}},\
  \bibinfo {pages} {483} (\bibinfo {year} {1997})}\BibitemShut {NoStop}%
\bibitem [{\citenamefont {Chakravarty}(2002)}]{Chakravarty}%
  \BibitemOpen
  \bibfield  {author} {\bibinfo {author} {\bibfnamefont {S.}~\bibnamefont
  {Chakravarty}},\ }\href {\doibase 10.1103/PhysRevB.66.224505} {\bibfield
  {journal} {\bibinfo  {journal} {Phys. Rev. B}\ }\textbf {\bibinfo {volume}
  {66}},\ \bibinfo {pages} {224505} (\bibinfo {year} {2002})}\BibitemShut
  {NoStop}%
\bibitem [{\citenamefont {Baxter}(1972)}]{Baxter8V}%
  \BibitemOpen
  \bibfield  {author} {\bibinfo {author} {\bibfnamefont {R.~J.}\ \bibnamefont
  {Baxter}},\ }\href {\doibase https://doi.org/10.1016/0003-4916(72)90335-1}
  {\bibfield  {journal} {\bibinfo  {journal} {Annals of Physics}\ }\textbf
  {\bibinfo {volume} {70}},\ \bibinfo {pages} {193} (\bibinfo {year}
  {1972})}\BibitemShut {NoStop}%
\bibitem [{\citenamefont {Baxter}(2016)}]{BaxterBook}%
  \BibitemOpen
  \bibfield  {author} {\bibinfo {author} {\bibfnamefont {R.~J.}\ \bibnamefont
  {Baxter}},\ }\href@noop {} {\emph {\bibinfo {title} {Exactly solved models in
  statistical mechanics}}}\ (\bibinfo  {publisher} {Elsevier},\ \bibinfo {year}
  {2016})\BibitemShut {NoStop}%
\bibitem [{\citenamefont {Runnels}(1976)}]{Runnels}%
  \BibitemOpen
  \bibfield  {author} {\bibinfo {author} {\bibfnamefont {L.}~\bibnamefont
  {Runnels}},\ }\href@noop {} {\bibfield  {journal} {\bibinfo  {journal}
  {Journal of Statistical Physics}\ }\textbf {\bibinfo {volume} {14}},\
  \bibinfo {pages} {39} (\bibinfo {year} {1976})}\BibitemShut {NoStop}%
\bibitem [{\citenamefont {Wu}\ and\ \citenamefont {Wu}(1989)}]{WuWu}%
  \BibitemOpen
  \bibfield  {author} {\bibinfo {author} {\bibfnamefont {X.~N.}\ \bibnamefont
  {Wu}}\ and\ \bibinfo {author} {\bibfnamefont {F.~Y.}\ \bibnamefont {Wu}},\
  }\href {\doibase 10.1088/0305-4470/22/2/003} {\bibfield  {journal} {\bibinfo
  {journal} {Journal of Physics A: Mathematical and General}\ }\textbf
  {\bibinfo {volume} {22}},\ \bibinfo {pages} {L55} (\bibinfo {year}
  {1989})}\BibitemShut {NoStop}%
\bibitem [{\citenamefont {Mouritsen}\ \emph {et~al.}(1981)\citenamefont
  {Mouritsen}, \citenamefont {Jensen},\ and\ \citenamefont {Frank}}]{MJF}%
  \BibitemOpen
  \bibfield  {author} {\bibinfo {author} {\bibfnamefont {O.~G.}\ \bibnamefont
  {Mouritsen}}, \bibinfo {author} {\bibfnamefont {S.~J.~K.}\ \bibnamefont
  {Jensen}}, \ and\ \bibinfo {author} {\bibfnamefont {B.}~\bibnamefont
  {Frank}},\ }\href {\doibase 10.1103/PhysRevB.23.976} {\bibfield  {journal}
  {\bibinfo  {journal} {Phys. Rev. B}\ }\textbf {\bibinfo {volume} {23}},\
  \bibinfo {pages} {976} (\bibinfo {year} {1981})}\BibitemShut {NoStop}%
\bibitem [{\citenamefont {Baxter}(1970)}]{baxtercoloring}%
  \BibitemOpen
  \bibfield  {author} {\bibinfo {author} {\bibfnamefont {R.}~\bibnamefont
  {Baxter}},\ }\href@noop {} {\bibfield  {journal} {\bibinfo  {journal}
  {Journal of Mathematical Physics}\ }\textbf {\bibinfo {volume} {11}},\
  \bibinfo {pages} {784} (\bibinfo {year} {1970})}\BibitemShut {NoStop}%
\bibitem [{\citenamefont {Lieb}(1967)}]{Lieb}%
  \BibitemOpen
  \bibfield  {author} {\bibinfo {author} {\bibfnamefont {E.~H.}\ \bibnamefont
  {Lieb}},\ }\href {\doibase 10.1103/PhysRev.162.162} {\bibfield  {journal}
  {\bibinfo  {journal} {Phys. Rev.}\ }\textbf {\bibinfo {volume} {162}},\
  \bibinfo {pages} {162} (\bibinfo {year} {1967})}\BibitemShut {NoStop}%
\bibitem [{\citenamefont {Hintermann}\ and\ \citenamefont
  {Merlini}(1972)}]{HintermannMerlini}%
  \BibitemOpen
  \bibfield  {author} {\bibinfo {author} {\bibfnamefont {A.}~\bibnamefont
  {Hintermann}}\ and\ \bibinfo {author} {\bibfnamefont {D.}~\bibnamefont
  {Merlini}},\ }\href {\doibase https://doi.org/10.1016/0375-9601(72)90261-7}
  {\bibfield  {journal} {\bibinfo  {journal} {Physics Letters A}\ }\textbf
  {\bibinfo {volume} {41}},\ \bibinfo {pages} {208 } (\bibinfo {year}
  {1972})}\BibitemShut {NoStop}%
\bibitem [{\citenamefont {Ding}\ \emph {et~al.}(2013)\citenamefont {Ding},
  \citenamefont {Wang}, \citenamefont {Zhang},\ and\ \citenamefont
  {Guo}}]{DWZG}%
  \BibitemOpen
  \bibfield  {author} {\bibinfo {author} {\bibfnamefont {C.}~\bibnamefont
  {Ding}}, \bibinfo {author} {\bibfnamefont {Y.}~\bibnamefont {Wang}}, \bibinfo
  {author} {\bibfnamefont {W.}~\bibnamefont {Zhang}}, \ and\ \bibinfo {author}
  {\bibfnamefont {W.}~\bibnamefont {Guo}},\ }\href {\doibase
  10.1103/PhysRevE.88.042117} {\bibfield  {journal} {\bibinfo  {journal} {Phys.
  Rev. E}\ }\textbf {\bibinfo {volume} {88}},\ \bibinfo {pages} {042117}
  (\bibinfo {year} {2013})}\BibitemShut {NoStop}%
\bibitem [{\citenamefont {Kelland}(1974)}]{Kelland}%
  \BibitemOpen
  \bibfield  {author} {\bibinfo {author} {\bibfnamefont {S.~B.}\ \bibnamefont
  {Kelland}},\ }\href {\doibase https://doi.org/10.1071/PH740813} {\bibfield
  {journal} {\bibinfo  {journal} {Aust. J. Phys.}\ }\textbf {\bibinfo {volume}
  {27}},\ \bibinfo {pages} {813} (\bibinfo {year} {1974})}\BibitemShut
  {NoStop}%
\bibitem [{\citenamefont {Siddharthan}\ and\ \citenamefont
  {Georges}(2001)}]{SG}%
  \BibitemOpen
  \bibfield  {author} {\bibinfo {author} {\bibfnamefont {R.}~\bibnamefont
  {Siddharthan}}\ and\ \bibinfo {author} {\bibfnamefont {A.}~\bibnamefont
  {Georges}},\ }\href {\doibase 10.1103/PhysRevB.65.014417} {\bibfield
  {journal} {\bibinfo  {journal} {Phys. Rev. B}\ }\textbf {\bibinfo {volume}
  {65}},\ \bibinfo {pages} {014417} (\bibinfo {year} {2001})}\BibitemShut
  {NoStop}%
\bibitem [{\citenamefont {Kitaev}(2003)}]{kitaev}%
  \BibitemOpen
  \bibfield  {author} {\bibinfo {author} {\bibfnamefont {A.}~\bibnamefont
  {Kitaev}},\ }\href {\doibase https://doi.org/10.1016/S0003-4916(02)00018-0}
  {\bibfield  {journal} {\bibinfo  {journal} {Annals of Physics}\ }\textbf
  {\bibinfo {volume} {303}},\ \bibinfo {pages} {2} (\bibinfo {year}
  {2003})}\BibitemShut {NoStop}%
\bibitem [{\citenamefont {Castelnovo}\ \emph {et~al.}(2005)\citenamefont
  {Castelnovo}, \citenamefont {Chamon}, \citenamefont {Mudry},\ and\
  \citenamefont {Pujol}}]{chamoncoloring}%
  \BibitemOpen
  \bibfield  {author} {\bibinfo {author} {\bibfnamefont {C.}~\bibnamefont
  {Castelnovo}}, \bibinfo {author} {\bibfnamefont {C.}~\bibnamefont {Chamon}},
  \bibinfo {author} {\bibfnamefont {C.}~\bibnamefont {Mudry}}, \ and\ \bibinfo
  {author} {\bibfnamefont {P.}~\bibnamefont {Pujol}},\ }\href@noop {}
  {\bibfield  {journal} {\bibinfo  {journal} {Physical review B}\ }\textbf
  {\bibinfo {volume} {72}},\ \bibinfo {pages} {104405} (\bibinfo {year}
  {2005})}\BibitemShut {NoStop}%
\bibitem [{\citenamefont {Balasubramanian}\ \emph {et~al.}(2022)\citenamefont
  {Balasubramanian}, \citenamefont {Bulmash}, \citenamefont {Galitski},\ and\
  \citenamefont {Vishwanath}}]{Bala}%
  \BibitemOpen
  \bibfield  {author} {\bibinfo {author} {\bibfnamefont {S.}~\bibnamefont
  {Balasubramanian}}, \bibinfo {author} {\bibfnamefont {D.}~\bibnamefont
  {Bulmash}}, \bibinfo {author} {\bibfnamefont {V.}~\bibnamefont {Galitski}}, \
  and\ \bibinfo {author} {\bibfnamefont {A.}~\bibnamefont {Vishwanath}},\
  }\href {https://arxiv.org/abs/2201.08856} {\bibfield  {journal} {\bibinfo
  {journal} {arXiv preprint arXiv:2201.08856}\ } (\bibinfo {year}
  {2022})}\BibitemShut {NoStop}%
\bibitem [{\citenamefont {Morampudi}\ \emph {et~al.}(2014)\citenamefont
  {Morampudi}, \citenamefont {von Keyserlingk},\ and\ \citenamefont
  {Pollmann}}]{Pollmann}%
  \BibitemOpen
  \bibfield  {author} {\bibinfo {author} {\bibfnamefont {S.~C.}\ \bibnamefont
  {Morampudi}}, \bibinfo {author} {\bibfnamefont {C.}~\bibnamefont {von
  Keyserlingk}}, \ and\ \bibinfo {author} {\bibfnamefont {F.}~\bibnamefont
  {Pollmann}},\ }\href {\doibase 10.1103/physrevb.90.035117} {\bibfield
  {journal} {\bibinfo  {journal} {Physical Review B}\ }\textbf {\bibinfo
  {volume} {90}} (\bibinfo {year} {2014}),\
  10.1103/physrevb.90.035117}\BibitemShut {NoStop}%
\bibitem [{\citenamefont {Verresen}\ \emph {et~al.}(2021)\citenamefont
  {Verresen}, \citenamefont {Lukin},\ and\ \citenamefont {Vishwanath}}]{Ruben}%
  \BibitemOpen
  \bibfield  {author} {\bibinfo {author} {\bibfnamefont {R.}~\bibnamefont
  {Verresen}}, \bibinfo {author} {\bibfnamefont {M.~D.}\ \bibnamefont {Lukin}},
  \ and\ \bibinfo {author} {\bibfnamefont {A.}~\bibnamefont {Vishwanath}},\
  }\href {\doibase 10.1103/physrevx.11.031005} {\bibfield  {journal} {\bibinfo
  {journal} {Physical Review X}\ }\textbf {\bibinfo {volume} {11}} (\bibinfo
  {year} {2021}),\ 10.1103/physrevx.11.031005}\BibitemShut {NoStop}%
\bibitem [{\citenamefont {Sacco}\ and\ \citenamefont {Wu}(1975)}]{SW}%
  \BibitemOpen
  \bibfield  {author} {\bibinfo {author} {\bibfnamefont {J.~E.}\ \bibnamefont
  {Sacco}}\ and\ \bibinfo {author} {\bibfnamefont {F.~Y.}\ \bibnamefont {Wu}},\
  }\href {\doibase 10.1088/0305-4470/8/11/013} {\bibfield  {journal} {\bibinfo
  {journal} {Journal of Physics A: Mathematical and General}\ }\textbf
  {\bibinfo {volume} {8}},\ \bibinfo {pages} {1780} (\bibinfo {year}
  {1975})}\BibitemShut {NoStop}%
\bibitem [{\citenamefont {Griffiths}(1967)}]{Griffiths}%
  \BibitemOpen
  \bibfield  {author} {\bibinfo {author} {\bibfnamefont {R.~B.}\ \bibnamefont
  {Griffiths}},\ }\href {\doibase 10.1063/1.1705219} {\bibfield  {journal}
  {\bibinfo  {journal} {Journal of Mathematical Physics}\ }\textbf {\bibinfo
  {volume} {8}},\ \bibinfo {pages} {478} (\bibinfo {year} {1967})},\ \Eprint
  {http://arxiv.org/abs/https://doi.org/10.1063/1.1705219}
  {https://doi.org/10.1063/1.1705219} \BibitemShut {NoStop}%
\bibitem [{\citenamefont {Kelly}\ and\ \citenamefont {Sherman}(1968)}]{KS}%
  \BibitemOpen
  \bibfield  {author} {\bibinfo {author} {\bibfnamefont {D.~G.}\ \bibnamefont
  {Kelly}}\ and\ \bibinfo {author} {\bibfnamefont {S.}~\bibnamefont
  {Sherman}},\ }\href {\doibase 10.1063/1.1664600} {\bibfield  {journal}
  {\bibinfo  {journal} {Journal of Mathematical Physics}\ }\textbf {\bibinfo
  {volume} {9}},\ \bibinfo {pages} {466} (\bibinfo {year} {1968})},\ \Eprint
  {http://arxiv.org/abs/https://doi.org/10.1063/1.1664600}
  {https://doi.org/10.1063/1.1664600} \BibitemShut {NoStop}%
\bibitem [{\citenamefont {Baxter}\ and\ \citenamefont {Wu}(1973)}]{BaxterWu}%
  \BibitemOpen
  \bibfield  {author} {\bibinfo {author} {\bibfnamefont {R.~J.}\ \bibnamefont
  {Baxter}}\ and\ \bibinfo {author} {\bibfnamefont {F.~Y.}\ \bibnamefont
  {Wu}},\ }\href {\doibase 10.1103/PhysRevLett.31.1294} {\bibfield  {journal}
  {\bibinfo  {journal} {Phys. Rev. Lett.}\ }\textbf {\bibinfo {volume} {31}},\
  \bibinfo {pages} {1294} (\bibinfo {year} {1973})}\BibitemShut {NoStop}%
\bibitem [{\citenamefont {Alcaraz}\ and\ \citenamefont {Xavier}(1997)}]{AX}%
  \BibitemOpen
  \bibfield  {author} {\bibinfo {author} {\bibfnamefont {F.}~\bibnamefont
  {Alcaraz}}\ and\ \bibinfo {author} {\bibfnamefont {J.}~\bibnamefont
  {Xavier}},\ }\href@noop {} {\bibfield  {journal} {\bibinfo  {journal}
  {Journal of Physics A: Mathematical and General}\ }\textbf {\bibinfo {volume}
  {30}},\ \bibinfo {pages} {L203} (\bibinfo {year} {1997})}\BibitemShut
  {NoStop}%
\bibitem [{\citenamefont {Yao}\ and\ \citenamefont {Kivelson}(2012)}]{YK}%
  \BibitemOpen
  \bibfield  {author} {\bibinfo {author} {\bibfnamefont {H.}~\bibnamefont
  {Yao}}\ and\ \bibinfo {author} {\bibfnamefont {S.~A.}\ \bibnamefont
  {Kivelson}},\ }\href {\doibase 10.1103/PhysRevLett.108.247206} {\bibfield
  {journal} {\bibinfo  {journal} {Phys. Rev. Lett.}\ }\textbf {\bibinfo
  {volume} {108}},\ \bibinfo {pages} {247206} (\bibinfo {year}
  {2012})}\BibitemShut {NoStop}%
\end{thebibliography}%

\appendix
\begin{widetext}

\section{Spin model mappings and analysis of phase transition}\label{spinmodelphase}
In this appendix, we discuss the less important, though still curious spin model mappings for the vertex models mentioned multiple times in the text.  The vertex models clearly exhibit $\mathbb{Z}_2$-symmetry breaking -- at zero temperature (where only $W(0)$ and $W(V_p)$ are nonzero), the classical ground states are the all dimer and no dimer configurations. 
In this section, we will explicitly construct classical spin models corresponding to these vertex models and argue for the existence of a phase transition for bipartite $G_D$.  We will also argue for a universal bound on the transition temperature of the spin models defined on nonbipartite lattices and develop a systematic high temperature expansion to estimate the transition temperatures for particular choices of models.  Numerical simulations indicated that there is no phase transition for nonbipartite $G_D$.

We first consider models where the dual lattice $G_D$ is bipartite. On the dual lattice $G_D$, we place two spins on the centers of each of the plaquettes to form the lattice $G_P$.  Therefore, for each edge of $G_D$, there are two spins on either side of it in $G_P$.  A dimer in $G_D$ corresponds to agreeing values of the spins, while no dimer corresponds to disagreeing values of the spins.  This construction was first pioneered by Baxter \cite{BaxterBook, SW, HintermannMerlini}.

It is clear that the even bond constraint is preserved under this mapping.  This follows from defining the bond variables $b_{ij} = s_i s_j$ with $i$ and $j$ surrounding a particular dimer.  Then the product of the bond variables around each vertex satisfies the gauge condition $\prod_{\circlearrowleft} b_{ij} = 1$, which implies the even dimer constraint.  Because of the global $\mathbb{Z}_2$ spin flip symmetry, each dimer configuration maps onto two spin configurations (this results in extra factors of 2 in the partition functions which we will neglect).

The effective Hamiltonian for the spin model will satisfy the property that the Boltzmann weights associated with each vertex configuration coincides with $W_p(k)$.  Denote the full Hamiltonian as $-\beta \mathcal{H} = \sum_{v} H_v$ where $v$ denotes a vertex in $G_D$ (i.e. center of plaquettes in $G_P$).  For a particular vertex $v$ with $p$ surrounding spins, denote by $S$ a subset of edges.  The operator that projects onto a state with edges in $S$ corresponding to bonds is
\begin{equation}
    P_v(S) = \frac{1}{2^p} \prod_{k \in S}(1+b_k) \prod_{k \notin S}(1-b_k).
\end{equation}
The operator projecting onto states with $n$ bonds is
\begin{equation}
    P_v(n) = \sum_{S \subset [p], |S| = n} P_v(S),
\end{equation}
and the total Hamiltonian is (assuming that dimer states projected onto $P_v(n)$ have normalized weights $W_v(n) > 1$)
\begin{equation}\label{eq:spinham}
    H_v = \sum_{n \in 2\mathbb{Z}} P_v(n) \log W_v(n),
\end{equation}
which therefore satisfies the even bond vertex rules imposed by the class of models we are studying.

Next, we note that since $W_v(n) = \langle \cos^n x \sin^{V_p - n} x \rangle$, we have $W_v(n) = W_v(V_p-n)$.  Therefore, the Hamiltonian can be written as
\begin{equation} \label{eq:classspinham}
    H_v = \frac{1}{2}\sum_{n \in 2\mathbb{Z}} \left(P_v(n) + P_v(V_p-n)\right) \log W_v(n).
\end{equation}
The above Hamiltonian is ferromagnetic (see Appendix \ref{bipphase}), and therefore the Griffiths-Kelly-Sherman (GKS) correlation inequalities \cite{Griffiths, KS} can be used to argue rather intuitively that the Hamiltonian spontaneously magnetizes at sufficiently low temperatures when $G_D$ is bipartite.  For a proof of this, the reader is referred to Appendix \ref{bipphase}. 

\subsection{Phase transition in non-bipartite models} \label{nonbipphase}

\begin{figure*}
    \centering
    \includegraphics[scale=0.5]{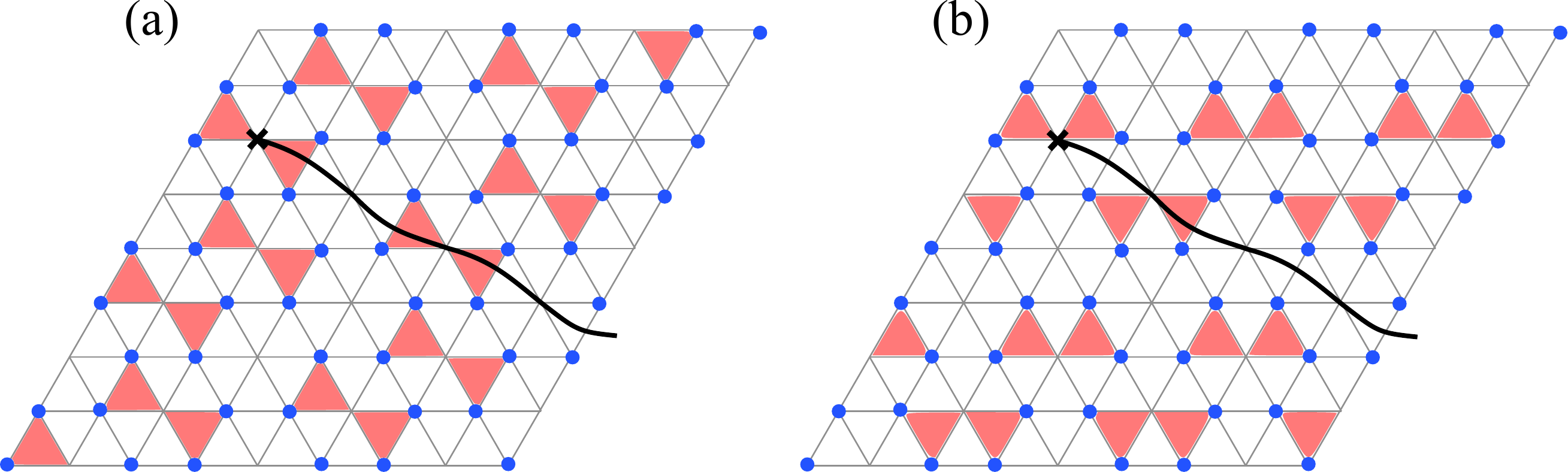}
    \caption{An example of two configurations which minimize the number of shaded triangles and satisfy the constraints listed in the text.  The string operator use to compute the magnetization is drawn.  There is an exponentially large number of such states, each of which can be related to one another by a loop of flips where the number of shaded and unshaded triangles around the loop are the same.}
   \label{fig:hightemp}
\end{figure*}

For non-bipartite dual lattices $G_D$, we argued that the vertex models are arrowed.  While there is no convenient mapping of these arrowed vertex models to spin models, the signed representation of the vertex weights is useful.  Recall that the partition function for the bipartite case can be written as
\begin{equation}
    Z_{\rm bip} = \sum_{s_v} e^{\sum_v H_v(s_v)}.
\end{equation}
As the vertex weights in the non-bipartite case are $W_p(n) = i^{p-n}\langle \cos^n x \sin^{p-n} x \rangle = i^{p-n} W_{p,\rm bip}(n)$, this can be represented by weighting each dimer by an additional factor of $-i$ at each vertex.  The partition function may be written using the bond variables as
\begin{equation}
    Z = \sum_{s_v} e^{\sum_v H_v(s_v)}e^{-i \frac{\pi}{4} \sum_v\left(b_1 + b_2 + \cdots + b_{p(v)}\right)},
\end{equation}
which satisfies the desired constraint since the relative weight between $b_i = 1$ and $b_i = -1$ is $-i$ at each vertex.  Because each dimer is counted twice, the partition function becomes
\begin{equation}
    Z = \sum_{s_v} e^{\sum_v H_v(s_v)}e^{-i \frac{\pi}{2} \sum_{e \in G_P} b_e},
\end{equation}
where $e$ denotes edges of $G_P$.  Writing this partition function in terms of $s_i$ variables, we find
\begin{equation} \label{eqn:partnonbip}
    Z \propto \sum_{s_v} \left(\prod_{i \in V_{\rm odd}} s_i\right) e^{\sum_v H_v(s_v)} = \left\langle  \prod_{i \in V_{\rm odd}} s_i \right \rangle_{\rm bip} Z_{\rm bip},
\end{equation}
where $V_{\rm odd}$ are the set of vertices of $G_P$ with odd degree and the subscript ``bip'' is to remind the reader that expectation values are taken under the thermal distribution of the bipartite model.  The magnetization of this model can be written as
\begin{equation}
    \langle s_k\rangle_{\rm non-bip} = \frac{\left\langle  s_k \prod_{i \in V_{\rm odd}} s_i \right \rangle_{\rm bip}}{\left\langle  \prod_{i \in V_{\rm odd}} s_i \right \rangle_{\rm bip}} \geq  \langle s_k\rangle_{\rm bip}
\end{equation}
where we used the GKS inequality \cite{Griffiths, KS}.  Because the magnetization in the non-bipartite model is greater than the magnetization in the bipartite model, the transition temperature must satisfy $T_{\rm non-bip} \geq T_{\rm bip}$.  If the equality was removed, this would show that the non-bipartite model was in an ordered phase of the extended statistical mechanical model given in Equation \ref{eqn:partnonbip}, thus implying that the system is gapped.

Next, we develop a high-temperature expansion which may be used to estimate the critical temperature for a particular model.  Namely, consider the case where $G_D$ is a triangular lattice.  Placing spins on the faces of the triangular lattice forms $G_P$, which is a honeycomb lattice.  We may construct the Hamiltonian in Equation \ref{eq:spinham} which is defined on $G_P$, but this particular model has an additional and convenient simplification.  In the next section, we show that a family of models shares this convenient property and can be described by similar high temperature expansions.  

We briefly outline this simplification, which has been previously used \cite{SW, BaxterBook}.  On the honeycomb lattice, the partition function of the vertex model is given by 
\begin{equation}
    Z = \sum_{\vec{s} = \{-1,1\}^N} \prod_{s_1^h,\ldots,s_6^h \in \hexagon} \cosh J\left(s_1^hs_2^h + s_2^hs_3^h + s_3^hs_4^h + s_4^hs_5^h + s_5^hs_6^h + s_6^hs_1^h\right),
\end{equation}
where $J$ is an adjustable constant.  In this representation, the no dimer and 6 dimer configurations have a weight $\cosh 6J$ while 2 and 4 dimer configurations have a weight $\cosh 2J$.  Adding additional ``ghost spins'' to the centers of the honeycomb lattice (forming a triangular lattice), we may write the partition function as
\begin{align}
    Z = \sum_{\vec{s} = \{-1,1\}^N} \exp\left(J \sum_{\triangle} s_i s_j s_k\right),
\end{align}
where the sum is over triangles of the resulting triangular lattice.  Since the ratio of the no bond/6 bond weights to that of the 2 bond/4 bond weights is $5$, we find that $\cosh 6J = 5 \cosh 2J$.  This model was solved by Baxter and Wu using Bethe ansatz methods \cite{BaxterWu, BaxterBook}.  The point $\cosh 6J = 5 \cosh 2J$ is a critical point of the model, which undergoes a second order phase transition associated with a global $\mathbb{Z}_4$ symmetry breaking.  This critical point is described by a four-state Potts model CFT \cite{AX}.

For the non-bipartite lattice, the partition function is now
\begin{align}
    Z = \sum_{\vec{s} = \{-1,1\}^N} \left(\prod_{m \in G_P} s_m\right) \exp\left(J \sum_{\triangle} s_i s_j s_k\right),
\end{align}
which is the expectation value of the product of spins over the honeycomb sublattice.  Using the identity
\begin{equation}
    \exp\left( J s_i s_j s_k \right) = \cosh J\left(1 + s_i s_j s_k \tanh J\right),
\end{equation}
We may then express this as a high-temperature expansion, in which we denote a configuration as a set of shaded triangles in the triangular lattice -- the partition function can be written as
\begin{equation}
    Z = \cosh^N J \sum_{C \in S} (\tanh J)^{N_C},
\end{equation}
where $S$ is the set of configurations where an \emph{odd} number of triangles are colored around each site of the honeycomb sublattice and an \emph{even} number of triangles are colored on the other sites.

A proxy for the ordering is the sublattice magnetization -- we compute $\langle s_i \rangle$, which gives
\begin{equation}
    \langle s_i \rangle = \frac{\sum_{C \in S_i} (\tanh J)^{N_C}}{\sum_{C \in S} (\tanh J)^{N_C}}
\end{equation}
where the set $S_i$ denotes the space of configurations with a defect at site $i$.  Note that we require the string to turn an even number of times over other sites apart from $i$.  This can be written conveniently as the expectation value of a string operator in the configuration space of the high temperature expansion.  Define the variable $\xi_i$ for triangle nearest to $i$, which equals $\tanh J$ if the triangle is uncolored and $\tanh^{-1} J$ if the triangle is colored.  Then,
\begin{equation}
    \langle s_i \rangle = \left \langle \prod_{k = i}^{\infty} \xi_k \right \rangle = \left \langle e^{(n_w - n_c) \log \tanh J}\right \rangle,
\end{equation}
where $n_w$ is the number of white triangles and $n_c$ is the number of colored triangles along the string.  Example of typical configurations favored at small $J$ are shown in Figure \ref{fig:hightemp}.

Let us compare this to the bipartite case.  The high temperature expansion is a sum over configurations which have an even number of colored triangles touching each site -- the only configuration favored at small $J$ is the configuration where no triangles are colored.  Then $n_w - n_c = N$ and the magnetization decays at $\langle s_i \rangle \sim e^{N \log \tanh^{-1} J}$.  In the non-bipartite case, the favored configurations at small $J$ have exactly $N/3$ triangles colored.  Therefore,we expect
\begin{equation}
    \langle s_i \rangle \sim e^{\left \langle n_w - n_c\right \rangle \log \tanh J} \sim e^{(N/3) \log \tanh^{-1} J}.
\end{equation}
which decays at a slower rate at large temperatures.  This points to the conclusion that the transition temperature should be strictly larger than that for the bipartite model.  In reality, numerics predicts the absence of a transition entirely.

\section{Phase transition in bipartite models} \label{bipphase}

In this appendix, we argue that the spin models (constructed in the previous appendix) corresponding to the dual vertex models spontaneously magnetize at sufficiently low temperatures.  The Hamiltonian $H_v$ (in Equation \ref{eq:classspinham}) is a polynomial in the variables $b_1, \cdots, b_p$.  To analyze this polynomial, we note that
\begin{align}
    P_v(n) + P_v(p-n) &= \sum_{S \subset [p], |S| = n} P_v(S) + P_v([p]\setminus S) \\
    &= \frac{1}{2^{V_p}} \sum_{S \subset [p], |S| = n} \prod_{k \in S}(1+b_k) \prod_{k \notin S}(1-b_k) + \prod_{k \in S}(1-b_k) \prod_{k \notin S}(1+b_k) \nonumber.
\end{align}
In the summand, any term involving a product of an odd number of $b_k$ variables will cancel, since the two terms in the summand are related by setting $b_k \to -b_k$.  Therefore, $H_v$ must be a polynomial containing terms with an even number of $b_k$ variables.  Next, invariance of the Hamiltonian under permutation of $b_k$ variables suggests that the coefficient of the term $\prod_{k \in S} b_k$ only depends on $|S|$.  It is more convenient to consider the polynomial setting $b_1 = b_2 = \cdots = b_p \triangleq b$,
\begin{equation}
    H_v(b_1 = b_2 = \cdots = b_p \triangleq b) = \frac{1}{2^p} \sum_{n \in 2\mathbb{Z}} \binom{p}{n} (1+b)^n (1-b)^{p-n} \log W_v(n).
\end{equation}
By the properties described above, if this polynomial has strictly positive coefficients, then the original polynomial $H_v$ has strictly positive coefficients (we ignore the constant term as it does not affect observables, as well as $\prod_{k \in [p]} b_k$, which is also an additive constant due to the gauge condition).  This is analytically verified using $W_v(n) = \langle \cos^n x \sin^{p - n} x \rangle$ up to $p = 50$ by a computer algebra system. 

We claim that this is sufficient to establish a phase transition in this class of models.  We utilize the well-known Griffiths-Kelly-Sherman (GKS) inequalities, which state that for a classical system of spins for a Hamiltonian $H = \sum_{S} J_S \prod_{i \in S} s_i$ with ferromagnetic interactions $J_S > 0$:
\begin{itemize}
    \item $\langle s_A \rangle \geq 0$ for $A$ any set of spins,
    \item $\langle s_A s_B \rangle - \langle s_A \rangle\langle s_B \rangle \geq 0$ for $A$ and $B$ any sets of spins.
\end{itemize}
In particular, the second condition implies that
\begin{equation}
    \frac{\partial \langle s_A \rangle}{ \partial J_B} = \langle s_A s_B \rangle - \langle s_A \rangle\langle s_B \rangle \geq 0.
\end{equation}
This means that the magnetization of the system $\langle s_i \rangle$ never decreases if additional ferromagnetic interactions are turned on.  This suggests that for a ferromagnetic Hamiltonian, we can tune some ferromagnetic interactions to zero until we reduce the system to a solvable model.  In the current context, $-\beta H = \beta \sum_v H_v$ is ferromagnetic in the original spin variables $s_i$, as $H_v$ is a polynomial in $s_i$ with positive coefficients.  Consider the Hamiltonian
\begin{equation}
    \widetilde{H}_v = c \sum_{i = 1}^{p} b_{i} b_{i+1} = c \sum_{i = 1}^{p} s_{i} s_{i+2},
\end{equation}
for $c>0$ and where we have indexed spins $s$ clockwise around the vertex and $b_i = s_i s_{i+1}$.   It is clear that $H_v$ can be deformed into $\widetilde{H}_v$ by decreasing ferromagnetic interactions in $H_v$, which follows from the fact that $H_v$ is a polynomial of products of even numbers of $b_k$, and the coefficients of $b_i b_j$ are positive for all $(i,j)$.

By the GKS inequality, the Hamiltonian $-\beta \widetilde{H} = \beta \sum_v \widetilde{H}_v$ now has the property that
\begin{equation}
    \langle s_i \rangle_{H} \geq \langle s_i \rangle_{\widetilde{H}}.
\end{equation}
Next, note that $G_P$ is bipartite.  To show this, suppose $G_P$ is not bipartite -- then it must contain an odd length cycle.  This odd length cycle encircles some number of nodes of $G_D$, and at least one of these nodes must have odd degree, contradicting the construction of $G_D$.  The spin-spin interactions $s_i s_{i+2}$ therefore connect nearest neighbors in one of the sublattices of $G_P$, which we henceforth write as $G_P = B_1 \cup B_2$.  The partition function of $\widetilde{H}$ is therefore
\begin{equation}
    Z(c, \widetilde{H}) = Z_{\rm{Ising}}(\beta c, B_1) \times Z_{\rm{Ising}}(\beta c, B_2).
\end{equation}
Thus calling $B$ the sublattice that $s_i$ is in,
\begin{equation}
    \langle s_i \rangle_{H} \geq \langle s_i \rangle_{\widetilde{H}} = \langle s_i \rangle_{B},
\end{equation}
At sufficiently low temperatures, a Peirels contour method can be applied to prove that the nearest-neighbor Ising model on $B$ magnetizes, indicating the presence of a symmetry broken phase at finite temperature.  Therefore $\langle s_i \rangle_{H} > 0$ at sufficiently low temperatures.  This establishes a more rigorous proof of the existence of a phase transition, which implies that the self dual point is at the phase transition point if there are only two phases.

\section{Class III: parent Hamiltonians for off-critical RK wavefunctions}\label{offcriticalHam}

The models in class I map onto balanced dimer models of statistical mechanics.  The models in class II are quantum vertex models where the number of dimers fluctuates.  In this appendix, we discuss what perturbing about the self dual point of the vertex model does to the original balanced dimer model.  We then construct frustration-free parent Hamiltonians corresponding to these statistical mechanics models.  In the process, we will need to construct a ``reverse BDVM duality.''  A similar approach has been used before to perturb about the quantum dimer model on a bipartite lattice and construct a gapped spin liquid phase, although the argument is field theoretical and not entirely microscopic \cite{YK}.

We start with a vertex model with vertex weights of the form $W_p(n) = \langle f_p(x)^n g_p(x)^{p-n} \rangle$.  Then the partition function of the vertex model is
\begin{equation}
    Z(k_1, \cdots, k_n) = \frac{2^N}{(2\pi)^n} \int_0^{2\pi} d^n\theta \prod_{p \in G_D} f_p(\theta_p)^{k_p} g_p(\theta_p)^{p-k_p}, 
\end{equation}
where $Z = \sum_{k_1,\cdots, k_n} Z(k_1, k_2, \cdots, k_n)$.  We now apply a reverse duality transformation.  Assume that $f$ and $g$ are $2\pi$-periodic and can be expanded in the Fourier series
\begin{equation}
    f_p(x) = \sum_{k \in \mathbb{Z}} \eta_p(k) e^{ikx} \hspace{0.4cm} g_p(x) = \sum_{q \in \mathbb{Z}} \xi_p(q) e^{iqx}.
\end{equation}
The partition function may be written in the form
\begin{equation}
    Z = \frac{2^N}{(2\pi)^n} \int_0^{2\pi} d^n\theta \prod_{(p,q) \in G_D} \left(f_p(\theta_p)f_q(\theta_q) + g_p(\theta_p)g_q(\theta_q)\right),
\end{equation}
Upon substituting the Fourier series representation, we find that
\begin{equation}
    Z = \frac{2^N}{(2\pi)^n} \int_0^{2\pi} d^n\theta \prod_{(p,q) \in G_D} \sum_{s_p, s_q} \left(\eta_p(s_p)\eta_q(s_q) + \xi_p(s_p)\xi_q(s_q)\right) e^{i \theta_p s_p + i \theta_q s_q}.
\end{equation}

We note that the variables $s_p$ are defined such that at node $p$, there are $V_p$ duplicate variables, one defined for each edge involving $p$.  To avoid notational ambiguity, we will add the subscript $s_p^{(e)}$ to denote the copy of the variable $s_p$ associated with edge $e = (p,q)$.  Notice that there will be two distinct variables $s_p^{(p,q)}$ and $s_q^{(p,q)}$ -- therefore, there are \emph{two} variables per edge. See Figure \ref{fig:inversedual} for clarity.  Then, the partition function may be regrouped into the form
\begin{equation}
    Z = \frac{2^N}{(2\pi)^n} \sum_{s_p^{(e)}}\int_0^{2\pi} d^n\theta \prod_{e = (p,q) \in G_D} \left(\eta_p(s_p^{(e)})\eta_q(s_q^{(e)}) + \xi_p(s_p^{(e)})\xi_q(s_q^{(e)})\right) e^{i \theta_p \sum_e s_p^{(e)}}.
\end{equation}
Performing the integral over $\theta_p$ gives
\begin{equation}
    Z = \frac{2^N}{(2\pi)^n} \sum_{s_p^{(e)}}\prod_{e = (p,q) \in G_D} \left(\eta_p(s_p^{(e)})\eta_q(s_q^{(e)}) + \xi_p(s_p^{(e)})\xi_q(s_q^{(e)})\right)\delta\left( \sum_e s_p^{(e)}\right).
\end{equation}

\begin{figure}
    \centering
    \includegraphics[scale=0.32]{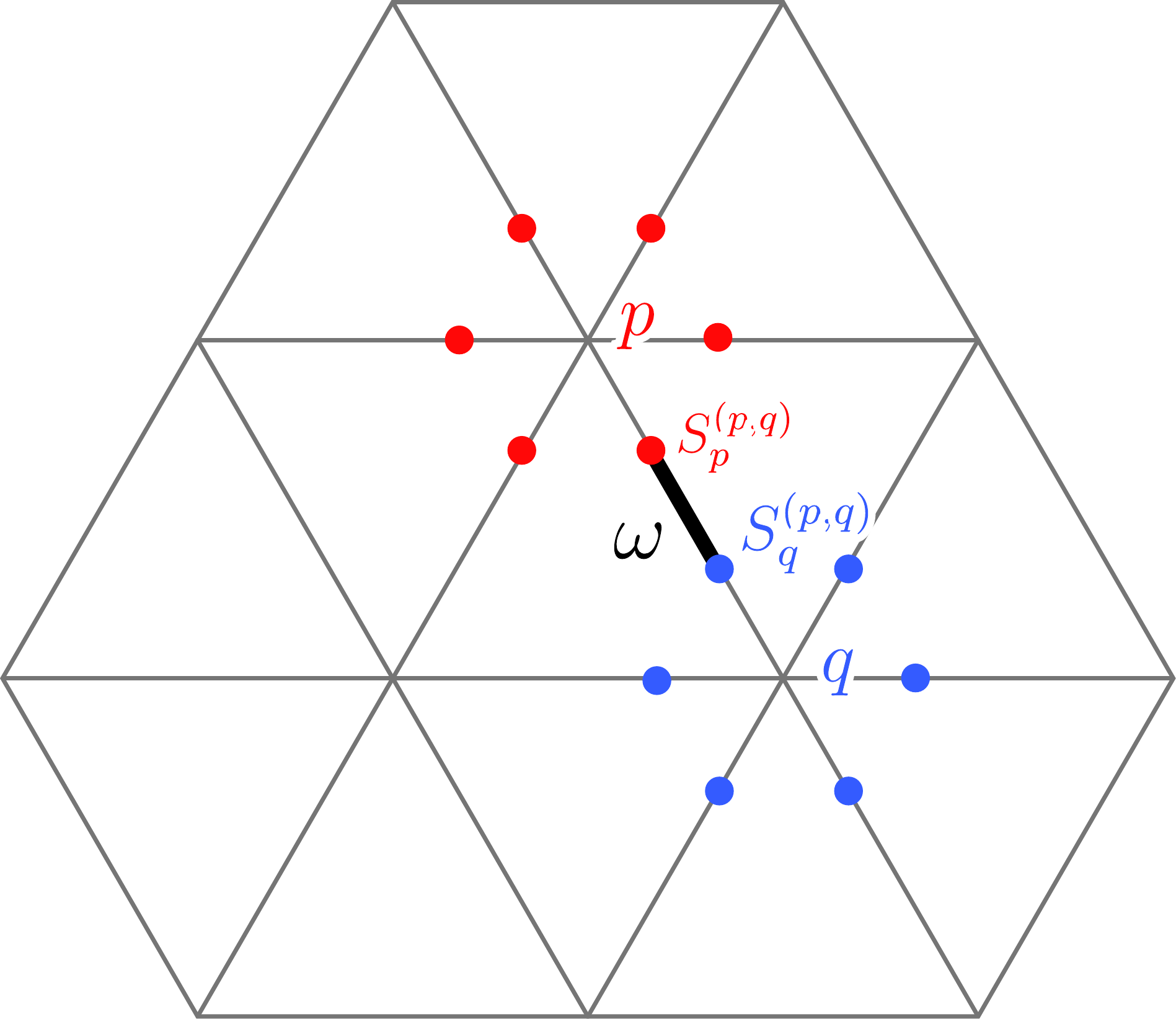}
    \caption{An illustration of the inverse duality mapping from a perturbed vertex model to a higher-spin model.  Each bond of the vertex is associated with two spins; each pair of spins carry a weight $\omega$.}
    \label{fig:inversedual}
\end{figure}

This dual spin model has a simple interpretation.  With each vertex on the original lattice we associate two spins, reminiscent of the spin doubling process introduced in the prior section.  The zero magnetization constraint is satisfied on the elementary plaquettes (i.e. for all variables $s_p^{(e)}$ with $p$ fixed), but the weighting is nonuniform.  For a fixed assignment of spins, the weight is 
\begin{align}
    W(s_p^{(e)}) &\propto \prod_{e = (p,q) \in G_D} \left(\eta_p(s_p^{(e)})\eta_q(s_q^{(e)}) + \xi_p(s_p^{(e)})\xi_q(s_q^{(e)})\right) \nonumber \\
    &\triangleq \prod_{e = (p,q) \in G_D} \omega (s_p^{(e)}, s_q^{(e)}),
\end{align}
where $\omega$ are the weights associated to each pair of doubled spins.  The remaining goal is to select Fourier coefficients of $f$ and $g$ such that the vertex model is perturbed away from the critical point and the corresponding model can be obtained as a low-energy effective Hamiltonian for a quantum spin model.

\subsection{Perturbing about Baxter-Wu criticality}\label{pertBW}

\begin{figure}
    \centering
    \includegraphics[scale=0.2]{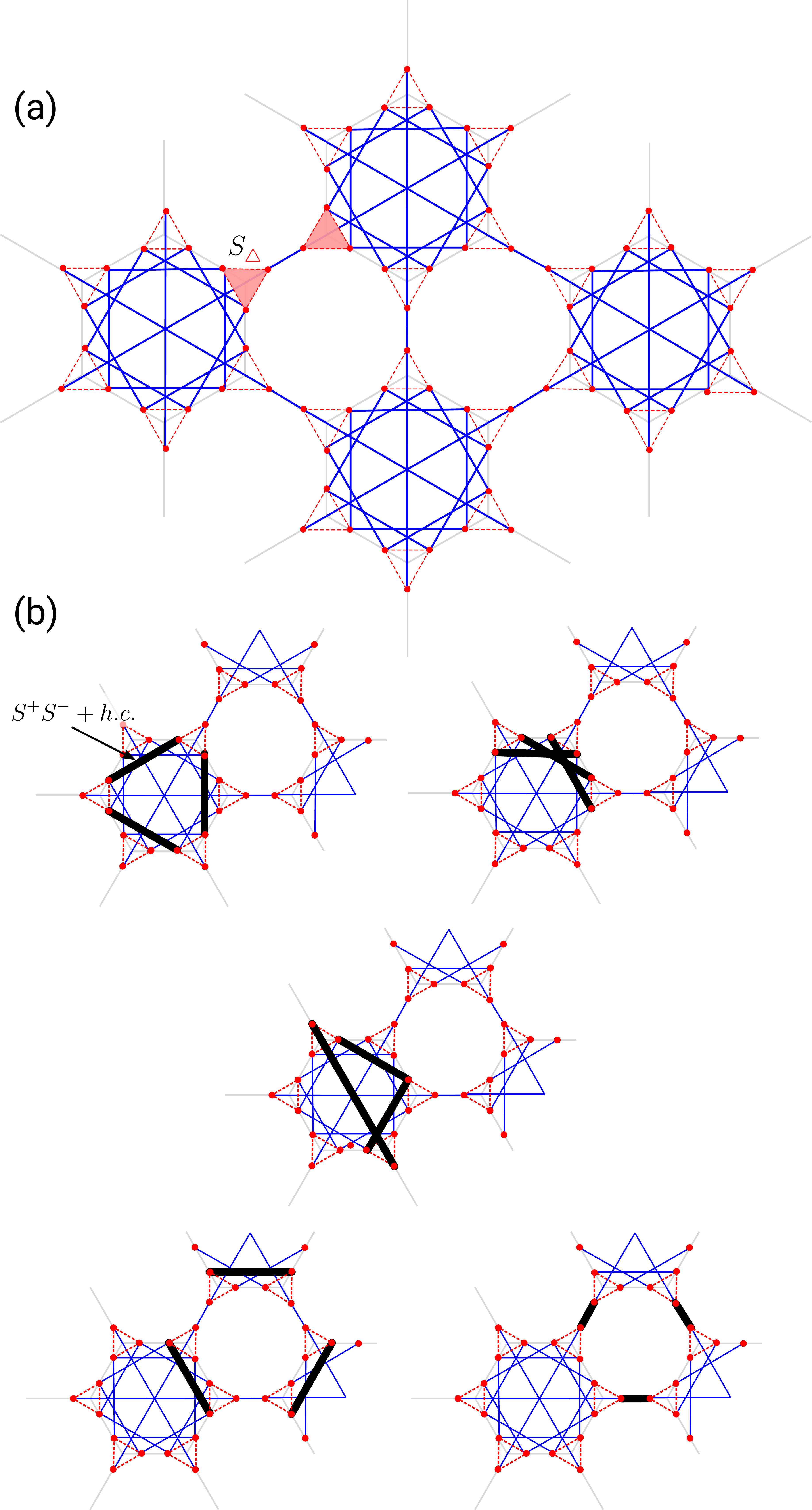}
    \caption{The lattice the quantum spin model is defined on (panel (a)) and spins are located on the red dots; the blue bonds indicate the terms in the Hamiltonian corresponding to boson hopping $S^+ S^- + \text{ h.c.}$ and the shaded triangles indicate locations of the pairwise constraints where the total $S_z$ restricted to $\pm 1, \pm 3$.  Additionally, the total $S_z$ of the 6 triangles around each star-shaped figure is zero.  The allowable ring exchange processes that occur at third order in perturbation theory (panel (b)) are indicated by blackened bonds.}
    \label{fig:spin32model}
\end{figure}
To illustrate this, we consider the triangular lattice vertex model.  We search for choices of $f$ and $g$ where the vertex weights are symmetric under dimer-no dimer exchange and have odd dimer weights equal to zero.  The simplest form of $f$ and $g$ that allows for this is
\begin{equation}
    f(x) = \cos x + \alpha \cos 3x \hspace{0.45cm} g(x) = \sin x - \alpha \sin 3x,
\end{equation}
where $\alpha$ is a constant.  The vertex weights as a function of $\alpha$ satisfy the desired conditions, as well as
\begin{equation}
    \frac{W_6(0)}{W_6(2)} = 5 \cdot \frac{1+3\alpha+9\alpha^2 + 6\alpha^3+9\alpha^4+\alpha^6}{1-5\alpha+9\alpha^2 - 10\alpha^3+9\alpha^4+\alpha^6}.
\end{equation}
This ratio is greater than or equal to $\approx 1.0223$ and less than or equal to $\approx 719$, so nearly any temperature in the effective Baxter-Wu spin model can be achieved by tuning the value of $\alpha$.  Note in particular that the ordered phase corresponds to $\alpha > 0$ and the disordered phase corresponds to $\alpha < 0$.  Writing $f(x) = \sum_s \eta_s e^{isx}$ and $g(x) = \sum_s \xi_s e^{isx}$, we see that the Fourier coefficients of $f$ and $g$ are non-zero for $s = -3, -1, 1, 3$; thus $s$ coincides with the values of $S_z$ for a spin-$3/2$ particle. The higher spin values ($3$ and $-3$) have Fourier components which are suppressed by factors of $\alpha$; in the limit $\alpha \to 0$, the spin-$3/2$ degree of freedom reduces to an effective spin-$1/2$ degree of freedom.  The non-zero values of the weights $\omega$ are 
\begin{align}
   \omega (-1, 1) &= \frac{1}{2} \hspace{1cm}  \omega (-3, -1) = \frac{\alpha}{2} \nonumber \\
   \omega (1, 3) &= \frac{\alpha}{2} \hspace{1cm}  \omega (-3, 3) = \frac{\alpha^2}{2}
\end{align}
These constraints encode rules that restrict the allowed types of exchange processes.  On the honeycomb lattice, the blue bonds are only allowed to have the above pairs of spin values.  Furthermore, on the remaining bonds, the sum of the spins must be left invariant under the exchange process.  

We first consider the Hamiltonian for a spin-$3/2$ easy-axis antiferromagnet on the honeycomb lattice.  A more convenient representation is to split each spin-$3/2$ particle into 3 spin-$1/2$ particles, each of which is placed on one of the edges touching the original site, forming a triangle.  The corresponding lattice with spins is shown in Figure \ref{fig:spin32model} panel (a).

The Hamiltonian we introduce is
\begin{equation}\label{eqn:ham32}
\mathcal{H} = \mathcal{H}_0 + J_\perp \sum_{\textcolor{blue}{\langle i, j \rangle_b}} \left(S_{\textcolor{blue}{i}}^+ S_{\textcolor{blue}{j}}^- + \text{h.c.}\right)
\end{equation}
with the unperturbed Hamiltonian
\begin{equation} \label{eqn:ham032}
    \mathcal{H}_0 = J_z \sum_{p} \left(\sum_{\textcolor{red}{\triangle} \in p} S_{\textcolor{red}{\triangle}}^z\right)^2 - J_z \sum_{\textcolor{red}{\triangle_i}, \textcolor{red}{\triangle_j}} \prod_{k \in \textcolor{red}{\triangle_i}} S_k^z \prod_{\ell \in \textcolor{red}{\triangle_j}} S_{\ell}^z.
\end{equation}

Let us first expound on the unperturbed Hamiltonian.  The first term imposes the constraint that the total spin around each hexagon is zero (as $S_\triangle$ corresponds to the z-component of an effective spin-$3/2$ ``superspin'' degree of freedom).  The second term imposes the constraint that the sum of the spin-$3/2$ ``superspins'' on the shaded red area shown in Figure \ref{fig:spin32model}[a] is either $0$ or $\pm 2$.  The energy cost of violating this constraint is always $J_z$.

When quantum fluctuations are added in, the extensive degeneracy is split by several possible ring exchange processes.  At third order in perturbation theory, the types of ring exchanges are illustrated in Figure \ref{fig:spin32model}[b].  The first ring exchange on the first row and the ring exchanges on the third row are equivalent to effective ring exchanges around a hexagon of spin-$3/2$ particles, if each triangle of spins is identified with a spin-$3/2$ degree of freedom.  The second ring exchange on the first row is equivalent to an exchange around a triangle of spin-$3/2$ particles, and the exchange on the second row is equivalent to two pairs of exchanges of spin-$3/2$ particles.  The amplitudes of the ring exchanges are calculated in Appendix \ref{ringexchcoeffs}.  Let us restrict the Hamiltonian to one particular ring exchange process:
\begin{equation}
 \mathcal{H} \supseteq  t \ket{C_1} \bra{C_2} + t \ket{C_2} \bra{C_1},
\end{equation}
where $t<0$ is the amplitude of this process in perturbation theory and $\ket{C_2}$ is related to $\ket{C_1}$ by a ring exchange.  Next, consider the \emph{classical} spin configurations $C_1$ and $C_2$ with Boltzmann weights $W(C_1)$ and $W(C_2)$.  If we add the following terms to the Hamiltonian (renaming it $\mathcal{H}'$):
\begin{equation}
 \mathcal{H}' \supseteq  t \ket{C_1} \bra{C_2} + t \ket{C_2} \bra{C_1} - t \sqrt{\frac{W(C_2)}{W(C_1)}}\ket{C_1} \bra{C_1} - t \sqrt{\frac{W(C_1)}{W(C_2)}}\ket{C_2} \bra{C_2}, 
\end{equation}
then the ground state in this two-state Hilbert space becomes
\begin{equation}
    \ket{\text{GS}} \propto \sqrt{W(C_1)} \ket{C_1} + \sqrt{W(C_2)} \ket{C_2}.
\end{equation}
If the Hamiltonian is expressed as a sum of ring exchange terms with additional projectors added, then the RK ground state satisfies
\begin{equation}
    \ket{\text{GS}} = \sum_i \sqrt{W(C_i)} \ket{C_i},
\end{equation}
where $C_i$ satisfies the easy-axis constraints.  This follows from the fact that the Hamiltonian is a sum of projectors, and the above state is annihilated by each of these projectors, rendering it the ground state.  The full Hamiltonian can be written in the form
\begin{equation}
    \mathcal{H} = \mathcal{H}_0 + J_\perp \sum_{\textcolor{blue}{\langle i, j \rangle_b}} \left(S_{\textcolor{blue}{i}}^+ S_{\textcolor{blue}{j}}^- + \text{h.c.}\right) + \frac{J_\perp^3}{J_z^2} \sum_{\hexagon, C} \gamma_C \prod_{j \in \hexagon} \prod_{k \in \triangle} P(C_{jk}),
\end{equation}
where the last term sums over projection operators onto easy-axis classical configurations $C$ around a given hexagon weighted by $\gamma_C$ (the value of the classical spin with index $k = 1,2,3$ in a triangle associated with a hexagonal cluster of triangles cluster $j = 1,\ldots,6$ is denoted by $C_{jk}$).  The expression for $\gamma_C$ is given by
\begin{equation}
    -\gamma_C = \sum_{C' \in R(C)} t_{C, C'} \sqrt{\frac{W(C')}{W(C)}} = \sum_{C' \in R(C)} t_{C, C'} \alpha^{\frac{1}{2}(n_{C'} - n_C)}
\end{equation}
where $R(C)$ is the set of all the configurations connected to $C$ via ring exchanges.  Furthermore, $n_C$ is the number of $S_z = 3/2$ ``superspins'' in $C$.  The second equality follows from the fact that the weights $\omega$ are proportional to $\alpha$ raised to the power of the number of $S_z = 3/2$ ``superspins''.  One can show that when the last two terms in the Hamiltonian above are treated perturbatively, the Hamiltonian can be expressed in the form
\begin{equation}
    \mathcal{H} \propto \sum_{C,C'} t_{C,C'}\left( \ket{C} \bra{C'} + \ket{C'} \bra{C} - \sqrt{\frac{W(C')}{W(C)}} \ket{C} \bra{C} - \sqrt{\frac{W(C)}{W(C')}} \ket{C'} \bra{C'}\right)
\end{equation}
where each pair $(C,C')$ are related by a single ring exchange.  This is frustration free as it is a sum of projectors, and the ground state is therefore $\ket{\text{GS}} = \sum_i \sqrt{W(C_i)} \ket{C_i}$.

There exists an alternative and more general construction of such a model. First, consider three sheets of a honeycomb lattice of spin-$1/2$ particles, such that each triple of points forms an effective spin-$3/2$ degree of freedom.  The easy axis Hamiltonian is $\mathcal{H}_0$ as before, but the interlayer hopping perturbation $J_\perp \sum_{\textcolor{blue}{\langle i, j \rangle_b}} \left(S_{\textcolor{blue}{i}}^+ S_{\textcolor{blue}{j}}^- + \text{h.c.}\right)$ is modified in the following way:  the bonds $\langle i, j \rangle_b$ are chosen so that a pair of triples share \emph{at most one bond}.  This constraint is required to avoid terms appearing at second order perturbation theory which corresponds to a ring exchange over a pair of bonds connecting the same two triplets.  These terms gap the model but do not introduce long-range entanglement since such processes lock the exchange of spin to within each hexagon.

Next, we would like to comment on the topological sectors in this model.  On the honeycomb lattice quantum dimer model, the number of topological sectors scales with the size of the system.  Here, due to additional exchange processes between non-nearest neighbor sites, we expect that the number of topological sectors be constant, mimicking the triangular lattice quantum dimer model.  However, there is an additional subtlety: it can be shown that the \emph{parity} of the number of $S_z = -3, 3$ particles cannot be changed by any local ring exchanges.  This parity constraint enhances the total number of topological sectors by a factor of $2$.  Therefore, we need to compute operator expectation values in the even and odd parity sectors: this can be facilitated by noting that
\begin{equation}
    Z_{e,o}(\alpha) = \frac{1}{2}\left(Z(\alpha) \pm Z(-\alpha)\right) = \frac{1}{2}\left(Z_{\text{BW}}(J_+) \pm Z_{\text{BW}}(J_-)\right)
\end{equation}
where $Z_{\text{BW}}$ is the partition function of the Baxter-Wu model and the couplings $J_{\pm}$ satisfy
\begin{equation}
    \frac{\cosh 6J_{\pm}}{\cosh 2J_{\pm}} = 5 \cdot \frac{1\pm 3\alpha+9\alpha^2 \pm 6\alpha^3+9\alpha^4+\alpha^6}{1\mp 5\alpha+9\alpha^2 \mp 10\alpha^3+9\alpha^4+\alpha^6}.
\end{equation}
Since $Z(\alpha) > Z(-\alpha)$ for $\alpha > 0$, expectation values of operators in the even and odd sectors are dominated by those in the ordered phase of the Baxter-Wu model.  Thus, the RK wavefunctions in the even and odd sectors are ordered.

Finally, we note that the choices of $f$ and $g$ presented above are not the only possibility (rather, just the simplest).  Surprisingly though, if we add higher harmonics to $f$ and $g$ this procedure does not properly perturb about the critical point, but rather corresponds to perturbations outside of the Baxter-Wu phase diagram.

\subsection{Other models}\label{pertBFG}

It is natural to wonder whether one can achieve topologically ordered phases as well.  We thus apply this method to the BFG model.  We make the choices $f(x) = \cos x + \alpha \cos 3x$ and $g(x) = i\sin x - i\alpha \sin 3x$.  The corresponding weights (symmetric under interchanging both arguments) are
\begin{align}
   \omega (1, 1) &= \frac{1}{2} \hspace{1cm}  \omega (3, -1) = \frac{\alpha}{2} \nonumber \\
   \omega (1, -3) &= \frac{\alpha}{2} \hspace{1cm}  \omega (-1, -1) = \frac{1}{2} \nonumber \\
   \omega (-3, 3) &= \frac{\alpha^2}{2} \hspace{1cm}  \omega (3, 3) = \frac{\alpha^2}{2}
\end{align}
Numerics from Section \ref{numerics} indicates that there is in fact no transition temperature, and the vertex model is always in a disordered phase.  This model is identical to the spin-$3/2$ model we previously studied, except that the second term in $\mathcal{H}_0$ from Equation \ref{eqn:ham032} constrains the easy axis spins on the bonds to sum to $\pm 1$ or $\pm 3$:
\begin{equation}
    \mathcal{H}_0 = J_z \sum_{p} \left(\sum_{\textcolor{red}{\triangle} \in p} S_{\textcolor{red}{\triangle}}^z\right)^2 + J_z \sum_{\textcolor{red}{\triangle_i}, \textcolor{red}{\triangle_j}} \prod_{k \in \textcolor{red}{\triangle_i}} S_k^z \prod_{\ell \in \textcolor{red}{\triangle_j}} S_{\ell}^z.
\end{equation}
However, instead of supporting an ordered phase for non-zero $\alpha$, this Hamiltonian supports a featureless disordered phase, which we believe to be a spin liquid phase.

Finally, we note that this procedure may allow one to construct quantum spin models with gapless phases.  For example, we consider a dual lattice with vertices of coordination number 8, an example of which is the union jack lattice.  For vertices of coordination number 8, the vertex weights can be described by two parameters $u = W_8(0)/W_8(4)$ and $v = W_8(2)/W_8(4)$.  Wegner's duality predicts a self-dual line in this model, which we conjecture is a critical line, corresponding to $u - 4v = 5$.  This line can be exactly accessed via the choice of $f$ and $g$
\begin{equation}
    f(x) = \cos x + \alpha \cos 7x \hspace{0.45cm} g(x) = \sin x - \alpha \sin 7x.
\end{equation}
The layered spin model associated with these deformations can be constructed in an analogous way to the triangular lattice vertex model, and will consist of 7 layers of planar spin-1/2 Hamiltonians with interlayer XY exchanges.  However, since this Hamiltonian explores the self dual line as a function of $\alpha$, we expect it to support a gapless phase, which is a subject of future work.

\section{Ring exchanges of spin-$3/2$ model} \label{ringexchcoeffs}

Here, we apply standard perturbation theory and present all of the amplitudes for the ring exchange processes of the spin-$3/2$ model defined in Appendix \ref{pertBW}.  Starting from a Hamiltonian $\mathcal{H} = \mathcal{H}_0 + V$ and defining the Green's function $G_0(E) = (E - \mathcal{H}_0)^{-1}$, the effective Hamiltonian for the ground state manifold at third order in perturbation theory is
\begin{equation}
    \mathcal{H}_{\text{eff}} = \mathcal{P} V G_0(E_0) V G_0(E_0) V \mathcal{P}
\end{equation}
where $\mathcal{P}$ is a Gutzwiller projection onto the ground state manifold.  Fortunately, for the models we consider, the Gutzwiller projection commutes with the ring exchanges, and thus the effective Hamiltonian is exact. Referring to Figure \ref{fig:spin32model} panel (b), we may write down the amplitudes of the ring exchange processes.  For the bottom right subpanel, the amplitude is
\begin{equation}
   \bra{C} \mathcal{H}_0 \ket{C'} = 6 \cdot \frac{J_\perp}{4} \cdot \frac{1}{2 J_z} \cdot  \frac{J_\perp}{4} \cdot \frac{1}{2 J_z} \cdot \frac{J_\perp}{4} = \frac{3 J_\perp ^3}{128 J_z^2},
\end{equation}
where the spins in $C$ and $C'$ associated with the blackened bonds in Figure \ref{fig:spin32model} panel (b) alternate as one traverses clockwise.  All other amplitudes are zero.  For the bottom left subpanel, the amplitude is also $\bra{C} \mathcal{H}_0 \ket{C'} = \frac{3  J_\perp ^3}{128 J_z^2}$, where each pair of spins associated with a blackened bond in $C$ need to be antiparallel.

Next we move to the top two subpanels, whose ring exchange amplitudes are equal.  The only requirement for the ring exchange to occur is that each pair of spins associated with a black bond be antiparallel; the amplitude is still $\bra{C} \mathcal{H}_0 \ket{C'} = \frac{3  J_\perp ^3}{128 J_z^2}$.  The ring exchange on the middle subpanel is of a slightly different nature: exchanges on two of the bonds create an effective furthest-neighbor exchange.  Thus, this process is equivalent to a pair of exchanges of spins which appears at third order.  The amplitude for this process is
\begin{equation}
   \bra{C} \mathcal{H}_0 \ket{C'} = 3 \cdot \frac{J_\perp}{4} \cdot \frac{1}{2 J_z} \cdot  \frac{J_\perp}{4} \cdot \frac{1}{2 J_z} \cdot \frac{J_\perp}{4} = \frac{3 J_\perp ^3}{256 J_z^2}.
\end{equation}
Next, we construct the following operator:
\begin{equation}
   \mathcal{P}_{\hexagon} = \prod_{\triangle} \left[\alpha (\mathcal{P}_{\triangle}(-3)+\mathcal{P}_{\triangle}(3)) + (\mathcal{P}_{\triangle}(-1)+\mathcal{P}_{\triangle}(1))\right],
\end{equation}
and the inverse operator
\begin{equation}
   \mathcal{P}^{-1}_{\hexagon} = \prod_{\triangle} \left[\alpha^{-1} (\mathcal{P}_{\triangle}(-3)+\mathcal{P}_{\triangle}(3)) + (\mathcal{P}_{\triangle}(-1)+\mathcal{P}_{\triangle}(1))\right].
\end{equation}
These operators are written in terms of $\mathcal{P}_{\triangle}(s)$ which projects onto spin configurations on a given triplet of spins whose z-components sum to $s$.  The full $\mathcal{P}_{\hexagon}$ operator computes the Boltzmann weight associated with a hexagon of spins, weighting each triplet of spins summing to $\pm 3$ by a factor of $\alpha$.

The additional terms we add to the Hamiltonian associated with the ring exchange process exchanging configurations $C$ and $C'$ is then
\begin{equation}
    \mathcal{H}' \supseteq \mathcal{P}(C)
    \left[\prod_{\boldsymbol{\langle i,j\rangle}}\left(S_i^+S_j^- + S_i^- S_j^+\right)\right]
    \mathcal{P}_{\hexagon}
    \mathcal{P}(C') \left[\prod_{\boldsymbol{\langle i,j\rangle}}\left(S_i^+S_j^- + S_i^- S_j^+\right)\right] \mathcal{P}^{-1}_{\hexagon} \mathcal{P}(C) + \left[C' \leftrightarrow C\right]
\end{equation}
where $\mathcal{P}(C)$ and $\mathcal{P}(C')$ project onto configurations $C$ and $C'$.  The notation $\boldsymbol{\langle i,j\rangle}$ indicates that the product is over the blackened bonds which connect the two configurations.  In generality, the full Hamiltonian can be written as
\begin{equation}
    \mathcal{H}' = \mathcal{H}_0 + J_\perp \sum_{\textcolor{blue}{\langle i, j \rangle_b}} \left(S_{\textcolor{blue}{i}}^+ S_{\textcolor{blue}{j}}^- + \text{h.c.}\right) + 
     \sum_{(C,C')} t_{C,C'} \mathcal{P}(C)
    \left[\prod_{\boldsymbol{\langle i,j\rangle}}\left(S_i^+S_j^- + \text{h.c.}\right)\right]
    \mathcal{P}_{\hexagon}
    \mathcal{P}(C') \left[\prod_{\boldsymbol{\langle i,j\rangle}}\left(S_i^+S_j^- + \text{h.c.}\right)\right] \mathcal{P}^{-1}_{\hexagon} \mathcal{P}(C). 
\end{equation}

\end{widetext}
\end{document}